\RequirePackage{pdf14}

\documentclass[a4paper,11pt]{article}
\pdfoutput=1

\usepackage{array,setspace,mathrsfs,amsfonts,amsmath,yfonts,dsfont,bbm,colonequals,amscd,overpic}
\usepackage{relsize,suffix,mathtools,cancel,bbm,tikz-cd}
\usepackage{subcaption}
\usepackage{multirow}

\usetikzlibrary{positioning}
\usetikzlibrary{chains}
\usetikzlibrary{arrows,fit,decorations.pathreplacing}
\tikzstyle{every picture}+=[remember picture]
\tikzstyle{na} = [baseline=-.5ex]

\usepackage{jheppub} 
\usepackage{enumerate}

\usepackage[T1]{fontenc} 
\usepackage{subcaption}
\graphicspath{ {graphs/} }

\makeatletter
\newcommand{\thickhline}{%
    \noalign {\ifnum 0=`}\fi \hrule height 1pt
    \futurelet \reserved@a \@xhline
}
\makeatother

\newcommand{\be}{\begin{equation}}
\newcommand{\ee}{\end{equation}}

\newcommand{\ba}{\begin{equation}\begin{aligned}}
\newcommand{\ea}{\end{aligned}\end{equation}}

\newcommand{\Df}{\Delta_{\phi}}
\newcommand{\cG}{\mathcal{G}}
\newcommand{\cGT}{\widetilde{\cG }}

\newcommand{\cO}{\mathcal{O}}

\newcommand{\zTr}{\mbox{$\frac{z}{z-1}$}}

\newcommand{\half}{\frac{1}{2}}

\newcommand{\btau}{\bar{\tau}}
\newcommand{\bq}{\bar{q}}
\newcommand{\bh}{\bar{h}}
\newcommand{\bc}{\bar{c}}
\newcommand{\DeltaVir}{\Delta_V}
\newcommand{\DeltaU}{\Delta_U }

\colorlet{darkblue}{blue!70!black}
\colorlet{darkgreen}{green!70!black}

\DeclareMathOperator*{\dDisc}{dDisc}

\newcommand{\Dmax}{\rho_{\rm max}}
\newcommand{\Density}{\rho}

\abstract{
We establish a precise relation between the modular bootstrap, used to constrain the spectrum of 2D CFTs, and the sphere packing problem in Euclidean geometry. The modular bootstrap bound for chiral algebra $U(1)^c$ maps exactly to the Cohn-Elkies linear programming bound on the sphere packing density in $d=2c$ dimensions. We also show that the analytic functionals developed earlier for the correlator conformal bootstrap can be adapted to this context. For $c=4$ and $c=12$, these functionals exactly reproduce the ``magic functions'' used recently by Viazovska \cite{viazovska8} and Cohn \textit{et~al.}~\cite{viazovska24} to solve the sphere packing problem in dimensions 8 and 24. The same functionals are also applied to general 2D CFTs, with only Virasoro symmetry. In the limit of large central charge, we relate sphere packing to bounds on the black hole spectrum in 3D quantum gravity, and prove analytically that any such theory must have a nontrivial primary state of dimension $\Delta_0 \lesssim c/8.503$.

}

\title{\boldmath \LARGE
Sphere Packing and Quantum Gravity
 }

\preprint{YITP-SB-19-11}

\author[a]{Thomas Hartman,}
\author[b,c]{Dalimil Maz\'a\v{c},}
\author[b,d]{Leonardo Rastelli}

\affiliation[a]{Department of Physics, Cornell University, Ithaca, New York, U.S.A.}
\affiliation[b]{C. N. Yang Institute for Theoretical Physics, Stony Brook University,\\
Stony Brook, NY 11794, U.S.A.}
\affiliation[c]{Simons Center for Geometry and Physics, Stony Brook University,\\
Stony Brook, NY 11794, U.S.A.}
\affiliation[d]{CERN, Theoretical Physics Department, 1211 Geneva 23, Switzerland}

{
\small

\begin{document}
\maketitle
\flushbottom

% !TEX root = ../ModularBootstrapV3.tex

\section{Introduction}\label{sec:Introduction}

Charting the space of  quantum field theories is a central task of theoretical physics, which has received renewed impetus with the modern resurgence of the conformal bootstrap program.  
Theories that live at the {\it boundary} of theory space  (because, {\it e.g}, they attain the largest allowed value of a certain operator dimension or central charge)
are prime targets for the bootstrap, as they are often amenable to precise analytical or numerical study. 

Thanks to holographic duality, statements about  the space of conformal field theories (CFTs) are equivalent to statements about the landscape of quantum gravity theories  in asymptotically anti de Sitter (AdS) space. 
The tight consistency conditions of CFT are expected to translate into non-obvious consistency requirements on the AdS side.  To wit,
a low-energy effective theory in AdS with arbitrarily prescribed matter content and symmetries  is in danger of belonging to the ``swampland'', that is,  of 
not admitting a non-perturbative completion. Can this intuition be made precise, leveraging the recent advances in the bootstrap program?

A fundamental question of this kind is
  whether  ``pure''  AdS gravity is a consistent theory, or whether instead 
 new degrees of freedom below the Planck scale (in addition to multi-gravitons) are needed  for non-perturbative consistency.  Via AdS/CFT, the quest for pure gravity can be phrased as follows.
One is looking  for a sequence of unitarity CFTs, with increasing number $N$ of degrees of freedom (as measured, for example, by the normalization $C_T$ of the stress tensor), such that for  $N \to \infty$ the only operators with finite scaling dimension are multi-trace composites of the stress tensor, and  correlation functions are of mean-field type, {\it i.e.}, they obey large $N$ factorization. One can also relax these assumptions slightly by allowing for a  {\it finite} number of additional single-trace operators whose dimensions remain bounded in the large $N$ limit. 

Black holes, and quantum gravity more generally, seem to live near the edge of theory space, as a consequence of the large hierarchy between the Planck scale and the low-energy effective theory. Pure gravity, or, if pure gravity does not exist, the theory of gravity with the largest gap, lives exactly at the edge, so it could be particularly amenable to bootstrap. Black holes also suggest a UV/IR connection in quantum gravity, whose implications on the CFT side are largely unexplored. For example, the weak gravity conjecture \cite{ArkaniHamed:2006dz}, motivated by properties of extremal black holes, translates into constraints on the spectrum of charged states in large-$N$ CFTs, with no known origin in quantum field theory.

These questions are particularly sharp for 3D AdS gravity, dual to 2D CFT, because multigraviton states in AdS$_3$ map to the Virasoro module of the identity in CFT$_2$,  which is a key technical simplification.
Consider again the question about  ``pure gravity''.  A natural strategy to look for 2D CFTs dual to pure gravity (or to rule them out) is to explore the {boundary} of theory space characterized by the largest allowed {\it gap} -- the largest dimension of the first non-trivial Virasoro primary.  The simplest (though by no means the only)  set of constraints on the gap arise from modular invariance of the CFT partition function. This is the ``modular bootstrap'' program pioneered by Hellerman \cite{Hellerman:2009bu} and pursued by several authors \cite{Friedan:2013cba,Collier:2016cls,Afkhami-Jeddi:2019zci}.  

In this work, we establish a precise connection between the modular bootstrap in 2D CFT  and the sphere packing problem in Euclidean geometry. The central question in sphere packing 
is finding the densest configuration of identical, non-overlapping spheres in $\mathbb{R}^d$. It is surprisingly deep, with connections to diverse areas from number theory to cryptography  \cite{ConwaySloane}.
For $d=2$, the answer is the honeycomb lattice, a result proved rigorously by T\'{o}th in 1940. For $d=3$, the answer is the face-centered cubic lattice. This was conjectured by Kepler four centuries ago, and famously proved by Hales in 1998 \cite{hales2005proof}. The epic proof fills hundreds of pages, relies on extensive use of computers to exhaustively check special cases, and took over a decade to be fully verified by a 22-person team \cite{hales2017formal}. 

In a remarkable paper in 2016, Viazovska solved the case $d=8$ \cite{viazovska8}, building on work of Cohn and Elkies \cite{cohn1,cohn2}. The answer is the $E_8$ root lattice, $\Lambda_8$, and the proof is simple and elegant. Viazovska's proof was immediately extended to $d=24$ \cite{viazovska24}, where the densest packing is the Leech lattice, $\Lambda_{24}$. The proofs of Hales and Viazovska both rely essentially on the method of linear programming, which is used (either on a computer or analytically) to rule out the existence of denser sphere packings. Cohn and Elkies conjectured the existence of `magic functions' which could be used to prove optimality of the $\Lambda_8$ and $\Lambda_{24}$ lattices, and gave overwhelming numerical evidence for their existence. Viazovska's breakthrough was to devise a method to construct magic functions analytically.

In hearing of a relation between sphere packings and modular bootstrap, one's 
first thought might be that lattices will provide the key connection.  After all, a $d$-dimensional lattice $\Lambda$ defines both a sphere packing  and a 2D CFT (the theory of $d$ free bosons compactified on $\Lambda$). However, this does not appear to be either a very natural or very useful connection. 
The compactified boson CFT depends on additional data -- it admits  a continuous moduli space,
more naturally related to the geometry of a $2d$-dimensional Lorentzian lattice than to the geometry of a $d$-dimensional Euclidean lattice. And
lattice sphere packings are very special, indeed it is widely believed that in sufficiently high dimension the densest packings are  not  lattice packings.

  The  relation that we establish in this work is more surprising. We relate the  spinless\footnote{This means that we set the angular  potential of the partition function to zero.} modular bootstrap
    of {\it general} CFTs with central charge $c$  to the {\it general} 
   sphere packing problem in $\mathbb{R}^{2c}$. The key fact is that both problems can be addressed by linear programming methods.
  The connection is  immediate  if one assumes that the CFT has chiral algebra $U(1)^c$. The  spinless modular bootstrap of such a CFT
    can be directly translated  to the Cohn-Elkies  linear programming approach to sphere packing in   $\mathbb{R}^{2c}$. If a certain technical conjecture holds (which can be verified in many cases), the two problems are fully mathematically equivalent. On the other hand, the modular bootstrap  of greater physical interest is for CFTs whose chiral algebra is just the Virasoro algebra.  The Virasoro modular bootstrap is 
    not directly equivalent to the  Cohn-Elkies problem, but can be formulated in a very similar way. 
    
Our main observation  is that the  {\it same} analytic functionals can be used to establish rigorous bounds for the Cohn-Elkies problem, the closely related $U(1)^c$ modular bootstrap 
{\it and} also the Virasoro modular bootstrap. What's more, these are precisely the analytic functionals previously constructed  in the context of the {\it four-point function} bootstrap in 1D CFT!
By a curious historical coincidence, in the same year that Viazovska found the magic functions that prove optimality of the $E_8$ lattice, one of us \cite{Mazac:2016qev} independently constructed analytic functionals for the crossing problem of four identical operators in CFT$_1$. These functionals  (further studied and generalized in \cite{Mazac:2018mdx, Mazac:2018ycv, Mazac:2018qmi,Mazac:2018biw,Kaviraj:2018tfd}) 
prove optimality of the generalized free fermion CFT$_1$, for arbitrary external dimension $\Delta_\phi$. In this context, this means attaining the largest allowed dimension of the first exchanged operator.

There is a simple dictionary that allows to apply the very same analytic functionals to the other linear programming problems that we have described.
Table 1 in Section 4 is our key relating the Cohn-Elkies approach to sphere packing, the spinless modular bootstrap and the CFT$_1$ four-point function bootstrap.
Our functionals turn out to be optimal for the modular bootstrap with central charges $c=4$ and $c=12$,   for both  the $U(1)^c$ case (which, as we have mentioned, is equivalent to the Cohn-Elkies problem)
and the Virasoro case.
Remarkably, when translated into sphere packing variables, they exactly reproduce the  magic functions used 
by Viazovska \cite{viazovska8} and by Cohn et al.~\cite{viazovska24} to prove that the $E_8$ and Leech lattices are the densest packings in dimensions 8 and 24, respectively. We also show how a complete basis of functionals for 1D CFTs, constructed in \cite{Mazac:2018ycv}, underlies the complete basis of sphere-packing functions found by mathematicians recently in \cite{InterpolationMath}, and generalizes their results to all dimensions.

 For $c \neq 4, 12$ our modular bootstrap functionals are not optimal, but still lead to the rigorous upper bound 
 $\Delta_0 (c) \leq c/8 + 1/2$  for  the largest scaling dimension of the first non-trivial primary, for any  $c \in (1, 4) \cup (12, \infty)$, for both Virasoro and  $U(1)^c$.
As we have emphasized, the $c \to \infty$ limit is especially interesting because large-$c$ CFTs with sparse spectrum are dual to $AdS_3$ quantum gravity.
In this limit, we are able to find an improved (though still suboptimal) functional that leads to the Virasoro analytic bound
$\Delta_0 (c) \lesssim c / 8.503$. This is the first analytic improvement over Hellerman's original bound \cite{Hellerman:2009bu} of $c/6$. It is not  quite as strong as the conjectured asymptotics $c/9.08$, based on extrapolating the numerics \cite{Afkhami-Jeddi:2019zci}.

Like the Hellerman bound, our bound constrains the spectrum of black holes in 3D quantum gravity. It is related (though somewhat indirectly due to the distinction between Virasoro and $U(1)^c$) to constraints on the density of sphere packing in high dimensions. In turn, dense sphere packings provide the most efficient classical error-correcting codes. Black holes are already known or conjectured to saturate bounds on entropy \cite{bekenstein1981universal}, scrambling \cite{Hayden:2007cs,Sekino:2008he}, chaos \cite{Maldacena:2015waa}, complexity \cite{Susskind:2014rva}, weak gravity \cite{ArkaniHamed:2006dz}, and more. It is intriguing to find yet another sense in which black holes live on the boundary of theory space.

\medskip
  
  The rest of the paper is organized as follows. Section 2 and Section 3 review modular bootstrap and  sphere packing, respectively, emphasizing the parallels between the linear programming approaches to both problems. In Section 4 we describe succinctly our main results, leaving most of the technical details for later sections. Section 5 reviews the construction of analytic functionals for the crossing problem in CFT$_1$.
It also contains some new results for a generalized crossing equation, which we need later to make full contact with the Cohn-Elkies problem. In Section 6 we present the technical details
of the analytic functionals for the modular bootstrap and reproduce the sphere packing magic functions. In Section 7 we study the large central charge limit and derive our asymptotic analytic bound of $c / 8.503$. In Section 8 we sketch the construction of a complete basis of functionals for the Cohn-Elkies problem in arbitrary dimension. We conclude in Section 9 with a discussion and some speculations.
Two appendices contain basic reference material on modular forms and further technical details on analytic functionals.

% !TEX root = ../ModularBootstrapV3.tex

\section{Review of Modular Bootstrap}\label{sec:ReviewModular}

\subsection{Overview of existing bounds}

The modular bootstrap is a method to constrain possible spectra of 2D CFTs. The simplest question it can address is: Given the central charge $c$, what is the largest allowed scaling dimension, $\Delta_0$, of the first non-trivial primary \cite{Hellerman:2009bu}?

The answer depends on the choice of the chiral algebra that we impose as a symmetry of the CFT. We will focus on two cases: $\mathrm{Vir}_c\times\mathrm{Vir}_c$, and the current algebra $U(1)^c \times U(1)^c$. Here $\mathrm{Vir}_c$ stands for the Virasoro algebra of central charge $c$, and the two copies of the algebra correspond to left and right movers. By the Sugawara construction, $\mathrm{Vir}_c$ is a subalgebra of $U(1)^c$. We will assume $c>1$ throughout this paper. The best possible upper bounds on the dimension of the first nontrivial primary $\Delta_0$ from the spinless modular bootstrap will be denoted $\DeltaVir(c)$ and $\DeltaU(c)$ for Virasoro and $U(1)^c$, respectively.

The bounds come from imposing invariance of the partition function $Z(\tau,\btau)$ under the modular group $PSL(2,\mathbb{Z})$. A restricted class of solutions of this problem comes from assuming that the partition function factorizes $Z(\tau,\btau)=Z_{\textrm{hol}}(\tau)Z_{\textrm{hol}}(-\btau)$, where $Z_{\textrm{hol}}(\tau)$ is a weakly holomorphic modular form of weight zero. In this case, the central charge must be an integer multiple of 24 and all scaling dimensions take integer values. Furthermore, it follows from the theory of modular forms that $\Delta_0 \leq \frac{c}{24} + 1$ for this class of partition functions \cite{hohn2007selbstduale,hohn2008conformal,Witten:2007kt}. We will not assume holomorphic factorization in this paper.

The first constraints valid for general unitary 2D CFTs with $c>1$ were obtained by Hellerman \cite{Hellerman:2009bu}, who proved $\DeltaVir(c) < \frac{c}{6} + 0.4737$. Using the linear programming method introduced by Rattazzi, Rychkov, Tonni, and Vichi \cite{Rattazzi:2008pe}, the bound on $\Delta_0$ has since been improved numerically \cite{Friedan:2013cba,Collier:2016cls,Afkhami-Jeddi:2019zci}. The strongest current numerical bounds, as a function of $c$, were found in \cite{Afkhami-Jeddi:2019zci}. There are two salient points. First, the bound is saturated by known partition functions at $c=4$ and $c=12$, to very high numerical accuracy. Second, the numerics become increasingly difficult at large $c$, so for $c \gtrsim 2000$, the numerical constraints are weak, and Hellerman's bound remains the strongest that has been established rigorously. Extrapolating the numerics from $c \lesssim 2000$ led to the conjectured asymptotics $\DeltaVir \approx c/9.08$ as $c \to \infty$ \cite{Afkhami-Jeddi:2019zci}.

\begin{figure}
\centering
\includegraphics[scale=1.0]{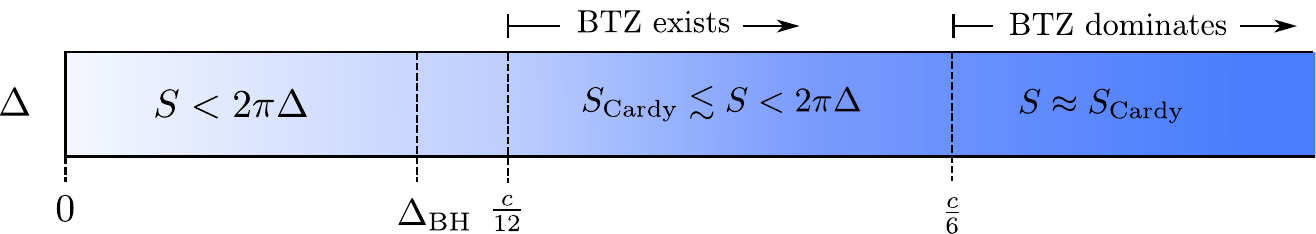}
\caption{\small Universal features of the spectrum of 3D gravity \label{fig:gravityfeatures}}
\end{figure}

The large-$c$ limit is especially interesting because 2D CFTs with large $c$ and a large gap are holographically dual to quantum gravity in three-dimensional anti-de Sitter space. All of the known, consistent constructions of 3D gravity come from string theory and share several features in the spectrum. This is summarized by the microcanonical entropy $S(\Delta)$, in Figure~\ref{fig:gravityfeatures}. There are black holes with entropy of order $c$ above some threshold, $\Delta_{\rm{BH}} \leq \frac{c}{12}$. The array of black hole solutions and the exact threshold depend on the particular theory, but at a minimum, all examples contain BTZ black holes for $\Delta \geq \Delta_{\rm BTZ} = \frac{c}{12}$. In the so-called enigma range, $\frac{c}{12} \leq \Delta \leq \frac{c}{6}$, BTZ black holes do not necessarily dominate the entropy, so $S$ is not universal, but it is bounded above by $S \leq 2\pi \Delta$. Finally, for $\Delta \geq \frac{c}{6}$, all theories have universal $S(\Delta)$ given by the Cardy entropy. These universal properties follow from modular invariance in the large-$c$ limit and the assumption of a sparse low-lying spectrum \cite{Hartman:2014oaa}.
For other applications and extensions of the modular bootstrap, see, {\it e.g.},~\cite{Keller:2012mr,Qualls:2013eha,Qualls:2014oea,Benjamin:2016fhe,Ashrafi:2016mns,Cho:2017fzo,Dyer:2017rul,Belin:2018oza,Anous:2018hjh,Bae:2018qym,Lin:2019kpn}.

The universality of BTZ black holes on the gravity side suggests that perhaps the bound from modular invariance is $c/12$, a factor of two stronger than the Hellerman bound. If so, and if there are theories with gap $\Delta_0  \approx \frac{c}{12}$ and arbitrarily large $c$, then these would be  theories of ``pure" 3D gravity, consisting only of gravitons and BTZ black holes. It is an open question whether pure gravity exists as a quantum theory \cite{Witten:2007kt,Maloney:2007ud,Keller:2014xba}. One goal of modular bootstrap is to settle this question. More generally, the eventual goal is to explore whether every consistent theory of quantum gravity in three dimensions comes from string theory, or to find alternative theories of 3D gravity. This may be possible if the bound on $\Delta_0$ can be pushed down to the BTZ black hole threshold, $c/12$. If the spectrum is allowed to be continuous then $\DeltaVir(c) \geq \frac{c-1}{12}$ \cite{Keller:2014xba}, but it is logically possible that the bound with a discrete spectrum could even surpass the BTZ black hole threshold.

\subsection{Partition functions}

The partition function of a unitary 2D CFT is
\be
Z(\tau,\btau) = \sum_{\mbox{\scriptsize states}} q^{h - c/24} \bq^{\bh - \bc/24} \ ,\qquad
q =e^{2\pi i \tau} \ , \quad \bq = e^{-2\pi i \btau} 
\label{eq:Zdefinition}
\ee
where the sum runs over all states of the theory on $S^1$, $(h,\bh)$ are their non-negative conformal weights and $(c,\bc)$ are the left and right central charges. There is a unique vacuum state with $h=\bh=0$. The partition function on a Euclidean torus of modulus $\tau$ equals $Z(\tau,\btau)$ restricted to $\btau=\tau^*$, where the star stands for complex conjugation. However, the sum \eqref{eq:Zdefinition} converges for any $\tau \in \mathbb{H}_{+}$ and $\btau \in \mathbb{H}_{-}$, with $\mathbb{H}_+$ and $\mathbb{H}_-$ the upper and lower half-planes, so we can treat $\tau$ and $\btau$ as independent and complex. In fact, $Z(\tau,\btau)$ is a holomorphic function in $\mathbb{H}_{+}\times\mathbb{H}_{-}$,
because each summand in \eqref{eq:Zdefinition} is holomorphic in this region and the sum converges uniformly in any compact set.

Modular invariance is the statement
\be
Z\left( \frac{a \tau + b}{c \tau + d} ,  \frac{a \btau + b}{c \btau + d} \right) = Z(\tau, \btau) , \quad \mbox{for all} \quad 
\begin{pmatrix}a & b\\ c& d \end{pmatrix} 
\in SL(2,\mathbb{Z}) \ .
\ee
States are organized under the symmetry algebra of the theory, so $Z$ can be expressed as a sum of characters,
\be
Z(\tau,\btau) = X_{\rm vac}(\tau,\btau) + \sum X_{h,\bh}(\tau,\btau) \ ,
\ee
where the sum runs over non-vacuum primary operators. The characters include the contribution of a primary and its descendants, so they take the form
\be
X_{h,\bh}(\tau,\btau) = q^{h - c/24} \bq^{\bh - \bc/24} (1 + \mbox{non-negative integer powers of\ }q,\bq ) \ .
\ee
$X_{\rm vac}$ may or may not be the analytic continuation of $X_{h,\bh}$ to $h=\bh=0$, hence the different name.

We will specialize to the one-complex-dimensional section $\tau = - \btau$, and denote the restricted partition function
\be
\mathcal{Z}(\tau) = Z(\tau, -\tau) \ .
\ee
This includes in particular all the rectangular Euclidean tori, for which $\tau = i\beta$ with real $\beta >0$. The partition function $\mathcal{Z}(i\beta)$ is the usual partition function of statistical mechanics at inverse temperature $\beta$. $\mathcal{Z}(\tau)$ is simply the analytic continuation of this function to the entire upper half-plane. By restricting to $\tau = -\btau$, we have set the angular potential to zero, and therefore dropped information about the spin $h - \bh$ in the partition function. $\mathcal{Z}(\tau)$ is holomorphic in $\mathbb{H}_+$, but it generally does not admit a series expansion into \emph{integer} powers of $q$ as $\tau\rightarrow i\infty$. The only non-identity element of $PSL(2,\mathbb{Z})$ which remains a symmetry of $\mathcal{Z}(\tau)$ is the $S$-transformation,
\be\label{strans}
\mathcal{Z}(\tau) = \mathcal{Z}(-1/\tau) \ .
\ee
The $T$-transformation $\tau \rightarrow \tau + 1$, $\btau \to \btau + 1$ does not respect the condition $\tau = - \btau$. 

$\mathcal{Z}(\tau)$ can be expanded into characters of the symmetry algebra
\be
\mathcal{Z}(\tau) = \chi_{\rm vac}(\tau) + \sum_{\Delta>0}\rho_\Delta \chi_{\Delta}(\tau) \ ,
\ee
where $\Delta = h + \bh$, and $\rho_\Delta$ is the integer degeneracy of primaries with this scaling dimension. The positivity condition $\rho_{\Delta} > 0$, which we assume throughout, is referred to as unitarity. We will primarily be interested in modular bootstrap with symmetry algebras $\mathrm{Vir}_c\times\mathrm{Vir}_c$ and $U(1)^c\times U(1)^c$. For $\mathrm{Vir}_c\times\mathrm{Vir}_c$, the characters take the form
\be
\chi_\Delta^V(\tau) = \frac{q^{\Delta - c/12}}{\prod_{k=1}^\infty(1-q^k)^2} = \frac{q^{\Delta - \frac{c-1}{12}}}{\eta(\tau)^2} \ , \quad
\chi_{\rm vac}^V(\tau) = (1-q)^2\chi_0^V(\tau) \ .
\ee
For the current algebra $U(1)^c \times U(1)^c$, we have
\be\label{u1char}
\chi_{\Delta}^U(\tau) = \frac{q^{\Delta-c/12}}{\prod_{k=1}^\infty(1-q^k)^{2c}}  = \frac{q^\Delta}{\eta(\tau)^{2c}} \ , \quad
\chi_{\rm vac}^U(\tau) = \chi_0^U(\tau) \ .
\ee

\subsection{Linear programming bounds}\label{ss:modularboundreview}
We can derive bounds on the gap by constructing linear functionals acting on functions of $\tau$ \cite{Rattazzi:2008pe,Hellerman:2009bu}. Let us write $S$-invariance \eqref{strans} as
\be
\Phi^{A}_{\rm vac}(\tau) + \sum_{\Delta >0 }\rho_\Delta \Phi^{A}_{\Delta}(\tau)  = 0
\ee
for all $\tau\in\mathbb{H}_{+}$, where
\be\label{defF}
\Phi^{A}_{\rm vac}(\tau) = \chi^{A}_{\rm vac}(\tau) - \chi^{A}_{\rm vac}(-1/\tau) \ ,\quad
\Phi^{A}_{\Delta}(\tau) = \chi^{A}_{\Delta}(\tau) - \chi^{A}_{\Delta}(-1/\tau)\,.
\ee
Here and in the following $A=U,V$ is a placeholder for the symmetry algebra of choice. If we can find a linear functional $\omega$ satisfying
\begin{align}
&\omega[\Phi^{A}_{\rm vac}] > 0\\
&\omega[\Phi^{A}_{\Delta}] \geq 0 \quad \mbox{for all}\quad \Delta\geq \Delta_*\,,
\end{align}
then all unitary partition functions must have a nontrivial primary with $\Delta < \Delta_*$. The infimum over all functionals with the above properties is the optimal linear programming bound on the gap. For the limiting functional the first condition is replaced with $\omega[\Phi^{A}_{\rm vac}] = 0$ and there is an associated modular-invariant partition function whose spectrum is annihilated by the optimal (also called extremal) functional.

The optimal bounds, over the space of all linear functionals acting on the spinless partition function $\mathcal{Z}(\tau)$, are denoted $\DeltaVir(c)$ for Virasoro and $\DeltaU(c)$ for $U(1)^c$. We will see that sphere packing is most directly connected to the modular bootstrap with a $U(1)^c$ chiral algebra, and that the linear programming bounds on sphere packing can be stated in terms of $\DeltaU(c)$.

\subsection{Functionals as eigenfunctions of the Fourier transform}
By construction, the antisymmetrized character $\Phi^{A}_{\Delta}$ is a $-1$ eigenfunction of the $S$ transformation. We 
It follows that the  function
\be
\omega(\Delta) = \omega[\Phi^{A}_{\Delta}]
\ee
can also be understood as a $-1$ eigenfunction of $S$, in the following sense \cite{Friedan:2013cba}. 
%$S$ acts as an integral transform on the characters,
%\be
%S\cdot  \chi^{A}_{\Delta}(\tau) \equiv \chi^{A}_\Delta(-1/\tau) = \int d\Delta' K_A(\Delta,\Delta') \chi^{A}_{\Delta'}(\tau)\,
%\ee
%where $K_A(\Delta,\Delta')$ is the crossing kernel for the characters of choice. The functional action is
%\be
%\omega(\Delta) = \omega[\chi^{A}_{\Delta} - S \cdot \chi^{A}_{\Delta}]  = f(\Delta) - \widetilde{f}(\Delta) \ ,
%\ee
%where $f(\Delta) = \omega[\chi^{A}_{\Delta}]$, and $\widetilde{f}(\Delta) = \int d\Delta' K_A(\Delta,\Delta') f(\Delta')$. Also, $S^2 = 1$. Therefore $\omega(\Delta)$ has eigenvalue $-1$ under the action of this integral transform.
For Virasoro symmetry, let us parametrize $\Delta$ by a vector $x\in\mathbb{R}^2$ as
\be
\Delta(x) = \frac{x^2}{2}  + \frac{c-1}{12} \ .
\ee
The crossing kernel takes the form of a 2D Fourier transform
\be
S \cdot \chi^V_{\Delta(x)} = \int\! d^2 y \, e^{-2\pi i x\cdot y} \chi^V_{\Delta(y)},
\ee
so  the function
\be
g(x) = \omega(\Delta(x))
\ee
is an eigenfunction of the 2D Fourier transform with eigenvalue $-1$.

For $U(1)^c$, the same argument demonstrates that $g(x)$ is a $-1$ eigenfunction of the $(2c)$-dimensional Fourier transform, with the identification
\be
\Delta(x) = \frac{1}{2}x^2 \ , \qquad x \in \mathbb{R}^{2c} \ .
\ee

A basis of $-1$ eigenfunctions for the Fourier transform in $\mathbb{R}^{d}$ is provided by the odd-degree associated Laguerre polynomials, 
\be\label{omegaibasis}
\omega_i(\Delta(x) ) = L_{2i-1}^{d/2-1}(4\pi x^2)e^{-2\pi x^2} \ .
\ee
The standard strategy for numerical bootstrap is to construct a basis of functionals by acting with derivatives $(\partial_\tau)^k$ at the crossing-symmetric point $\tau = i$. For modular bootstrap, this of course produces the same basis, \eqref{omegaibasis}. The corresponding derivative operator for Virasoro can be found in \cite{Afkhami-Jeddi:2017idc} and easily generalizes to $U(1)^c$.

\subsection{Saturation at $c=4,12$}
The Virasoro bootstrap converges to known, $S$-invariant functions for $c=4,12$ \cite{Collier:2016cls,Bae:2017kcl,Afkhami-Jeddi:2017idc}. The numerical bound at $c=12$, obtained by truncating to the first 2000 Laguerre polynomials, is \cite{Afkhami-Jeddi:2017idc}
\be
\DeltaVir(12) \leq 2 + 10^{-30} \ .
\ee
The zeros of the numerical functional appear to converge toward the non-negative integers, $\Delta = 0,1,2,3,\dots$. There are single roots at $0,1,2$ and double roots at the higher integers. The numerical bootstrap also produces a candidate partition function saturating this bound. To very high accuracy, it appears to be related to the theta function\footnote{The theta function of a lattice $\Lambda$ in $\mathbb{R}^d$ is defined as $\Theta_{\Lambda}(\tau) = \sum\limits_{x\in\Lambda}e^{i\pi\tau x^2}$.}
 for the Leech lattice $\Lambda_{24}$,
\be
\mathcal{Z}_{12}(\tau) = \frac{1}{\eta(\tau)^{24}} \Theta_{\Lambda_{24}}(\tau) -24 \ .
\ee
This can also be written in terms of the modular $j$-function, $\mathcal{Z}_{12}(\tau)= j(\tau) - 744$. 
This also happens to be the partition function of the chiral monster CFT \cite{frenkel1984natural}, with $c=24, \bc = 0$, but the appearance of the $j$-function in the present context is a surprise. Recall that we did not impose $T$-invariance; it appears for free in the optimal partition function at $c=12$. In other words, there is no obvious reason \textit{a priori} to expect an integer spectrum in a non-chiral CFT.

For $c=4$, the situation is similar. The numerics converge towards $\DeltaVir(4) = 1$, with zeroes at the nonnegative integers. The candidate partition function with this spectrum is built from the theta function for the the $E_8$ lattice,
\be
\mathcal{Z}_{4}(\tau) = \frac{1}{\eta(\tau)^8} \Theta_{\Lambda_8}(\tau)  = (j(\tau))^{1/3}\ .
\ee

\bigskip

% !TEX root = ../ModularBootstrapV3.tex

\section{Review of the Sphere Packing Problem}\label{sec:ReviewSphere}

A thorough introduction to sphere packing can be found in the short book by Thompson \cite{thompson1983error}, the long book by Conway and Sloane \cite{ConwaySloane}, and, for recent developments, the review articles \cite{cohnReviewAMS,2016arXiv160702111D,tothpacking}. Here we will review just enough background to explain the linear programming method of Cohn and Elkies \cite{cohn1,cohn2},  Viazovska's proof for $E_8$ \cite{viazovska8}, and its extension to the Leech lattice \cite{viazovska24}. All of the results reviewed in this section are mathematically rigorous in the original papers, including the numerics, which are done in rational or interval arithmetic to control numerical errors.

\subsection{Basics}

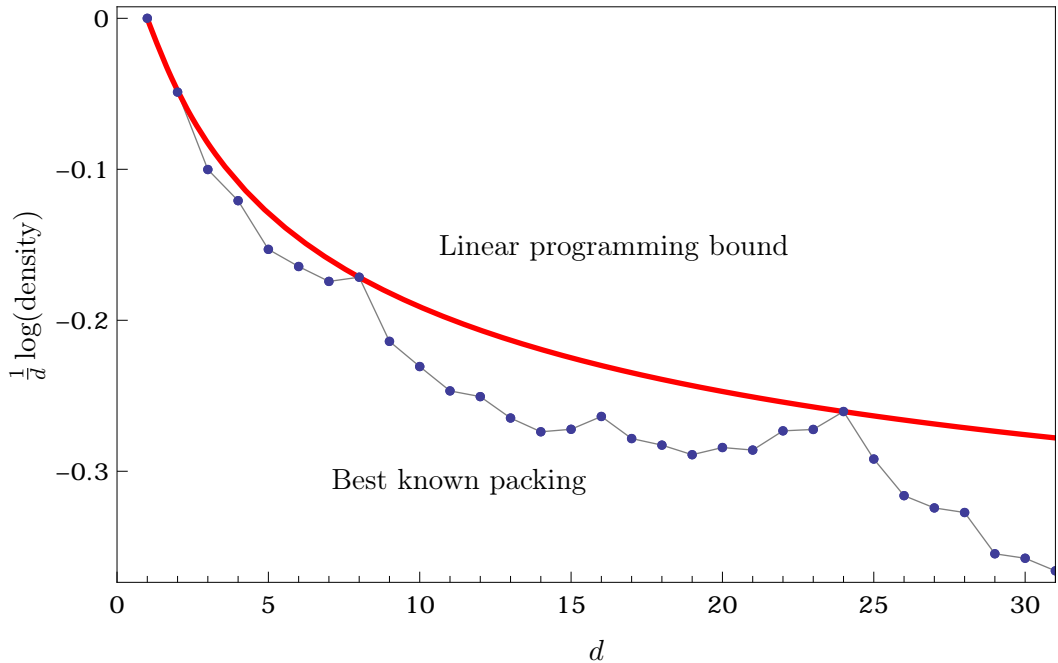
\begin{figure}
\begin{center}
\begin{overpic}[scale=1.0]{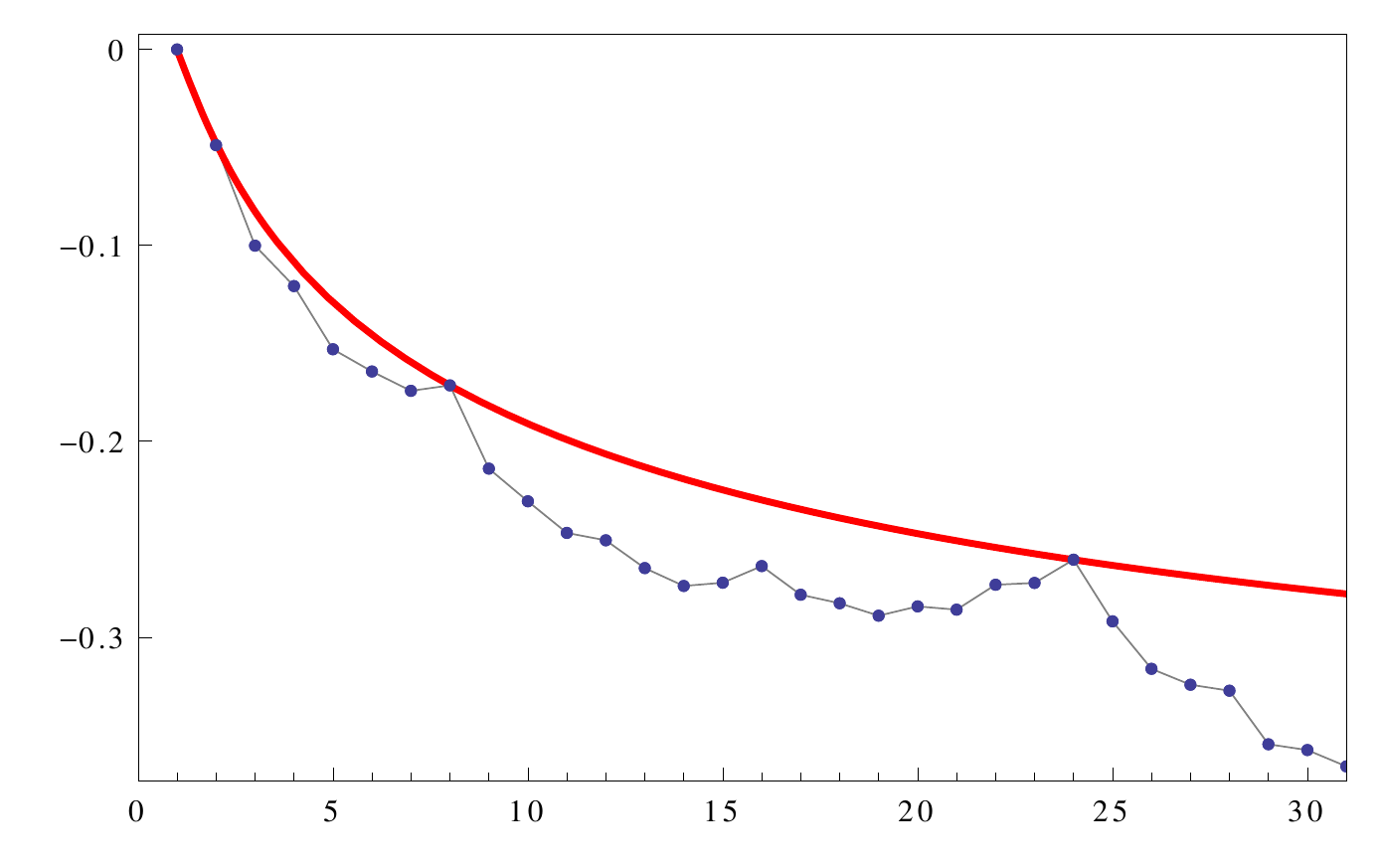}
\put(40,37){Linear programming bound}
\put(30,15){Best known packing}
\put(54,-1){$d$}
\put(0,25){\rotatebox{90}{$\frac{1}{d}\log(\mbox{density})$}}
\end{overpic}
\end{center}
\caption{Density of sphere packings in $\mathbb{R}^d$. For the best known packings, see table I.1 of \cite{ConwaySloane}.\label{fig:bestpackings}}
\end{figure}

The simplest packings are \textit{lattice packings}, where a sphere is centered at each point on a lattice $\Lambda \subset \mathbb{R}^d$. The sphere diameter is equal to the length of the shortest lattice vector, so the problem of finding a dense lattice packing is one of constructing lattices with no short vectors and fixed volume of the unit cell. This is already reminiscent of the conformal bootstrap, if we were to restrict to compactified free theories, where the spectrum is specified by a lattice.

The solved cases, $d=1,2,3,8,24$, are all lattice packings, but in general, lattice packings are not optimal.  A more general configuration is a \textit{periodic packing}, which is a crystal, having one or more spheres per unit cell. Not all sphere packings are periodic, but by taking the unit cell very large, any packing can be well approximated by a periodic one, so it suffices to restrict to this case for the purposes of proving bounds on the density.

The \textit{density} of a periodic packing is the fraction of the unit cell occuppied by spheres. If there are $N$ spheres in the unit cell, each with radius $r$, then this fraction is
\be
\Density_d = \frac{N\, \mbox{vol}(B^d)r^{d}}{|\Lambda|} \ ,
\ee
where $\mbox{vol}(B^d) = \pi^{d/2}/\Gamma(d/2+1)$ is the volume of the unit ball in $\mathbb{R}^d$, and the denominator $|\Lambda| = \mbox{vol}(\mathbb{R}^{d}/\Lambda)$ is given by the determinant of the lattice basis. The highest achievable packing density for a given $d$ will be denoted $\Dmax$. 

In most cases, the best known upper bounds on $\Dmax$ come from linear programming \cite{cohn1}. These bounds, together with the densest known packings, are plotted for small $d$ in fig.~\ref{fig:bestpackings}. The bounds are saturated in the dimensions where the sphere packing problem has been solved, with the exception of the Kepler problem, $d=3$.

At large $d$, some general upper and lower bounds are known. Early arguments by Minkowski (see \cite{Hlawka1943}) and Blichfeldt \cite{Blichfeldt1929} led to the allowed asymptotics $ 2^{-d} \lesssim \Dmax \lesssim 2^{-d/2}$. The lower bound has since been improved by a linear prefactor. The current best upper bound at large $d$ is by Kabatyanski and Levenshtein, $\Dmax \leq 2^{-0.599d + o(d)}$ \cite{kabatiansky1978bounds} (with a prefactor improved by Cohn and Zhao \cite{cohn2014sphere}). 

\subsection{Linear programming method}\label{ssec:LProgramming}

\subsubsection*{The Cohn-Elkies Theorem}

Linear programming has long been used in coding theory, starting with the work of Delsarte in 1972 \cite{delsarte1972bounds}. It can be used to bound, for example, the number of codewords in an error-correcting code. Bounds on error-correcting codes can be translated into sphere packing, to place rigorous upper bounds on $\Dmax$ \cite{kabatiansky1978bounds, gorbachev2000}. (See the discussion section for further comments on this connection.) Linear programming was later applied directly to the sphere packing problem, without going through an intermediate coding problem, by Cohn and Elkies. We will now review the main theorem of Cohn and Elkies \cite{cohn1,cohn2}. We reformulate their proof in the language of linear functionals familiar from the bootstrap.

Consider a periodic packing, specified by a lattice $\Lambda \subset \mathbb{R}^d$ and vectors $v_1, \dots, v_N$. A sphere is centered at each $v_i$ and its translations by $\Lambda$. The distances between centers of spheres are given by $|x+v_i-v_j|$ for $x\in\Lambda$ and $i,j=1,\ldots,N$. The density of a packing is invariant under an overall rescaling. Therefore, we can assume without loss of generality that the shortest distance between the centers of distinct spheres in the packing is equal to $1$, and set the sphere radius to $1/2$. The density of the packing is then given by
\be
\Density_d = \frac{N \mbox{vol}(B^d)}{2^d|\Lambda|}\,.
\label{eq:densityPeriodic}
\ee
Proving an upper bound on $\rho_d$ thus amounts to proving a universal upper bound on $N/|\Lambda|$ for all periodic sets of vectors which have unit minimal distance between different vectors.

In order to prove such a bound, we start by defining the averaged theta function as a weighted sum over all distances between centers of spheres in the packing \cite{odlyzko1980theta},
\be
\Theta(\tau) =  \sum\limits_{i,j=1}^{N}  \sum_{x \in \Lambda} e^{\pi i \tau (x + v_i - v_j)^2} \ .
\ee
For later convenience, we divide this by a power of the Dedekind $\eta$-function to define the partition function of the packing 
\be\label{zsphere}
\mathcal{Z}(\tau) = \frac{\Theta(\tau)}{\eta(\tau)^{d}} =
\sum\limits_{i,j=1}^{N}  \sum_{x \in \Lambda} \chi^{U}_{(x+v_i-v_j)^2/2}(\tau)
 \ ,
\ee
where $\chi^{U}_{\Delta}(\tau)$ are the $U(1)^c$ characters \eqref{u1char} for $c=d/2$. It turns out that there exists a version of the modular bootstrap equation \eqref{strans} satisfied by every periodic packing. While $\mathcal{Z}(\tau)$ is not necessarily invariant under the $S$ transformation, $\mathcal{Z}(\tau)$ can be expanded in the crossed-channel characters $\chi^U_{\Delta}(-1/\tau)$ with positive coefficients. The precise equation follows directly from the Poisson summation formula with respect to the lattice $\Lambda$ and reads
\be\label{spherecrossing}
\sum\limits_{i,j=1}^{N}  \sum_{x \in \Lambda} \chi^{U}_{(x+v_i-v_j)^2/2}(\tau)
= \frac{1}{|\Lambda|}\sum_{y \in \Lambda^*} \left|\sum\limits_{i=1}^{N} e^{2\pi i v_i\cdot y} \right|^2 \, ,
\chi^{U}_{y^2/2}(-1/\tau)\,.
\ee
where $\Lambda^*$ stands for the dual lattice.% and we used the fact that the Fourier transform of $\chi^{U}_{x^2/2}(\tau)$ in $x\in\mathbb{R}^{2c}$ is $\chi^{U}_{y^2/2}(-1/\tau)$.

Let us consider a linear functional $\omega$ acting on functions of $\tau$ and define $f : \mathbb{R}^{d} \to \mathbb{R}$ as the functional action 
\be\label{ffromw}
f(x) = \omega \left[ \chi^{U}_{x^2/2}(\tau) \right] \ .
\ee
The action of $\omega$ on the crossed-channel characters is given by the Fourier transform of $f(x)$ 
\begin{align}\label{fourier}
\omega\left[ \chi^{U}_{y^2/2} (-1/\tau)\right]=\widehat{f}(y)= \int\limits_{\mathbb{R}^d}\!d^dx\, e^{-2\pi i x\cdot y} f(x)
\ .
\end{align}
When we apply $\omega$ to \eqref{spherecrossing}, we get
\be\label{posum}
\sum_{i,j=1}^N \sum_{x\in \Lambda} f(x+v_i-v_j) =
\frac{1}{|\Lambda|} \sum_{y \in \Lambda^*}\left| \sum_{j=1}^N e^{2\pi i v_j \cdot y} \right|^2\widehat{f}(y) \ .
\ee
Actually, we could have obtained this equation more simply by applying Poisson summation directly to the left-hand side of \eqref{posum}, without introducing the linear functional $\omega$. This more direct route is taken by Cohn and Elkies \cite{cohn1}. We have rephrased the proof in terms of the action of a linear functional to draw a parallel to the conformal bootstrap, and because this point of view is useful in constructing the optimal functionals analytically. 

We will see that \eqref{posum} plays the same role as the crossing equation in conformal field theory. This equation also has an analogue in coding theory, known as the MacWilliams identities.

The function $f(x)$ constructed in \eqref{ffromw} is spherically symmetric, $f(x)=f(|x|)$. The argument can be generalized so that $f(x)$ in \eqref{posum} is a more general function on $\mathbb{R}^d$, but this does not improve the bounds \cite{cohn1}.

In order to derive a bound on the density from equation \eqref{posum}, we proceed by extracting the ``identity" ($\Delta=0$) contributions from both sides. On the LHS, these come from terms with $x=0$ and $i=j$, while on the RHS from the term with $y=0$. Moving all identity terms to the left, we arrive at
\be\label{nfaa}
Nf(0)-\frac{N^2}{|\Lambda|}\widehat{f}(0)=-\sum_{
\substack{x\neq 0\\ \rm{or\ } i\neq j}} f(x+v_i-v_j)+
 \frac{1}{|\Lambda|} \sum_{y \in \Lambda^*\backslash\{0\} }
 \left| \sum_{j=1}^N e^{2\pi i v_j \cdot y} \right|^2\widehat{f}(y) \ .
\ee
Suppose now that we can find a functional $\omega$ such that $f$ satisfies
\ba\label{cond}
\widehat{f}(y) &\geq 0  \quad \mbox{for all\ } y\in\mathbb{R}^{d}\\
f(x) &\leq 0 \quad \mbox{for all\ } |x| \geq 1\,.
\ea
Since the minimal distance between centers of distinct spheres in the packing is $1$ by assumption, it follows that all terms on the RHS of \eqref{nfaa} are non-negative. Therefore $Nf(0) \geq  N^2 \widehat{f}(0) / |\Lambda| $. This produces the desired upper bound on $N/\Lambda$ and thus from \eqref{eq:densityPeriodic} a general bound on the sphere-packing density\footnote{Note that $f(0),\widehat{f}(0)>0$ by construction since $\widehat{f}(y)\geq0$ for all $y\in\mathbb{R}^{d}$ and $f(0) = \int_{\mathbb{R}^d}d^dy \widehat{f}(y)$.}
\be
\Density_d \leq \frac{\mbox{vol}(B^d)}{2^d}\frac{f(0)}{\widehat{f}(0)}\,.
\ee

It is sometimes convenient to restate the theorem as follows \cite{cohn1,cohn2}. Suppose that instead of imposing unit shortest distance between sphere centers, we normalize the packing by $|\Lambda|=N$, so that the density becomes
\be
\rho_d =\mbox{vol}(B^d)R^{d}\,,
\ee
where $2R$ is the shortest distance between sphere centers. Thus to prove an upper bound on $\rho_d$, we seek an upper bound on $R$ valid for all packings with $|\Lambda|=N$. To prove the bound $R\leq R_{*}$ from \eqref{nfaa}, we need a function $f(x)$ satisfying
\ba
\textrm{(i)}\quad  &f(0) = \widehat{f}(0) > 0\\
\textrm{(ii)}\quad &\widehat{f}(y) \geq 0 \quad \textrm{for all}\quad  y\in\mathbb{R}^{d}\\
\textrm{(iii)}\quad &f(x) \leq 0  \quad\textrm{for all}\quad x\geq 2R_{*}
\label{eq:CE2}
\ea
If such a function exists, we get the universal bound $\rho_d \leq \mbox{vol}(B^d)R_{*}^{d}$. To see that the two formulations of the Cohn-Elkies theorem lead to precisely the same bound on the density, we can rescale the argument of $f(x)$ in the second formulation by $2R_*$ to produce $f(x)$ of the first formulation.

It is important to note that for the Possion summation formula \eqref{posum} to hold, $f(x)$ needs to be sufficiently smooth and decay sufficiently fast at infinity. This is equivalent to saying that not every linear functional $\omega$ can be commuted with the infinite sums over $x$ and $y$ in \eqref{spherecrossing}.\footnote{Conditions under which a linear functional can be commuted with the sum over operators and thus gives a correct bootstrap equation were analyzed for the $sl(2,\mathbb{R})$ four-point bootstrap in \cite{Rychkov:2017tpc}.} For \eqref{posum} to hold, it is sufficient that $f(x)$ is a Schwartz function in $\mathbb{R}^d$. Although \eqref{posum} holds for more general functions, in practice the optimal bounds arise from Schwartz functions so we can restrict to them in the following. When $\omega$ is a finite linear combination of derivatives in $\tau$ evaluated at $\tau=i$, then $f(x)$ is a Schwartz function. We will see that the functionals which lead to optimal bounds are given by contour integrals in $\mathbb{H}_{+}$ which still lead to Schwartz functions.

Positivity in Fourier space can also be understood geometrically. A function $f(x)$ is said to be \textit{positive definite} if, for any $x_1,x_2,\dots,x_N$, the matrix $f(x_i-x_j)$ is positive semidefinite. Assuming fast enough fall-off, $f$ is positive definite if and only if it has non-negative Fourier transform. This point of view, explained further in \cite{cohn1,viazovska2018sharp}, gives another simple proof of the Cohn-Elkies theorem, and relates it to the results of Delsarte.

\subsubsection*{Linear programming}

Clearly the best bound on $\Dmax$ coming from the Cohn-Eliies theorem is obtained by finding the $f$ in the second formulation with minimal $R_{*}$. Alternatively, in the first formulation, we normalize $f(0) = 1$, and solve the infinite-dimensional linear programming (LP) problem:
\be\label{lp1}
\mbox{  maximize $\widehat{f}(0)$ subject to \eqref{cond} \ .}
\ee
This setup does not completely exhaust the constraints on sphere packing, so even a complete solution of the LP problem does not generally solve the packing problem in $d$ dimensions. However, in dimensions $d=1,2,8,24$, miraculously, the LP bound becomes sharp. This was first observed numerically \cite{kabatiansky1978bounds,cohn1} where it was found that the LP bound is very nearly saturated by the best known packings in these dimensions. For other $d$, the LP bound is not optimal; it might still be possible to solve the packing problem by optimization, but only by replacing positive definiteness by the more general notion of a \textit{geometrically positive} function or by including higher-point correlations on the packing (\textit{e.g.}, \cite{de2015semidefinite,viazovska2018sharp}).

The direct solution of \eqref{lp1} by linear programming is possible but cumbersome. In practice, Cohn and Elkies trade it for a simpler optimization problem, which produces the same optimum.  Let us work with the second formulation \eqref{eq:CE2} and set
\be\label{fhg1}
f = h - g , \qquad \widehat{f} = h + g\,.
\ee
$h$ and $g$ are radial Schwartz functions which are respectively even and odd eigenfunctions of the Fourier transform in $\mathbb{R}^{d}$,
\be\label{fhg2}
\widehat{h} = h , \quad \widehat{g} = -g \ .
\ee
These can be decomposed into sums of even or odd degree Laguerre functions, respectively,
\be
g(x) = \sum_{i=1}^M \beta_i L_{2i-1}^\nu(2\pi|x|^2)e^{-\pi|x|^2} \ , \quad
h(x) = \sum_{i=1}^M \alpha_i L_{2i-2}^\nu(2\pi|x|^2)e^{-\pi|x|^2}
\ee
where $\nu = d/2-1$. We have truncated the expansion at some even integer $M$ in order to render the problem finite dimensional. The bound is rigorous at any $M$, and improves as $M$ is increased. To fix the $M$ coefficients $\beta_i$, up to an overall scaling, impose the $M-1$ equations
\ba\label{gzeroes}
g(r_\mu) &= g'(r_\mu) = 0 \\
g(0) &= 0
\ea
for $\mu = 1 \dots P$, with
\be
 P=M/2-1 \ .
\ee
Denote by $r_0$ the position of the last sign change of $g$, with $g(r_0) = 0$. 

The nonlinear optimization problem is to choose the double zeroes $r_\mu$ in order to minimize the single zero $r_0$.  This step relied on a computer, until Viazovska's proof. 

Once this optimization is done, for any given $M$, we have now completely determined $g$ (up to a a multiplicative constant) and $r_0, \dots, r_P$. The even eigenfunction $h$ is fixed by imposing
\begin{align}\label{hzeroes}
h(r_\mu) &= h'(r_\mu) = 0\\
g(r_0) + h(r_0)  &= g'(r_0) + h'(r_0)  = 0\notag
\end{align}
for $\mu = 1 \dots P$.

Although unproven, it was conjectured by Cohn and Elkies, and checked numerically, that for any $d\geq1$ this procedure gives a function $f$ that satisfies the assumptions of the Cohn-Elkies theorem and therefore places an upper bound on $\Dmax$.  For future reference, we record this observation as: 

\textbf{Conjecture 3.1}  (Cohn and Elkies \cite{cohn1}). The function $f = h-g$, constructed by forcing the single and double zeroes as in \eqref{gzeroes} and \eqref{hzeroes} and minizing $r_0$, satisfies the assumptions of the Cohn-Elkies theorem \eqref{eq:CE2}.

Unlike the original linear program \eqref{lp1}, the problem of choosing $r_\mu$ to maximize $r_0$ is not globally convex, and is not guaranteed to agree with \eqref{lp1}. However, in practice, a local optimum is easy to find, and this is good enough --- once a candidate $f$ is identified by this method, it can be checked that it satisfies the assumptions of the theorem, and therefore leads to rigorous sphere-packing bounds.

The difficult step in the procedure is of course to choose $r_\mu$ to minimize $r_0$. This was implemented numerically in \cite{cohn1,cohn2,cohn2009optimality}, by guessing an initial point $r_\mu$ and using Newton's method. As illustrated in figure \ref{fig:bestpackings}, the numerical upper bound for $d=8,24$ is extremely close to the packing density of the $\Lambda_8$ and $\Lambda_{24}$ lattices. The strongest numerical bounds were obtained in  \cite{cohn2009optimality},  keeping $P = 200$ roots (i.e.,~Laguerre polynomials through degree 803):
\begin{align}
d&=8: \quad\  \frac{\Dmax}{\Density(\Lambda_8)} \leq 1 + 10^{-14}\\
d&=24: \quad \frac{\Dmax}{\Density({\Lambda_{24}})} \leq 1 +  10^{-29} \notag
\end{align}
Cohn and Elkies conjectured that as $M \to \infty$, the procedure converges to a `magic function' $f_8(x)$ or $f_{24}(x)$ capable of solving the sphere packing problem in these dimensions. The magic function must have zeroes on the actual length spectrum of the $\Lambda_8$ or $\Lambda_{24}$ lattice, so that the right-hand side of \eqref{nfaa} vanishes. This implies, in $d=8$,
\be\label{roots8}
f_8(\sqrt{2k}) = \widehat{f}_8(\sqrt{2k}) = 0  , \qquad  k = 1,2,3,\dots \infty
\ee
and all of these should be double zeroes except for $f(r_0 = \sqrt{2})$. (Here the spheres have radius $1/\sqrt{2}$, since the length of the shortest vector in $\Lambda_8$ is $\sqrt{2}$.) Similarly, for $d=24$, the magic function must have
\be\label{roots24}
f_{24}(\sqrt{2k}) = \widehat{f}_{24}(\sqrt{2k}) = 0  , \qquad  k = 2,3,\dots \infty
\ee
with all zeroes of multiplicity two except $f(r_0 = \sqrt{4})$.

\subsection{Viazovska's proof}\label{ssec:ViazovskaProof}
The numerics left little doubt that the magic functions exist, but they were difficult to find. They were recently found by Viazovska in $d=8$ \cite{viazovska8} and by Cohn, Kumar, Miller, Radchenko and Viazovska in $d=24$ \cite{viazovska24}.
The key idea of \cite{viazovska8} is to start from the ansatz
\ba\label{vansatz}
h(r) &= i\sin^2(\pi r^2/2) \int_0^{i \infty} d\tau H(\tau) e^{i \pi r^2 \tau} \\
g(r) &= i\sin^2(\pi r^2/2) \int_0^{i \infty} d\tau G(\tau) e^{i \pi r^2 \tau} \,.
\ea
As in \eqref{fhg1}-\eqref{fhg2}, the magic function is $f= h-g$ with $h,g$ the $\pm 1$ eigenfunctions of the Fourier transform. The integrands $G$ and $H$ will be designed to give $h,g$ all the requisite properties, using quasimodular forms as building blocks.  

The ansatz builds in by hand the desired zeroes at $r=\sqrt{2},\sqrt{4},\sqrt{6},\dots$. However it also produces some extraneous zeros. In particular, $h(r)$ should be positive for $r=0$. Furthermore, the double root at $r = \sqrt{2}$ should be a simple root for $d=8$, and similarly the double root at $r=2$ should be a simple root in $d=24$. These extraneous zeroes must be canceled  by singularities from the integral.

We must find conditions on $G,H$ that will ensure $g,h$ are eigenfunctions of the $d$-dimensional Fourier transform. This is achieved by taking the Fourier transform of \eqref{vansatz} with the help of some judicious contour deformations, which can be found in Viazovska's paper and will be reviewed in section \ref{ss:mod2sphere} when we reproduce these results from the bootstrap. In the end, we find the following conditions on $G,H$:
\begin{subequations}\label{hgcond}
\begin{align}
H(-1/\tau) &= -\frac{1}{2}(-i\tau)^{2-d/2}\left[ H(\tau+1) + H(\tau-1) - 2 H(\tau)\right] \label{hcond1}\\
H(\tau+1) &= -(-i\tau)^{d/2-2} H(-1-1/\tau) \label{hcond2}\\
G(-1/\tau) &= \frac{1}{2}(-i\tau)^{2-d/2}\left[ G(\tau+1) + G(\tau-1) - 2 G(\tau)\right] \label{gcond1}\\
G(\tau+1) &= (-i\tau)^{d/2-2} G(-1-1/\tau) \ .\label{gcond2}
\end{align}
\end{subequations}
These transformation rules, together with some information about the functions' singular behavior, are enough to find $H$ and $G$. To proceed, we specialize to $d=8$ and $d=24$ in turn. Relevant background on modular forms is reviewed in appendix \ref{s:modforms}.

\subsubsection{The magic function in $\mathbb{R}^8$} \label{sssec:magicFunction8}
Let $\phi(\tau) = \tau^2 H(-1/\tau)$. For $d=8$, \eqref{hcond2} implies
\be
\phi(\tau) = \phi(\tau + 1) \ .
\ee
Therefore, $\phi$ has an expansion in $q = e^{2\pi i  \tau}$, and we can hope to build it using modular forms.  The other condition \eqref{hcond1} is 
\be\label{phibb}
\phi(\tau) = \Delta^{(2)} [ \tau^2 \phi(-1/\tau) ]
\ee
where $\Delta^{(2)}$ is the second finite difference operator: 
$
\Delta^{(2)}[ X(\tau) ] = \frac{1}{2}[X(\tau+1) + X(\tau-1) - 2 X(\tau)] 
$.
Eq \eqref{phibb} is satisfied if we pick $\phi$ to be a weakly holomorphic quasimodular form for $SL(2,\mathbb{Z})$, with weight 0 and depth 2. That is, 
\be\label{phitrans}
 \phi(-1/\tau) =  \phi(\tau) + \tau^{-1} \psi_1 + \tau^{-2} \psi_2 \ ,
\ee
where $\psi_1$ and $\psi_2$ are periodic under $\tau \sim \tau+1$.  Functions of this type can be built from the second Eisenstein series, $E_2$. 
 In analogy with the construction of the $j$-function as the ratio of two weight-12 modular forms, a natural guess is that $\Delta \phi$, with $\Delta$ the modular discriminant, is a weight 12, depth 2 quasimodular form. This suggests the ansatz
\be
\phi = \frac{1}{\Delta}\times ( E_2^2 p_8(E_4, E_6) + E_2 p_{10}(E_4,E_6) + p_{12}(E_4, E_6) ) 
\ee
with $p_k$ a weight-$k$ polynomial. Some restrictions on the fall-off behavior, or fixing the most singular terms by comparing to numerics, then leads to 
\be
\phi = \frac{4\pi (E_2 E_4 - E_6)^2 }{5(E_6^2 - E_4^3)} \ .
\ee
This defines the $+1$ eigenfunction in the integral ansatz \eqref{vansatz}, with $H(\tau) = \tau^2\phi(-1/\tau)$.

The integral converges for $r>\sqrt{2}$ and is otherwise defined by analytic continuation. At large imaginary $\tau$, 
\be
H(\tau) = \frac{1}{60\pi q} + \frac{42}{5\pi} + 4 i \tau + O(\tau^2 q)
\ee
These terms produce singularities that cancel the extraneous zeros in $\sin^2(\pi r^2/2)$ discussed above; it is straightforward to integrate them and see that
\be
h(0) = 1 \ , \quad h(\sqrt{2}) = 0   \ , \quad h'(\sqrt{2}) =  -\frac{1}{60\sqrt{2}} \ .
\ee

We now turn to the $-1$ eigenfunction, $g$. The conditions \eqref{hgcond} show that $\tilde{\phi} = \tau^2 G(-1/\tau)$ is antiperiodic under $\tau \to \tau +1$, so that it can be expanded in odd powers of $q^{1/2}$. This suggests that the relevant modular group is the congruence subgroup $\Gamma(2)$. Indeed, by an argument similar to the above, it suffices to choose $G$ to be a weakly holomorphic modular form of weight -2 for $\Gamma(2)$, and impose
\be\label{extragg}
\tau^2 G(-1/\tau) = G(\tau) - G(\tau+1) \ .
\ee
A natural guess is that $G$ is a weight-10 modular form divided by the discriminant, $\Delta \propto (\theta_2 \theta_3 \theta_4)^8 $. Modular forms for $\Gamma(2)$ are polynomials in $\theta_3^4$ and $\theta_4^4$, which are both weight 2. This gives 6 weight-10 forms that can appear in the numerator; imposing \eqref{extragg} fixes some of the coefficients, and the singular behavior fixes the rest, leading eventually to
\be
G(\tau) = - \frac{32 \theta_4^4(5\theta_3^8-5 \theta_3^4 \theta_4^4 + 2 \theta_4^8)}{15\pi \theta_3^8 \theta_2^8} \ .
\ee
The expansion at large imaginary $\tau$ is\be
G = - \frac{1}{60\pi q} - \frac{12}{5\pi} + \frac{256 q^{1/2}}{3\pi}    - \frac{5877 q}{5\pi} + O(q^{3/2}) \ .
\ee
Upon doing the integral in \eqref{vansatz} we see that the extraneous zeros are canceled as desired,
\be
g(0) = g'(0) = 0 , \quad g(\sqrt{2}) = 0 , \quad g'(\sqrt{2}) = \frac{1}{60\sqrt{2}} \ .
\ee
It follows that $f = g - h$ has exactly the properties \eqref{roots8} required of the magic function. This completes the proof that the densest packing in 8 dimensions is the $E_8$ root lattice. The rigorous proof \cite{viazovska8} is not much more difficult than what we have just sketched;  the only extra steps are checking the integral manipulations more carefully, and a straightforward proof that subtracting $h-g$ does not produce any new roots.

\subsubsection{The magic function in $\mathbb{R}^{24}$}
The details in $d=24$ are a bit different due to the weights in \eqref{hgcond}, but the extension is straightforward \cite{viazovska24}. For the +1 eigenfunction, take $H(\tau) = \tau^{10} \phi(-1/\tau)$ with
\be
\phi = \frac{65536\pi^{25}}{110565} \frac{1}{\Delta^2}(25 E_4^4- 49 E_6^2 E_4 + 48 E_6 E_4^2 E_2 + (-49E_4^3 + 25 E_6^2)E_2^2) \ .
\ee
For the -1 eigenfunction, take
\be
G = -\frac{1048576\pi^{23}}{4095}\frac{1}{\Delta^2}( 7 \theta_4^{20} \theta_2^8 + 7 \theta_4^{24} \theta_2^4 + 2 \theta_4^{28} ) \ .
\ee
The resulting $h,g$ have the desired roots \eqref{roots24}. This proves that the densest packing in 24 dimensions is the Leech lattice.

\bigskip

% !TEX root = ../ModularBootstrapV3.tex

\section{The relation between sphere packing and modular bootstrap}\label{sec:Relation}

There are clear similarities between the modular bootstrap and the Cohn-Elkies method for bounding the sphere packing density. In this section, we will describe the precise relation, and summarize how the same analytic functionals can be applied to both problems. 
%The rest of the paper will be fill in all the technical details.
%The , and give a preview of results from the more technical sections that follow.

\subsection{The case of isodual lattices}
As a warm-up, the modular bootstrap problem with $U(1)^c$ characters can be directly used to constrain isodual lattices in $\mathbb{R}^{2c}$. An isodual lattice is one which is isometric (geometrically congruent) to its dual lattice. In particular,  the vector norms and their multiplicities are the same for $\Lambda$ and $\Lambda^*$, and we have $|\Lambda| = |\Lambda^*|=1$. We define the partition function of an isodual lattice $\Lambda$ as a special case of \eqref{zsphere} with $N=1$ and $v_1=0$,
\be\label{ziso}
\mathcal{Z}(\tau) = \sum_{x \in \Lambda} \chi_{x^2/2}^U(\tau) \ .
\ee
%where as usual the the space dimension $d$  and ghe central charge $c$ and are related by
%\be
%d = 2 c \ .
%\ee
It follows from the Poisson summation formula \eqref{spherecrossing} and isoduality of $\Lambda$ that $\mathcal{Z}(\tau)$ is $S$-invariant,
\be\label{zisos}
\mathcal{Z}(\tau) = \mathcal{Z}(-1/\tau) \ .
\ee
The problem of maximizing the length of the shortest nonzero lattice vector in the sum \eqref{ziso} subject to S-invariance \eqref{zisos} is manifestly identical to the problem of maximizing $\Delta_0$ in the $U(1)^c$ modular bootstrap discussed in section \ref{ss:modularboundreview}. Thus for any isodual lattice in $\mathbb{R}^{2c}$, the length $L_{\rm min}$ of the shortest non-zero vector obeys
\be
L_{\rm min} %\colonequals \min_{x \in \Lambda \backslash \{ 0 \} } |x| 
\leq \sqrt{2 \DeltaU(c) } \ .
\ee
Since in a lattice packing the sphere radius  is bounded above by  $L_{\rm min}/2$, we immediately obtain an upper bound on the sphere-packing density among all isodual lattice packings,
\be\label{isodualbound}
\rho_d^{\rm isodual} \leq    \mbox{vol}(B^d) \left( \frac{L_{\rm  min}}{2} \right)^d =   \mbox{vol}(B^d) \left[ \frac{\DeltaU ( \frac{d}{2} ) }{2}
\right]^{\frac{d}{2}}\,.
\ee
%where   $\mbox{vol}(B^d) = $ is the volume of the $d$-dimensional unit ball.

\subsection{A bound on arbitrary sphere packings}

Now we turn to general  sphere packings. Let $\omega$ be the optimal functional for the $U(1)^c$ bootstrap, leading to the bound $\DeltaU(c)$. Consider the radial function $g: \mathbb{R}^{2c} \to \mathbb{R}$ obtained by acting with $\omega$ on the crossing equation,
\begin{align}
&g(x) = \omega[ \Phi^{U}_{x^2/2} ]
\end{align}
where $\Phi^{U}_\Delta(\tau)$ is defined in \eqref{defF}. As explained above, $g$ obeys $g(0) = 0$, is odd under the Fourier transform in $x \in \mathbb{R}^{2c}$ and satisfies the positivity condition
\be
g(x) \geq 0 \quad \mbox{for all} \quad |x|  \geq \sqrt{2\DeltaU(c)} \ .
\ee
Suppose that one can construct {\it another} radial function $h(x): \mathbb{R}^{2c} \to \mathbb{R}$, which is even under the Fourier transform, and satisfies
\begin{align}
&(1) \quad h(0) > 0  \notag\\
&(2) \quad h(x) - g(x) \leq 0 \quad \textrm{for all} \quad |x| \geq \sqrt{2\Delta_U(c)}\\
&(3) \quad h(x) + g(x) \geq 0 \quad \textrm{for all}\quad x \in \mathbb{R}^{2c}\notag
\end{align}
The Cohn-Elkies theorem states that if such $h(x)$ exists, then the bound on the density \eqref{isodualbound} actually applies to arbitrary sphere packings. Experimentally, both with numerics and with the analytic functionals described below, we find that given $g(x)$, it is always possible to find $h(x)$ satisfying the above properties. This is essentially equivalent to the observation of Cohn and Elkies, stated above as conjecture 3.1, that $h(x)$ always exists given some choice of single and double zeroes inherited from $g(x)$.

\bigskip

Thus we have the conjecture that for all sphere packings in $\mathbb{R}^{d}$,
\be\label{generalbound}
\rho_d \leq \mathrm{vol}(B^d)\left[ \frac{\DeltaU(\frac{d}{2})}{2}\right]^{\frac{d}{2}} \ .
\ee
For all of the analytic and numerical functionals that have been constructed explicitly, this upper bound is actually a theorem, because in these cases we also have $h$.

\subsection{Asymptotics of the sphere-packing bounds at large $d$}

The large-$c$ (or equivalently large-$d$) limit is of great interest. In modular bootstrap, this limit is related to holographic theories of 3d gravity. In sphere packing the large-$d$ limit has applications to the construction of efficient codes \cite{ConwaySloane}.

The Minkowski lower bound on the sphere packing density, $\rho_d \geq 2^{-d}$, combined with the conjecture \eqref{generalbound}, leads to a lower bound on $\DeltaU(c)$,
\be
\DeltaU(c) \geq \frac{\Gamma(c+1)^{\frac{1}{c}}}{2\pi} \ .
\ee
For $c \gg 1$, this becomes
\be
\DeltaU(c) \geq \frac{c}{2\pi e} + o(c) \ .
\ee
Numerically, $\frac{1}{2\pi e} \approx 0.05855 \approx \frac{1}{17.08}$. At large $d$, the densest known sphere packings have these same asymptotics. It has been proved in \cite{torquato2006new} that the linear programming bound of Cohn and Elkies on the density cannot be better than $2^{-(0.7786\dots + o(1))d}$, which (assuming conjecture \eqref{generalbound}) translates into
\be
\DeltaU(c) \geq \frac{c}{12.57} + o(c) \ .
\ee 
The best upper bound on the density at large $d$ is that of Kabatiansky and Levenshtein (KL) \cite{kabatiansky1978bounds},
\be
 \rho_{d} \leq 2^{-(a + o(1))d} \ , 
 \ee
 with $a  = 0.5990\dots$. Cohn and Zhao \cite{cohn2014sphere} proved that linear programming is at least as strong as the KL bound (and improved this bound by a linear prefactor). Since the linear programming bound on isodual lattice packings is at least as strong as the linear programming bound on general packings, we find the rigorous inequality (not relying on the conjecture \eqref{generalbound})
 \be
 \frac{\pi^c}{\Gamma(c+1)}\left[ \frac{\DeltaU(c)}{2} \right]^{c} \leq 2^{-2(a+o(1))c} \ .
 \ee
 For $c\gg 1$, this gives
 \be
 \Delta_U(c) \leq \frac{c}{9.795} + o(c) .
 \label{eq:KLBound}
 \ee
 In summary, sphere packing bounds lead to
 \be
 \frac{c}{12.57} \lesssim \DeltaU(c) \lesssim \frac{c}{9.795}
 \ee
 at large $c$, with the lower bound conditional on conjecture \eqref{generalbound}. 
 The $U(1)^c$ modular bootstrap bound is at least as good as the best upper bound on general sphere packings, 
and it is of great interest to determine the asymptotic slope of $\DeltaU(c)$, either analytically or numerically.

We can compare these asymptotic bounds on $\DeltaU(c)$ with what is known about the bound in the Virasoro case at large $c$. It was shown in \cite{Friedan:2013cba} that $\Delta_V(c)$ can never drop below $(c-1)/12$ and prior to the present work, the best asymptotic upper bound on $\Delta_V(c)$ was that of Hellerman \cite{Hellerman:2009bu}
\be
\frac{c}{12} \lesssim \Delta_V(c) \lesssim \frac{c}{6}\,.
\ee
In this paper, we will improve the slope of the upper bound to $\Delta_V(c) \lesssim \frac{c}{8.503}$. The same technique also applies to $\Delta_U(c)$, but only leads to $\Delta_U(c) \lesssim \frac{c}{8.856}$, which is weaker than the KL bound \eqref{eq:KLBound}.

\subsection{Preview of analytic functionals}
So far, we have described the relation between sphere packing and the $U(1)^c$ bootstrap. Now we will connect both of these problems to the Virasoro bootstrap, relevant to generic CFTs and to 3D quantum gravity.  The Virasoro bootstrap is not equivalent to $U(1)^c$, but the key point is that exactly the same analytic functionals can be applied to both problems. These functionals have already been constructed in the bootstrap literature in the context of the four-point function bootstrap on a line \cite{Mazac:2016qev,Mazac:2018mdx}. This will reproduce the results of Viazovska et al.~in 8 and 24 dimensions \cite{viazovska8,viazovska24}, and unify the sphere packing solutions with new bounds on black holes from the Virasoro bootstrap. In this section we introduce the functionals, and preview some results of the more technical sections that follow.

\begin{figure}
\begin{center}
\begin{overpic}[scale=1.0]{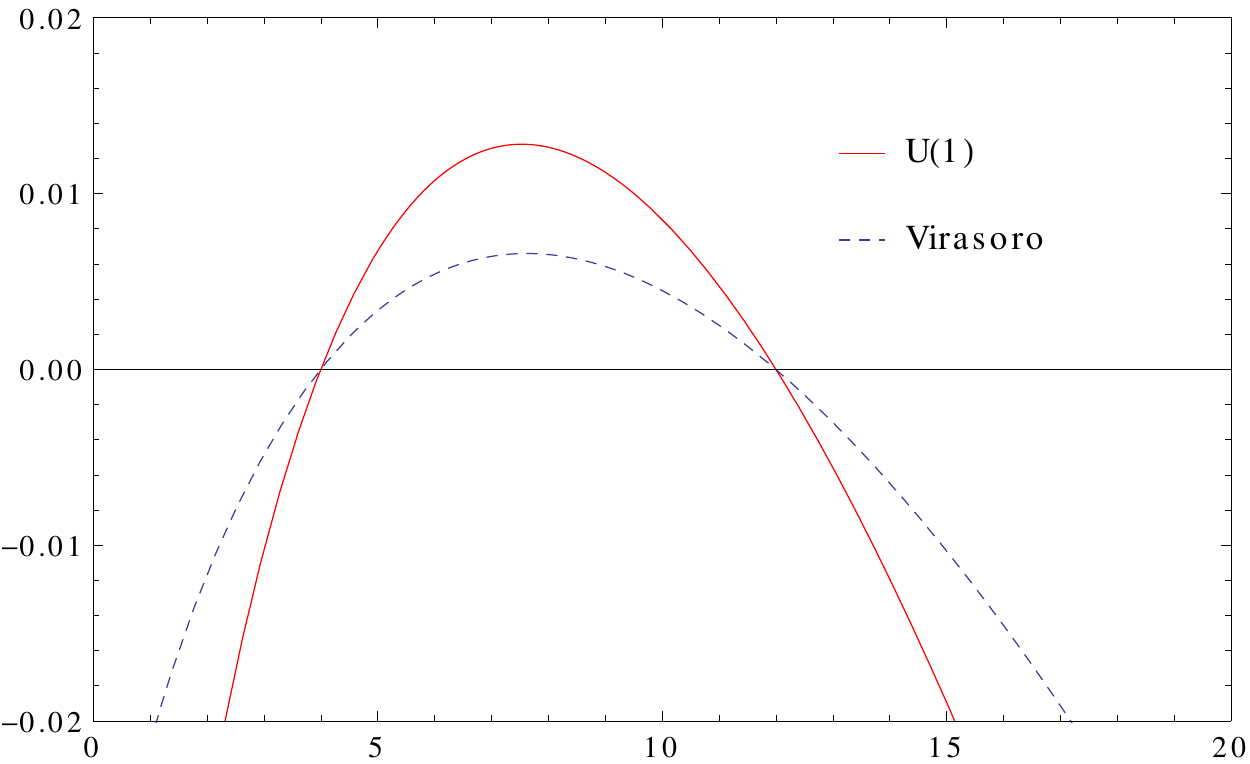}
\put(57,-3){$c$}
\put(-5,24){\rotatebox{90}{$\Delta_{U,V}(c) - (c+4)/8$}}
\end{overpic}
\end{center}
\caption{\small Upper bounds from linear programming on the gap for $U(1)^c$ and $\mathrm{Vir}_c$, denoted $\Delta_{U}(c)$ and $\Delta_V(c)$. We have subtracted $\frac{c+4}{8}$ from both bounds. $\frac{c+4}{8}$ is the optimal bound in the case of the four-point function bootstrap in 1D, translated to modular bootstrap variables. \label{fig:comparebounds}}
\end{figure}

First, let us directly compare the linear programming bounds for Virasoro and $U(1)^c$. Numerical results are shown in Figure \ref{fig:comparebounds}. We plotted $\Delta_{U}(c) - \frac{c+4}{8}$ and $\Delta_{V}(c) - \frac{c+4}{8}$. We can see that the bounds coincide with each other and with $\frac{c+4}{8}$ at $c=4$ and $c=12$. In other words
\ba
&\Delta_U(4) = \Delta_V(4) = 1\\
&\Delta_U(12) = \Delta_V(12) = 2\,.
\ea
Note that the $U(1)^c$ problem with $c=4,12$ maps precisely to the sphere-packing problem in $d=8,24$.

We provide the following explanation for the above behaviour of $\Delta_{U}(c)$ and $\Delta_{V}(c)$. Firstly, the torus partition function of a 2D CFT can be computed as the sphere four-point function of twist operators in the symmetric product orbifold of two copies of said theory \cite{Lunin:2000yv}. When the theory has central charge $c$, the twist operator has total scaling dimension
\be
\Df=\frac{c}{8}\,.
\ee
The Euclidean partition function on a torus of modulus $\tau$ maps to the Euclidean four-point function with cross-ratio
\be
z = \lambda(\tau)\,,
\label{eq:zFromTau1}
\ee
where $\lambda(\tau)=\frac{\theta_2(\tau)^4}{\theta_3(\tau)^4}$ is the modular lambda function, and the cross-ratio is related to the location of the twist operators $w_i\in\mathbb{R}^{2}$, $i=1,\ldots,4$ by
\be
|z|^2 = \frac{w^2_{12}w^2_{34}}{w^2_{13}w^2_{24}}\,,\quad
|1-z|^2 = \frac{w^2_{14}w^2_{23}}{w^2_{13}w^2_{24}}\,,
\ee
where $w_{ij}=w_i-w_j$. The $S$-transformation on the torus $\tau\leftrightarrow-1/\tau$ maps to the standard crossing transformation of four points $z\leftrightarrow 1-z$. The configurations which are relevant for the spinless modular bootstrap correspond to rectangular tori, i.e. $\tau\in i\,\mathbb{R}_{>0}$. Under \eqref{eq:zFromTau1}, this maps to the locus $z\in(0,1)$, which corresponds to configurations where the four twist operators are collinear. Therefore, spinless modular bootstrap at central charge $c$ is almost equivalent to the four-point function bootstrap in 1D with external operators of dimension $c/8$. The only difference is that the torus characters $\chi^{U}_{\Delta}(\tau)$ and $\chi^{V}_{\Delta}(\tau)$ do not map exactly to the conformal blocks of the 1D conformal algebra $sl(2,\mathbb{R})$. In fact, thanks to the large symmetry algebra of the symmetric product orbifold, a torus character of dimension $\Delta$ maps to a positive linear combination of the 1D conformal blocks of dimensions $2\Delta,\,2\Delta+2,\,2\Delta+4,\ldots$. The prefactor $2$ is present because a single primary of the original theory maps to two copies of that primary in the doubled theory.

The merit of the mapping from the torus partition function to the sphere four-point function is that the optimal upper bound on the gap coming from the four-point function bootstrap in 1D with $sl(2,\mathbb{R})$ blocks is known exactly, since the extremal functionals have been constructed in this case \cite{Mazac:2016qev,Mazac:2018mdx}. For any value of the external dimension $\Df>0$, the bound is saturated by fermionic mean field theory, where the gap in the spectrum of $sl(2,\mathbb{R})$ primaries is
\be
\Delta_{\textrm{1D}}(\Df) = 2\Df+1\,.
\ee
Therefore, if we could ignore the difference between the torus character $\chi^{U,V}_{\Delta}(\tau)$ and $sl(2,\mathbb{R})$ block of dimension $2\Delta$, the modular bootstrap bounds $\Delta_{U,V}(c)$ would both be equal to
\be
\frac{\Delta_{\textrm{1D}}(c/8)}{2} = \frac{c+4}{8}\,.
\label{eq:naiveBound}
\ee
It is precisely for $c=4$ and $c=12$ that the difference between torus characters and $sl(2,\mathbb{R})$ blocks plays no role, and \eqref{eq:naiveBound} becomes the correct bound for both $U(1)^c$ and $\mathrm{Vir}_c$. Furthermore, we can argue that $\Delta_{U,V}(c)<\frac{c+4}{8}$ for $c\in(1,4)$ and $c\in(12,\infty)$, as suggested by Figure \ref{fig:comparebounds}.

The claims of the previous paragraph can be proven simply by taking the extremal functionals for the 1D four-point function bootstrap, and applying them to modular bootstrap using the inverse of the mapping \eqref{eq:zFromTau1}
\be
\tau(z) = i \frac{K(1-z)}{K(z)} \ ,
\ee
where $K(z)$ is the complete elliptic integral of the first kind. For any $c>1$, this gives a linear functional $\beta_c$ acting on functions $\mathcal{F}(\tau)$ of $\tau\in\mathbb{H}_{+}$. As we will explain in the following sections, $\beta_c$ takes the form
\be
\beta_c[{\cal F}] = \int_{\frac{1}{2}}^1 dz  \mu_c(z) Q^{\beta}_{\frac{c}{8}}(z) {\cal F}(\tau (z)) + \frac{1}{2} \int_{\frac{1}{2}}^{\frac{1}{2} + i \infty} dz \mu_c(z) R^{\beta}_{\frac{c}{8}}(z) {\cal F}(\tau(z)) \ .
\ee
Here $\mu_c(z) = [2^8 z(1-z)]^{-\frac{c}{12}}$ is a measure arising from the Weyl transformation needed to go from the torus partition function to the sphere four-point function. The functional is specified by kernels $Q^{\beta}_{\Df}(z)$, $R^{\beta}_{\Df}(z)$. For general $\Df$, the kernels are given in terms of generalized hypergeometric functions. $R^{\beta}_{\Df}(z)$ takes the form \eqref{eq:fBeta}, and $Q^{\beta}_{\Df}(z) = - (1-z)^{2\Df-2}R^{\beta}_{\Df}\left(\mbox{$\frac{z}{z-1}$}\right)$. The special cases $c=4$ and $c=12$ map to $\Df=\frac{1}{2}$ and $\Df=\frac{3}{2}$. The kernels reduce to rational functions of $z$ at these points, see \eqref{eq:gBetaHalfs}.

We will see that $\beta_c$ has the following properties in the context of modular bootstrap:
\begin{itemize}
\item For any $c>1$, the functions of $\Delta$ given by $\beta_c[\Phi^{U}_{\Delta}]$ and $\beta_c[\Phi^{V}_{\Delta}]$ have a simple zero at
\be
\Delta_0 = \frac{c+4}{8}
\ee
and double zeroes at
\be
\Delta_n = \frac{c+4}{8} + n , \qquad n =1,2,\dots \ .
\ee
\item $\beta_c[\Phi^{U}_{\Delta}]$ and $\beta_c[\Phi^{V}_{\Delta}]$ are non-negative for $\Delta>\Delta_0$.
\item $\beta_c[\Phi^{U}_{\textrm{vac}}]$ and $\beta_c[\Phi^{V}_{\textrm{vac}}]$ both vanish for $c=4,12$ and are positive for $c\in(1,4)$ and $c\in(12,\infty)$.
\end{itemize}
It follows that for $c=4$ and $c=12$, $\beta_c$ is the optimal functional for both the $U(1)^c$ and $\mathrm{Vir}_c$ gap maximization problem. Moreover, for these values of $c$, the functions of $x\in\mathbb{R}^{2c}$ given by $\beta_c[\Phi^U_{x^2/2}]$ are precisely the Fourier-odd parts of the magic functions for sphere packing found by Viazovska \cite{viazovska8} and Cohn et al.~\cite{viazovska24}. We will also exhibit the linear functionals giving the Fourier-even part.

A result worth highlighting is that from the bootstrap point of view, the prefactor $\sin^2(\pi r^2/2)$ in Viazovska's ansatz has a very natural origin. We will see that it comes from a double discontinuity, an object which has played a central role in recent analytic approaches to the conformal bootstrap \cite{Hartman:2015lfa,Caron-Huot:2017vep}.

For $c\in(1,4)$, and $c\in(12,\infty)$, functional $\beta_c$ proves the upper bounds
\be
\DeltaU(c) <\frac{c+4}{8} , \qquad \DeltaVir(c) < \frac{c+4}{8} \ .
\ee
At large $c$, this improves on the Hellerman bound, $\DeltaVir(c) \lesssim \frac{c}{6}$. Away from $c=4,12$, these bounds are not optimal because the functional is positive, rather than zero, when acting on the modular bootstrap vacuum characters. (It vanishes on the $sl(2,\mathbb{R})$ vacuum conformal blocks, but the vacuum torus characters contain additional contributions from $sl(2,\mathbb{R})$ blocks of dimensions $2,4,\ldots$.) With a bit more work, in section \ref{ssec:ImprovedLargeC}, we will also find a functional that vanishes on the vacuum at large $c$, and leads to the asymptotic bounds 
\be
\DeltaU(c) \lesssim \frac{c}{8.856}\,,\qquad\DeltaVir(c) \lesssim \frac{c}{8.503} \ .
\ee
As noted in the introduction, the latter is slightly weaker than the conjectured true asymptotics based on extrapolating the numerical bound, $\DeltaVir(c) \sim c/9.08$ \cite{Afkhami-Jeddi:2019zci}.

The relations between various variables and objects entering the three problems  (sphere packing bounds, Virasoro modular bootstrap and the four-point bootstrap on a line) are summarized in Table \ref{tab:dictionary}.

\begin{table}
\centering
  \begin{tabular}{l l l l}
\thickhline
Sphere packing& Virasoro mod.~boot.  & 1D 4-point boot. & Relationship \\\hline
$d$: dimension of $\mathbb{R}^{d}$ & $c$: central charge& $\Df$: external dim. & $\Df=\frac{c}{8}=\frac{d}{16}$\\
$\tau$: torus modulus & $\tau$: torus modulus & $z$: 4-point cross ratio& $z(\tau)=\frac{\theta_2(\tau)^4}{\theta_3(\tau)^4}$\\
\multirow{2}{*}{$\Theta(\tau)$: theta fn.} & \multirow{2}{*}{$\mathcal{Z}(\tau)$: partition fn.} & \multirow{2}{*}{$\mathcal{G}(z)$: 4-point fn.} & $\mbox{\small$\mathcal{G}(z)$} \!=\! \mbox{\tiny$\left[\frac{2^{-8}z^2}{(1-z)}\right]^{\frac{c}{12}}$}\mbox{\small$\mathcal{Z}(\tau)$}$\\
&&& $\mbox{\small$\mathcal{Z}(\tau)$} \!=\! \frac{\Theta(\tau)}{\eta(\tau)^d}$\\
$r$: distance in $\mathbb{R}^d$ &$\Delta$: exchanged dim. & $\Delta_{\textrm{1D}}$: exchanged dim. & $\Delta_{\textrm{1D}}=2\Delta = r^2$\\
$\chi^{U}_{\frac{r^2}{2}}(\tau) = \frac{e^{i\pi r^2 \tau}}{\eta(\tau)^{d}}$ & $\chi^{V}_{\Delta}(\tau) =\frac{e^{2\pi i\Delta\tau}}{\eta(\tau)^2}$ & $G_{\Delta_{\textrm{1D}}}(z)$: eqn. \eqref{eq:sl2Block} & eqn. \eqref{eq:characterExpansion}\\
\vspace{-0.5cm}\\
\thickhline
  \end{tabular}
  \caption{Dictionary relating the bounds on sphere packing, the Virasoro modular bootstrap and the four-point function bootstrap on a line.}
  \label{tab:dictionary}
\end{table}

% !TEX root = ../ModularBootstrapV3.tex

\section{Analytic Extremal Functionals}\label{sec:ReviewFunctionals}
\subsection{Four-point function bootstrap with $sl(2,\mathbb{R})$}
We will now go through the arguments of the previous section in detail. Our strategy for understanding the modular bootstrap bounds $\Delta_{U}(c)$ and $\Delta_{V}(c)$ will be to relate them to another conformal bootstrap setup, for which the optimal bound and the extremal functionals are known analytically. This exactly-solved case is the conformal bootstrap of the four-point function of identical primaries restricted to a line. We will now review this setup and the analytic construction of the corresponding extremal functionals. Most of this section is a review of results that can be found in \cite{Mazac:2016qev} and \cite{Mazac:2018mdx}, where the reader can find more details. Only the construction of functionals for the generalized crossing equation in Section \ref{ssec:omegaPlus} is new. We use the language closer to the second reference because it is closer to Viazovska's construction of the magic functions.

Let us consider a local operator $\phi(x)$ in a unitary CFT in $D\geq1$ dimensions. We will study the correlation functions of $\phi(x)$ restricted to a spacelike line. The subalgebra of the $D$-dimensional conformal algebra which maps such a line to itself is $sl(2,\mathbb{R}) = so(1,2)$. We will take $\phi(x)$ to be primary under this $sl(2,\mathbb{R})$. Thus $\phi(x)$ can be e.g. a scalar primary or a component of a spinning primary of the full CFT. If $\phi(x)$ has bosonic statistics, its two-point function takes the form
\be
\langle\phi(x_1)\phi(x_2)\rangle = \frac{1}{|x_{12}|^{2\Delta_{\phi}}}\,,
\ee
where $x_i$ are the positions along the line, $x_{ij}\equiv x_i-x_j$, and $\Delta_\phi$ is the scaling dimension of $\phi(x)$. If $\phi(x)$ has fermionic statistics, we have instead
\be
\langle\phi(x_1)\phi(x_2)\rangle = \frac{\mathrm{sgn}(x_{12})}{|x_{12}|^{2\Delta_{\phi}}}\,.
\ee
Thanks to conformal symmetry, the four-point function can be written in terms of a single function $\cG (z)$ as follows:
\be
\langle\phi(x_1)\phi(x_2)\phi(x_3)\phi(x_4)\rangle = 
\langle\phi(x_1)\phi(x_2)\rangle\langle\phi(x_3)\phi(x_4)\rangle\cG (z)\,.
\ee
Here
\be
z \equiv \frac{x_{12}x_{34}}{x_{13}x_{24}}
\ee
is the unique cross-ratio of four points on a line. For the Euclidean correlator, $z$ ranges over real numbers. If $\phi(x)$ is a scalar primary in a $D>1$ CFT, we can write the four-point function in a general (i.e. not necessarily collinear) configuration as follows:
\be
\langle\phi(w_1)\phi(w_2)\phi(w_3)\phi(w_4)\rangle = 
\langle\phi(w_1)\phi(w_2)\rangle\langle\phi(w_3)\phi(w_4)\rangle\cG (z,\bar{z}) \ ,
\ee
where $w_i$ are positions in $\mathbb{R}^{D}$ and $z,\bar{z}$ are defined by
\be
z\bar{z} = \frac{w_{12}^2w_{34}^2}{w_{13}^2w_{24}^2}\qquad
(1-z)(1-\bar{z}) = \frac{w_{14}^2w_{23}^2}{w_{13}^2w_{24}^2}\,.
\ee
In the Euclidean signature, we have $z^*=\bar{z}$, where the star denotes complex conjugation. Restricting to collinear configurations is equivalent to setting $z=\bar{z}\in\mathbb{R}$
\be
\cG (z) = \cG (z,z)\,.
\ee
$\cG(z,\bar{z})$ is analogous to the torus partition function $Z(\tau,\bar{\tau})$ for general shape of the torus, while $\cG (z)$ is analogous to $\mathcal{Z}(\tau)=Z(\tau,-\tau)$, which describes only rectangular tori.

$\cG (z)$ has singularities at $z=0,1,\infty$, corresponding to the coincident point limits $x_2\rightarrow x_{1,3,4}$. Symmetry under permutations of the external operators fixes $\cG (z)$ in the intervals $z\in(-\infty,0)$ and $z\in(1,\infty)$ in terms of $\cG (z)$ in the interval $z\in(0,1)$. We will thus focus on $\cG (z)$ for $z\in(0,1)$ without loss of generality from now on. Symmetry under $x_1\leftrightarrow x_3$ implies the crossing symmetry of $\cG (z)$
\be
z^{-2\Df}\cG (z) = (1-z)^{-2\Df}\cG (1-z)\,.
\label{eq:crossing}
\ee
This equation holds no matter whether $\phi$ has bosonic or fermionic statistics. It is analogous to the symmetry of the torus partition function under the $S$ transformation.

The four-point function can be expanded using the OPE applied to the product $\phi(x_1)\phi(x_2)$. The OPE can be organized into irreducible representations of $sl(2,\mathbb{R})$, giving rise to
\be
\cG (z) =\!\! \sum\limits_{\mathcal{O}\in\times\phi\times\phi}\!\!(c_{\phi\phi\mathcal{O}})^2G_{\Delta_{\mathcal{O}}}(z)\quad\textrm{for}\quad z\in(0,1)\,.
\label{eq:4ptOPE}
\ee
The sum runs over the $sl(2,\mathbb{R})$ primaries appearing in the OPE and $c_{\phi\phi\mathcal{O}}$ is the corresponding OPE coefficient. $G_{\Delta_{\mathcal{O}}}(z)$ stands for the conformal block capturing the total contribution of $\mathcal{O}$ and its $sl(2,\mathbb{R})$ descendants to the four-point function,
\be
G_{\Delta}(z) = z^{\Delta}{}_2F_1(\Delta,\Delta;2\Delta;z)\,.
\label{eq:sl2Block}
\ee
We assume the theory is unitarity, so we can choose a basis of primary operators such that $c_{\phi\phi\mathcal{O}}\in\mathbb{R}$, and $(c_{\phi\phi\mathcal{O}})^2>0$. The existence of the OPE \eqref{eq:4ptOPE} guarantees that $\cG (z)$ can be analytically continued from $z\in(0,1)$ to the upper and lower half-plane and that the result is a function holomorphic in $\mathcal{R}\equiv\mathbb{C}\backslash(-\infty,0]\cup[1,\infty)$. We will denote this analytic continuation simply $\cG (z)$. $\cG (z)$ has a pair of branch cuts at $z\in(-\infty,0]$ and $z\in[1,\infty)$. The OPE \eqref{eq:4ptOPE} converges to $\cG (z)$ uniformly in any compact subset of $\mathcal{R}$.

It is natural to ask what is the physical meaning of the limit $z\rightarrow\infty$ of $\cG (z)$ taken in the upper or lower half-plane. This limit is known as the Regge, or chaos limit \cite{Costa:2012cb,Maldacena:2015waa}. To see this, we can consider the contour $z=\frac{1}{2}+ i t$ for $t\in[0,\infty)$. $\cG (1/2)$ is equal to the thermal four-point function of the CFT$_{D}$, quantized on the hyperbolic space $H_{D-1}$, with the four operators inserted at equal distances along the thermal circle. For $t>0$, operators $\phi(x_2)$ and $\phi(x_4)$ are evolved by a Lorentzian time. $\cG (1/2+i t)$ thus computes the out-of-time-order four-point function and $t\rightarrow\infty$ probes its late-time behaviour. It follows from positivity of $(c_{\phi\phi\mathcal{O}})^2$ that $\cG (z)$ satisfies a boundedness condition in this limit\footnote{This is not the `bound on chaos' proved in \cite{Maldacena:2015waa} but a simpler bound coming from a Cauchy-Schwartz inequality. See \cite{Hartman:2015lfa} for further discussion.} 
\be
|\cG (z)| = O(|z|^{2\Df})\quad\textrm{as}\quad|z|\rightarrow\infty\,.
\ee

The OPE can be combined with the crossing equation \eqref{eq:crossing} to get a sum rule known as the conformal bootstrap equation
\be
\sum\limits_{\mathcal{O}\in\times\phi\times\phi}\!\!(c_{\phi\phi\mathcal{O}})^2F_{\Delta_{\mathcal{O}}}(z) = 0\,,
\label{eq:4ptbootstrap}
\ee
where
\be
F_{\Delta}(z) = z^{-2\Df}G_{\Delta}(z) - (1-z)^{-2\Df}G_{\Delta}(1-z)\,.
\ee
Equation \eqref{eq:4ptbootstrap} holds everywhere in $\mathcal{R}$, with the sum converging uniformly in any compact subset of $\mathcal{R}$.

We can ask what is the maximal gap above the identity in the spectrum of scaling dimensions appearing in the OPE \eqref{eq:4ptOPE} compatible with unitarity and the bootstrap equation. As explained in \ref{ss:modularboundreview}, upper bounds on the gap can be produced by exhibiting suitable linear functionals $\omega$ acting on holomorphic functions in $\mathcal{R}$ \cite{Rattazzi:2008pe}. Specifically, if we can find $\omega$ satisfying
\ba
\omega[F_{0}]&>0\\
\omega[F_{\Delta}]&\geq0\quad\textrm{for all}\quad \Delta\geq\Delta_{\textrm{*}}\,,
\label{eq:functionalConditions}
\ea
then every unitary solution of \eqref{eq:4ptbootstrap} must contain a primary operator distinct from identity with $\Delta<\Delta_{*}$. The conclusion is derived by applying $\omega$ to \eqref{eq:4ptbootstrap} and swapping the action of $\omega$ with the infinite sum over $\mathcal{O}$. The swapping is automatically allowed for all functionals acting only in the interior of $\mathcal{R}$, such as is the case for the numerical bootstrap functionals consisting of a finite sum of derivatives at $z=\frac{1}{2}$. However, the requirement that the swapping is allowed is an important constraint on functionals whose support touches the boundary of $\mathcal{R}$, as will be the case for the extremal functionals constructed shortly \cite{Rychkov:2017tpc}.

The infimum of values $\Delta_{*}$ for which a functional satisfying \eqref{eq:functionalConditions} exists is the optimal upper bound on the gap, denoted $\Delta_{\textrm{1D}}(\Df)$. At the same time, $\Delta_{\textrm{1D}}(\Df)$ is the maximal gap above the identity among all solutions to \eqref{eq:4ptbootstrap}. Furthermore, there exists a functional $\beta_{\Df}$, called the extremal functional, satisfying \eqref{eq:functionalConditions} with $\Delta_{*}=\Delta_{\textrm{1D}}(\Df)$ and the first condition replaced by $\beta_{\Df}[F_{0}]=0$. This is because $\beta_{\Df}$ must annihilate all $F_{\Delta}$ present in the optimal solution. Indeed, suppose that the optimal solution contains non-identity primaries with dimensions $\Delta_n$ for $n=0,1,\ldots$, where $\Delta_0=\Delta_{\textrm{1D}}(\Df)$. Then $\beta_{\Df}[F_{\Delta}]$ must have a zero of odd order at $\Delta=\Delta_0$ and zeros of even order at $\Delta=\Delta_n$ for $n=1,2,\ldots$ to ensure $\beta[F_{\Delta}]\geq 0$ for all $\Delta\geq\Delta_0$. Typically, the zero at $\Delta_0$ is first-order and the zeros at the higher $\Delta_n$ are second-order.

It turns out that for any $\Df>0$, the solution of \eqref{eq:4ptbootstrap} with maximal gap is the four-point function of the elementary fields in fermionic mean field theory:
\be
\cG (z) = 1 + \left(\mbox{$\frac{z}{1-z}$}\right)^{2\Df}-z^{2\Df}\,.
\label{eq:gGFF}
\ee
The primary operators appearing in the OPE are the identity and double-trace operators of dimensions $\Delta_n = 2\Df + 2n+1$ for $n=0,1,\ldots$
\be
\cG (z) = 1 + \sum\limits_{n=0}^{\infty}(c_n)^2 G_{\Delta_n}(z)\,,
\ee
where the OPE coefficients take the form
\be
(c_n)^2 = \frac{2(2\Df)^2_{2 n+1}}{(2n+1)!(4\Df+2 n)_{2 n+1}}\,.
\ee
Clearly, the gap is
\be
\Delta_{\textrm{1D}}(\Df) = 2\Df + 1\,.
\ee
The strategy of the proof of optimality of this solution is to construct the extremal functional $\beta_{\Df}$ with the correct structure of zeros and positivity properties. We will now review the construction in more detail.

\subsection{Construction of the extremal functionals}\label{ssec:funConstruction}
In order to prove that the fermionic mean field theory is indeed the optimal solution to \eqref{eq:4ptbootstrap}, we must construct (for each $\Df>0$) a functional $\beta$ such that\footnote{Note the slight change of notation with respect to Section 4. There we found it clearer to label the $\beta$ functional  with the subscript $c$, here we use the subscript $\Delta_\phi = c/8$.}
\begin{enumerate}
\item $\beta_{\Df}[F_\Delta]$ has a simple zero at $\Delta=2\Df+1$.
\item $\beta_{\Df}[F_\Delta]$ has double zeros at $\Delta=2\Df+2n+1$ for $n=1,2,\ldots$.
\item $\beta_{\Df}[F_{\Delta}]\geq 0$ for all $\Delta\geq2\Df+1$.
\end{enumerate}
These properties are reminiscent of another object familiar from recent developments in the analytic conformal bootstrap: the double discontinuity \cite{Hartman:2015lfa,Caron-Huot:2017vep}. For our purposes, we will define the double discontinuity around $z=0$ as follows. Firstly, let us define $\cGT (z) \equiv z^{-2\Df}\cG (z)$, so that crossing symmetry reads $\cGT (z) = \cGT (1-z)$. The (fermionic) double discontinuity of $\cGT (z)$ is then given by
\be
\dDisc[\cGT (z)] = \cGT (z) +
(1-z)^{-2\Df}\frac{\cGT ^{\curvearrowleft}\!\left(\zTr\right)+\cGT ^{\text{\rotatebox[origin=c]{180}{\reflectbox{$\curvearrowleft$}}}}\!\left(\zTr\right)}{2}\quad\textrm{for}\quad z\in(0,1)\,.
\label{eq:dDiscDef}
\ee
This definition agrees with the standard double discontinuity in $D>1$, restricted to $z=\bar{z}$, when the external operators are fermions. See \cite{Mazac:2018qmi} for a more detailed discussion.
The symbols $\cGT ^{\curvearrowleft}(z)$, $\cGT ^{\text{\rotatebox[origin=c]{180}{\reflectbox{$\curvearrowleft$}}}}(z)$ denote the analytic continuation of $\cGT (z)$ from $z\in(0,1)$ to $z\in(-\infty,0)$ above and below the branch point $z=0$. The transformation $z\mapsto\zTr$ appears because it is a symmetry of the s-channel $sl(2,\mathbb{R})$ Casimir. Crucially, this implies the s-channel conformal blocks are invariant up to a phase,
\ba
G_{\Delta}^{\curvearrowleft}\!\left(\zTr\right) &= e^{i \pi \Delta} G_{\Delta}(z)\\
G_{\Delta}^{\text{\rotatebox[origin=c]{180}{\reflectbox{$\curvearrowleft$}}}}\!\left(\zTr\right) &= e^{-i \pi \Delta} G_{\Delta}(z)\,.
\label{eq:gTr}
\ea
Let us now apply $\dDisc$ to the contribution of a single s-channel conformal block of dimension $\Delta$, i.e. to $z^{-2\Df}G_{\Delta}(z)$. Using \eqref{eq:gTr}, we find that the three terms in \eqref{eq:dDiscDef} nicely combine to give
\be
\dDisc[z^{-2\Df}G_{\Delta}(z)] = 2\sin^2\left[\frac{\pi}{2}(\Delta-2\Df-1)\right] z^{-2\Df} G_{\Delta}(z)\,.
\ee
We conclude that for any $z\in(0,1)$, $\dDisc[z^{-2\Df}G_{\Delta}(z)]$ is non-negative for all $\Delta\geq0$ and has double zeros in $\Delta$ at $2\Df+1+2\mathbb{Z}$, which includes the spectrum of the fermionic mean field theory. It is therefore natural to write an ansatz for $\beta_{\Df}[F_{\Delta}]$ in the form of an integral of the double discontinuity times a weight-function $Q^{\beta}_{\Df}(z)$
\ba
\beta_{\Df}[F_{\Delta}] &= \int\limits_{0}^{1}\!\!dz\,Q^{\beta}_{\Df}(z)\dDisc[z^{-2\Df}G_{\Delta}(z)] =\\
&= 2\sin^2\left[\frac{\pi}{2}(\Delta-2\Df-1)\right] \int\limits_{0}^{1}\!\!dz\,Q^{\beta}_{\Df}(z)z^{-2\Df}G_{\Delta}(z)\,.
\label{eq:funAns}
\ea
Provided $Q^{\beta}_{\Df}(z)\geq 0$ for all $z\in(0,1)$ and provided the integral converges, this ansatz gives a non-negative function of $\Delta$ with double zeros at the fermionic double-trace dimensions, as required by properties 2 and 3 above. In order for $\Delta=2\Df+1$ to be a simple zero of $\beta_{\Df}[F_{\Delta}]$ rather than a double zero, we need to impose $Q^{\beta}_{\Df}(z) \sim d\,z^{-2}$ as $z\rightarrow 0^+$ for some $d> 0$. The integral then has a simple pole at $\Delta=2\Df+1$ since
\be
\int\limits_0^1\!\!dz\, z^{\Delta-2\Df-2} = \frac{1}{\Delta-2\Df-1}\,.
\ee
This simple pole combines with the double zero of $\sin^2\left[\frac{\pi}{2}(\Delta-2\Df-1)\right]$ to give a simple zero, as needed. %spacing
In order to complete the construction, we need to realize the ansatz \eqref{eq:funAns} as a linear functional acting on $F_\Delta(z)$. We will see that the requirement that this is possible uniquely fixes $Q^{\beta}_{\Df}(z)$. We claim that the following linear functional does the job\footnote{In reference \cite{Mazac:2018mdx}, $Q^{\beta}_{\Df}(z)$ and $R^{\beta}_{\Df}(z)$ were denoted respectively $g_{\beta}(z)$ and $f_{\beta}(z)$.}
\be
\beta_{\Df}[\mathcal{F}] = \int\limits_{\frac{1}{2}}^{1}\!\!dz\,Q^{\beta}_{\Df}(z) \mathcal{F}(z)
+ \frac{1}{2}\!\!\!\int\limits_{\frac{1}{2}}^{\frac{1}{2}+i\infty}\!\!\!dz\,R^{\beta}_{\Df}(z)\mathcal{F}(z)\,,
\label{eq:funDef}
\ee
where $R^{\beta}_{\Df}(z)$ is defined from $Q^{\beta}_{\Df}(z)$ by
\be
R^{\beta}_{\Df}(z) = -(1-z)^{2\Df-2}Q^{\beta}_{\Df}\left(\mbox{$\frac{z}{z-1}$}\right)
\ee
and $Q^{\beta}_{\Df}(z)$ is required to satisfy several constraints discussed below. Here $\mathcal{F}(z)$ is an arbitrary function holomorphic in $\mathcal{R}$ and satisfying $\mathcal{F}(z) = \mathcal{F}(1-z)$. Note that the first contour integral in \eqref{eq:funDef} probes the Euclidean region, including the Euclidean OPE limit $z\rightarrow 1$, while the second contour probes the out-of-time-order region $\frac{1}{2}+i t$, including the Regge/chaos limit $z\rightarrow i\infty$. When $\Delta>2\Df+1$, there is a contour deformation which takes $\beta_{\Df}[F_\Delta]$ from the form \eqref{eq:funDef} to the desired form \eqref{eq:funAns}. The strategy is to deform the contour in the second term in \eqref{eq:funDef} so that it lies on the real axis. The contour deformation is possible if and only if $Q^{\beta}_{\Df}(z)$ satisfies the following two constraints:
\begin{enumerate}
\item $R^{\beta}_{\Df}(z)=-(1-z)^{2\Df-2}Q^{\beta}_{\Df}\left(\mbox{$\frac{z}{z-1}$}\right)$ is a holomorphic function in $\mathbb{C}\backslash[0,1]$ satisfying
\be
R^{\beta}_{\Df}(z) = R^{\beta}_{\Df}(1-z)\,.
\label{eq:fSymmetry}
\ee
\item $Q^{\beta}_{\Df}(z)$ satisfies the functional equation
\be
Q^{\beta}_{\Df}(z) + Q^{\beta}_{\Df}(1-z) -
(1-z)^{2\Df-2}\frac{Q_{\Df}^{\beta\curvearrowleft}\!\left(\zTr\right)+Q_{\Df}^{\beta\text{\rotatebox[origin=c]{180}{\reflectbox{$\curvearrowleft$}}}}\!\left(\zTr\right)}{2} = 0
\label{eq:dDiscZero}
\ee
for $z\in(0,1)$.
\end{enumerate}
It follows from constraint 1 that $Q_{\Df}^{\beta}(z)$ is a holomorphic function in $\mathcal{R}$. Constraint 2 essentially says that the double discontinuity of $Q_{\Df}^{\beta}(z)$ around $z=0$ must vanish. Finally, we must impose constraints on the asymptotic behaviour of $Q_{\Df}^{\beta}(z)$ as $z\rightarrow0,1$:
\begin{enumerate}
\item[3.] $Q^{\beta}_{\Df}(z)=O((1-z)^{2\Df})$ as $z\rightarrow 1$ or equivalently $R^{\beta}_{\Df}(z)=O(z^{-2})$ as $z\rightarrow\infty$.
\item[4.] $Q^{\beta}_{\Df}(z) \sim \frac{2}{\pi^2} z^{-2}$ as $z\rightarrow0$.
\end{enumerate}
Constraint 3 is needed to ensure that $\beta_{\Df}[F_\Delta]$ is finite for all $\Delta\geq0$ and also that $\beta_{\Df}$ can be swapped with the OPE sum although the contours in the definition \eqref{eq:funDef} reach the boundary of $\mathcal{R}$. As discussed above, constraint 4 guarantees that $\beta_{\Df}[F_\Delta]$ has a simple zero at $\Delta=2\Df+1$ and unit slope there.

$Q^{\beta}_{\Df}(z)$ is uniquely fixed by constraints 1--4. For $\Df=1/2$ and $\Df=3/2$, $Q^{\beta}_{\Df}(z)$ is a rational function of $z$:
\ba
&\Df=\frac{1}{2}:\qquad Q^{\beta}_{\frac{1}{2}}(z) = \frac{(1-z) \left(2 z^2+z+2\right)}{\pi ^2 z^2}\,,\quad\quad
R^{\beta}_{\frac{1}{2}}(z) = -\frac{5 (z-1) z+2}{\pi ^2 (z-1)^2 z^2}\\
&\Df=\frac{3}{2}:\qquad Q^{\beta}_{\frac{3}{2}}(z) = \frac{(1-z)^3 \left(2 z^2+3 z+2\right)}{\pi ^2 z^2}\,,\quad\,
R^{\beta}_{\frac{3}{2}}(z) = -\frac{7 (z-1) z+2}{\pi ^2 (z-1)^2 z^2}\,.
\label{eq:gBetaHalfs}
\ea
For general $\Df$, $R^{\beta}_{\Df}(z)$ is given in terms of generalized hypergeometric functions
\ba
R^{\beta}_{\Df}(z) =
-\kappa(\Df)
&\frac{2z-1}{w^{3/2}}\left[\, _3\widetilde{F}_2\left(-\frac{1}{2},\frac{3}{2},2
   \Df+\frac{3}{2};\Df+1,\Df+2;-\frac{1}{4 w}\right)+\right.\\
	&\,\;\;\left.+\frac{9}{16 w} \,
   _3\widetilde{F}_2\left(\frac{1}{2},\frac{5}{2},2 \Df+\frac{5}{2};\Df+2,\Df+3;-\frac{1}{4w}\right)\right]\,,
\label{eq:fBeta}
\ea
where $w\equiv z(z-1)$, ${}_3\widetilde{F}_2$ stands for the \emph{regularized} hypergeometric function\footnote{The regularized hypergeometric function is defined as ${}_p\widetilde{F}_q(a_1,\ldots,a_p;b_1,\ldots,b_q;z)=[\Gamma(b_1)\cdots\Gamma(b_q)]^{-1}{}_p F_q(a_1,\ldots,a_p;b_1,\ldots,b_q;z)$} and
\be
\kappa(\Df) = \frac{\Gamma(4\Df+4)}{2^{8\Df+5}\Gamma(\Df+1)^2}\,.
\ee
It can be checked that $Q^{\beta}_{\Df}(z)>0$ for all $z\in(0,1)$ and all $\Df>0$, which guarantees $\beta_{\Df}[F_\Delta]\geq 0$ for all $\Delta>2\Df+1$. Finally, note that our $\beta_{\Df}$ automatically annihilates the identity vector $F_0(z)$ as a result of crossing symmetry of the fermionic mean-field four-point function \eqref{eq:gGFF} since it already annihilates $F_{2\Df+2n+1}(z)$ for all $n=0,1,\ldots$.

\subsection{The $\alpha$ functional}\label{ssec:alpha}
It will be useful to review another interesting functional introduced in \cite{Mazac:2018mdx}, called $\alpha$ in that reference. Just like $\beta_{\Df}[F_{\Delta}]$, also $\alpha_{\Df}[F_{\Delta}]$ has double zeros at $\Delta=2\Df+2n+1$ for $n=1,2,\ldots$, but instead of
\be
\beta_{\Df}[F_{2\Df+1}]=0\,,\qquad \beta_{\Df}[\partial_{\Delta}F_{2\Df+1}] = 1\,,
\ee
we have
\be
\alpha_{\Df}[F_{2\Df+1}]=1\,,\qquad \alpha_{\Df}[\partial_{\Delta}F_{2\Df+1}] = 0\,.
\ee
Again, there is a unique functional with these properties. Several expressions simplify if we work instead with the following linear combination of $\alpha_{\Df}$ and $\beta_{\Df}$
\be
\widetilde{\alpha}_{\Df} \equiv \alpha_{\Df} - \left[\frac{3}{2}H\!\left(\Df+\frac{1}{2}\right)-\frac{1}{2}H(\Delta _{\phi})-\log (2)\right] \beta_{\Df}\,,
\ee
where $H(x)$ is the harmonic number. $\widetilde{\alpha}_{\Df}$ is also defined as in \eqref{eq:funDef} using a suitable weight-function $Q^{\widetilde{\alpha}}_{\Df}(z)$. $Q^{\widetilde{\alpha}}_{\Df}(z)$ now needs to satisfy the same properties 1--3 above, with property 4 replaced by
\be
Q^{\widetilde{\alpha}}_{\Df}(z) = -\frac{2}{\pi^2}\left[\log(z)+\frac{3}{2}H\!\left(\Df+\frac{1}{2}\right)-\frac{1}{2}H(\Delta _{\phi})-\log (2) + o(1)\right]z^{-2}
\ee
as $z\rightarrow 0^+$. The extra $\log(z)$ means that the integral in \eqref{eq:funAns} develops a double pole at $\Delta=2\Df+1$, which cancels against the double zero of the prefactor, leading to nonzero $\widetilde{\alpha}_{\Df}[F_{2\Df+1}]$. The weight-function $R^{\widetilde{\alpha}}_{\Df}(z)\equiv -(1-z)^{2\Df-2}Q^{\widetilde{\alpha}}_{\Df}\left(\zTr\right)$ which solves all these constraints reads
\ba
R^{\widetilde{\alpha}}_{\Df}(z)=
\kappa(\Df)\frac{2(z-2)(z+1)}{(2z-1)w^{3/2}}
&\left[
{}_3\widetilde{F}_2\left(-\frac{1}{2},-\frac{1}{2},2\Df +\frac{3}{2};\Df +2,\Df +2;-\frac{1}{4 w}\right)+\right.\\
+\frac{(2 \Df +3) (2 \Df +5)}{16 w} &{}_3\widetilde{F}_2\left(\frac{1}{2},\frac{1}{2},2 \Df +\frac{5}{2};\Df +3,\Df +3;-\frac{1}{4 w}\right)-\\
-\frac{3 (4 \Df +5)}{256 w^2}&\left.{}_3\widetilde{F}_2\left(\frac{3}{2},\frac{3}{2},2 \Df +\frac{7}{2};\Df +4,\Df +4;-\frac{1}{4 w}\right)\right].
\label{eq:fAlphaTilde}
\ea
Unlike $\beta_{\Df}$, $\widetilde{\alpha}_{\Df}$ does not annihilate the identity vector $F_0$. Indeed, if we apply $\widetilde{\alpha}_{\Df}$ to the crossing equation for the fermionic mean field, we get
\be
\widetilde{\alpha}_{\Df}[F_0] = -(c_1)^2\widetilde{\alpha}_{\Df}[F_{2\Df+1}] = - 2\Df\,,
\ee
Among other things, $\widetilde{\alpha}_{\Df}$ can be used to make the proof of extremality of the fermionic mean field fully rigorous. Indeed, the extremal functional $\beta_{\Df}$ does not rule out the existence of potential solutions to crossing with spectrum consisting of identity and a subset of the fermionic mean field spectrum not including the operator at $\Delta=2\Df+1$. To fix this, we can consider the family of functionals $\beta_{\Df}-\epsilon\widetilde{\alpha}_{\Df}$ for small and positive $\epsilon$. This functional is positive when acting on identity, and non-negative for $\Delta\geq2\Df+1+\delta(\epsilon)$, where $\delta(\epsilon)\rightarrow 0^+$ as $\epsilon\rightarrow 0^+$. Thus, every unitary solution to crossing \eqref{eq:4ptbootstrap} must have a gap at most $2\Df+1$.

\subsection{Functionals for a generalized crossing equation}\label{ssec:omegaPlus}
The functionals $\widetilde{\alpha}_{\Df}$ and $\beta_{\Df}$ constructed above are useful when both sides of the crossing equation involve the same set of operators with identical coefficients. For the applications to the sphere-packing problem, it will be important to consider a generalization of the crossing equation where the s- and t-channel sum are allowed to be independent
\be
\sum\limits_{\mathcal{O}}c_{\mathcal{O}}^2 z^{-2\Df}G_{\Delta_{\mathcal{O}}}(z) =
\sum\limits_{\mathcal{P}}c_{\mathcal{P}}^2 (1-z)^{-2\Df}G_{\Delta_{\mathcal{P}}}(1-z)\,.
\ee
We will assume that both sums start with the identity operator $\Delta=0$, which appears with unit coefficient on both sides. We would like to maximize the gap in the s-channel with no constraint on the t-channel spectrum, assuming $c_{\mathcal{O}}^2,c_{\mathcal{P}}^2>0$. This problem is the $sl(2,\mathbb{R})$ analogue of the sphere-packing bootstrap problem discussed in Section \ref{ssec:LProgramming}, while the standard crossing equation \eqref{eq:4ptbootstrap} is the analogue of the sphere-packing bootstrap restricted to isodual lattices.

Let us denote
\ba
\widetilde{G}^{(s)}_{\Delta}(z) &\equiv z^{-2\Df}G_{\Delta}(z)\\
\widetilde{G}^{(t)}_{\Delta}(z) &\equiv (1-z)^{-2\Df}G_{\Delta}(1-z)\,.
\ea
In order to prove an upper bound $\Delta_{*}$ on the s-channel gap, we need to construct a linear functional $\omega$ such that
\begin{enumerate}
\item $\omega[\widetilde{G}^{(s)}_{0}]>\omega[\widetilde{G}^{(t)}_{0}]$
\item $\omega[\widetilde{G}^{(s)}_{\Delta}]\geq 0$ for all $\Delta\geq\Delta_*$
\item $\omega[\widetilde{G}^{(t)}_{\Delta}]\leq 0$ for all $\Delta\geq0$
\end{enumerate}
For the extremal functional, the first condition is replaced by $\omega[\widetilde{G}^{(s)}_{0}]=\omega[\widetilde{G}^{(t)}_{0}]$. We conjecture that the s-channel gap in this more general problem is still maximized by the fermionic mean field theory, with the same spectrum and OPE coefficients in both channels. This means that the extremal functional annihilates $\widetilde{G}^{(s)}_{2\Df+2n+1}$ and $\widetilde{G}^{(t)}_{2\Df+2n+1}$ for $n=0,1,\ldots$. We should then anticipate the following structure of zeros
\begin{enumerate}
\item $\omega[\widetilde{G}^{(s)}_{\Delta}]$ has a simple zero at $\Delta=2\Df+1$ with positive slope.
\item $\omega[\widetilde{G}^{(s)}_{\Delta}]$ has double zeros for all $\Delta=2\Df+2n+1$ where $n=1,2,\ldots$.
\item $\omega[\widetilde{G}^{(t)}_{\Delta}]$ has double zeros for all $\Delta=2\Df+2n+1$ where $n=0,1,\ldots$.
\end{enumerate}
We will construct the extremal functional by decomposing it into parts symmetric and antisymmetric under $z\leftrightarrow 1-z$
\be
\omega[\mathcal{F}(z)] = \omega_+\left[\frac{\mathcal{F}(z)+\mathcal{F}(1-z)}{2}\right] + \omega_-\left[\frac{\mathcal{F}(z)-\mathcal{F}(1-z)}{2}\right]\,.
\label{eq:omegaFull}
\ee
Let us denote
\be
F^{\pm}_{\Delta}(z) = z^{-2\Df}G_{\Delta}(z) \pm (1-z)^{-2\Df}G_{\Delta}(1-z)\,,
\ee
so that
\ba
\omega[\widetilde{G}^{(s)}_{\Delta}] &= \frac{\omega_+[F^{+}_{\Delta}(z)]+\omega_-[F^{-}_{\Delta}(z)]}{2}\\
\omega[\widetilde{G}^{(t)}_{\Delta}] &= \frac{\omega_+[F^{+}_{\Delta}(z)]-\omega_-[F^{-}_{\Delta}(z)]}{2}\,.
\ea
We will take $\omega_- = \beta_{\Df}$, i.e. the extremal functional for the standard crossing problem discussed above. It remains to fix $\omega_{+}$, which we will denote $\beta^{+}_{\Df}=\omega_{+}$ from now on.  $\beta^{+}_{\Df}[F^{+}_{\Delta}]$ must have the following structure of zeros
\begin{enumerate}
\item $\beta^{+}_{\Df}[F^{+}_{\Delta}]$ has a simple zero at $\Delta=2\Df+1$ with unit slope.
\item $\beta^{+}_{\Df}[F^{+}_{\Delta}]$ has double zeros for all $\Delta=2\Df+2n+1$ where $n=1,2,\ldots$.
\end{enumerate}
The construction of $\beta^{+}_{\Df}$ proceeds along the same lines as the construction of $\beta_{\Df}$. We first make the ansatz
\be
\beta^{+}_{\Df}[F^{+}_{\Delta}] =
2\sin^2\left[\frac{\pi}{2}(\Delta-2\Df-1)\right] \int\limits_{0}^{1}\!\!dz\,Q^{\beta+}_{\Df}(z)z^{-2\Df}G_{\Delta}(z)
\ee
and demand $Q^{\beta+}_{\Df}(z)\sim 2\pi^{-2}z^{-2}$ as $z\rightarrow 0^+$ so that the pole of the integral at $\Delta=2\Df+1$ produces a simple zero of $\beta^{+}_{\Df}[F^{+}_{\Delta}]$ with a unit derivative. Again, this ansatz arises from a genuine linear functional of the form
\be
\beta^{+}_{\Df}[\mathcal{F}_+] = \int\limits_{\frac{1}{2}}^{1}\!\!dz\,Q^{\beta+}_{\Df}(z) \mathcal{F}_+(z)
+ \frac{1}{2}\!\!\!\int\limits_{\frac{1}{2}}^{\frac{1}{2}+i\infty}\!\!\!dz\,R^{\beta+}_{\Df}(z)\mathcal{F}_+(z)\,,
\label{eq:omegaPlus}
\ee
where $R^{\beta+}_{\Df}(z) \equiv -(1-z)^{2\Df-2}Q^{\beta+}_{\Df}\left(\zTr\right)$, and where $Q^{\beta+}_{\Df}(z)$ is required to satisfy a set of functional constraints. Here $\mathcal{F}_+(z)$ is any function holomorphic in $\mathcal{R}$ satisfying $\mathcal{F}_+(z)=\mathcal{F}_+(1-z)$. The constraints on $Q^{\beta+}_{\Df}(z)$ are essentially the same as those on $Q^{\beta}_{\Df}(z)$, with a few extra minus signs sprinkled in
\begin{enumerate}
\item $R^{\beta+}_{\Df}(z) \equiv -(1-z)^{2\Df-2}Q^{\beta+}_{\Df}\left(\zTr\right)$ is a holomorphic function in $\mathbb{C}\backslash[0,1]$ satisfying
\be
R^{\beta+}_{\Df}(z) = -R^{\beta+}_{\Df}(1-z)\,.
\ee
\item $Q^{\beta+}_{\Df}(z)$ satisfies the functional equation
\be
Q^{\beta+}_{\Df}(z) - Q^{\beta+}_{\Df}(1-z) -
(1-z)^{2\Df-2}\frac{Q_{\Df}^{\beta+\curvearrowleft}\!\left(\zTr\right)+Q_{\Df}^{\beta+\text{\rotatebox[origin=c]{180}{\reflectbox{$\curvearrowleft$}}}}\!\left(\zTr\right)}{2} = 0
\ee
for $z\in(0,1)$.
\item[3.] $Q^{\beta+}_{\Df}(z)=O((1-z)^{2\Df})$ as $z\rightarrow 1$ or equivalently $R^{\beta+}_{\Df}(z)=O(z^{-2})$ as $z\rightarrow\infty$.
\item[4.] $Q^{\beta+}_{\Df}(z) \sim \frac{2}{\pi^2} z^{-2}$ as $z\rightarrow0^+$.
\end{enumerate}
We can solve for $Q^{\beta+}_{\Df}(z)$ for general $\Df>0$ using a similar procedure as used in \cite{Mazac:2018mdx} to find $Q^{\beta}_{\Df}(z)$. The details are in Appendix \ref{sec:kernelEven}. We find the unique solution
\be
R^{\beta+}_{\Df}(z) =
\frac{(4\Df+8) \kappa \left(\Df+\frac{1}{2}\right)}{(4\Df+5)w^{\frac{3}{2}}}
{}_3\widetilde{F}_2\left(-\frac{1}{2},\frac{3}{2},2\Df+\frac{5}{2};\Df+\frac{3}{2},\Df+\frac{5}{2};-\frac{1}{4w}\right)\,,
\label{eq:fEven}
\ee
where $w=z(z-1)$. The formula is valid for $z>1$ and extended to $\mathbb{C}\backslash[0,1]$ by analytic continuation from there. For example for $\Df=1/2$, this reduces to
\be
Q^{\beta+}_{\frac{1}{2}}(z) = \frac{\left\{\left(\sqrt{z}+1\right)^2 \left(2 z-3 \sqrt{z}+2\right) K\!\left[-\frac{\left(1-\sqrt{z}\right)^2}{4 \sqrt{z}}\right]-4 (z^2-z+1) E\!\left[-\frac{\left(1-\sqrt{z}\right)^2}{4 \sqrt{z}}\right]\right\}^2}{2 \pi ^2 (z-1)^2 z^{3/2}}\,,
\label{eq:gOmegaPlusHalf}
\ee
where $K(z)$ and $E(z)$ are respectively the complete elliptic integrals of the first and second kind.

Our contruction guarantees that $\omega$, given by \eqref{eq:omegaFull} has the right structure of simple and double zeros on the fermionic mean field operators in both channels. However, we still need to check that it has the right positivity properties. While we could not find a general proof of the correct positivity properties for arbitrary $\Df>0$, we checked that they are satisfied for many different values of $\Df$ and expect that they are valid for all $\Df>0$.

% !TEX root = ../ModularBootstrapV3.tex

\section{Analytic Functionals for Modular Bootstrap and Sphere Packing}\label{sec:Bounds}
\subsection{From $\tau$ to $z$}
Although the functionals $\alpha$ and $\beta$ of the previous section were derived in the context of the 1D four-point function bootstrap, they are also very useful for the modular bootstrap. This is because the torus partition function of a 2D CFT can be equivalently thought of as the sphere four-point function of twist operators in the $\mathbb{Z}_2$ symmetric product orbifold of the CFT. To see this, note that a complex torus of modulus $\tau$ can be presented as the following curve in $\mathbb{C}^2$
\be
y^2 = x(x-\lambda(\tau))(x-1)\,,
\label{eq:curve}
\ee
where $(x,y)\in\mathbb{C}^2$ and
\be
\lambda(\tau) = \frac{\theta_2(\tau)^4}{\theta_3(\tau)^4}\,.
\label{eq:lambda}
\ee
Here $\theta_2(\tau)$ and $\theta_3(\tau)$ are theta functions reviewed in Appendix \ref{s:modforms}. In other words, the torus is a double cover of the four-punctured sphere with punctures at $0,\lambda(\tau),1$ and $\infty$, where the covering map sends $(x,y)\mapsto x$.

Let us denote the original CFT by $\mathcal{T}$ and the product orbifold of two copies of $\mathcal{T}$ by $\mathrm{Sym}^2(\mathcal{T})$. The above covering gives us the following recipe for computing the torus partition function of $\mathcal{T}$. $\mathrm{Sym}^2(\mathcal{T})$ contains the $\mathbb{Z}_2$ twist operator $\sigma_2$, which has the property that going once around it is equivalent to switching the two copies of $\mathcal{T}$.\footnote{More precisely, we take $\sigma_2$ to be the vacuum in the twisted sector.} When $\mathcal{T}$ has central charge $c$, $\sigma_2$ is a scalar conformal primary of scaling dimension
\be
\Delta_{\sigma} = \frac{c}{8}\,.
\ee
Consider $\mathrm{Sym}^2(\mathcal{T})$ on $S^2$. If we place four twist operators at $x=0,\lambda(\tau),1,\infty$, the two copies of $\mathcal{T}$ will be connected in the right way to act as a single copy of $\mathcal{T}$ living on the curve \eqref{eq:curve}. This means that the torus partition function of $\mathcal{T}$ is simply related to the $S^2$ four-point function of $\sigma_2$. The only subtlety is that the partition function $Z(\tau,\bar{\tau})$ is defined on the torus with the flat metric, which is equal to the pull-back metric from flat $\mathbb{C}\backslash\{0,\lambda(\tau),1\}$ only up to a Weyl transformation. Let us write the four-point function of twist operators in flat space as
\be
\langle\sigma_2(w_1) \sigma_2(w_2)\sigma_2(w_3)\sigma_2(w_4)\rangle =
\langle\sigma_2(w_1) \sigma_2(w_2)\rangle\langle\sigma_2(w_3)\sigma_2(w_4)\rangle \cG(z,\bar{z})\,.
\ee
Then $\mathcal{G}(z,\bar{z})$ is related to $Z(\tau,\bar{\tau})$ as follows   \cite{Lunin:2000yv}
\be
\mathcal{G}(z,\bar{z}) = \left|\frac{z^2}{2^8(1-z)}\right|^{\frac{c}{12}}Z(\tau,\bar{\tau})\,,
\label{eq:GFromZ}
\ee
where $z=\lambda(\tau)$ and $\bar{z}=\lambda(-\bar{\tau})$. The prefactor comes from the Weyl transformation between the two metrics and nonzero conformal weight of the twist operators.

For the Euclidean partition function computed on physical tori, we have $\bar{\tau}=\tau^*$, which maps to the four-point function $\cG(z,\bar{z})$ computed in Euclidean signature, i.e. for $\bar{z}=z^*$. As discussed earlier, $Z(\tau,\bar{\tau})$ can in fact be analytically continued to a function of independent \emph{complex} variables $\tau$ and $\bar{\tau}$, which is holomorphic in $\mathbb{H}_{+}\times\mathbb{H}_{-}$. This produces an analytic continuation of $\cG(z,\bar{z})$ to arbitrary independent complex $z$ and $\bar{z}$.

\begin{figure}[ht]%
\begin{center}
\includegraphics[width=\textwidth]{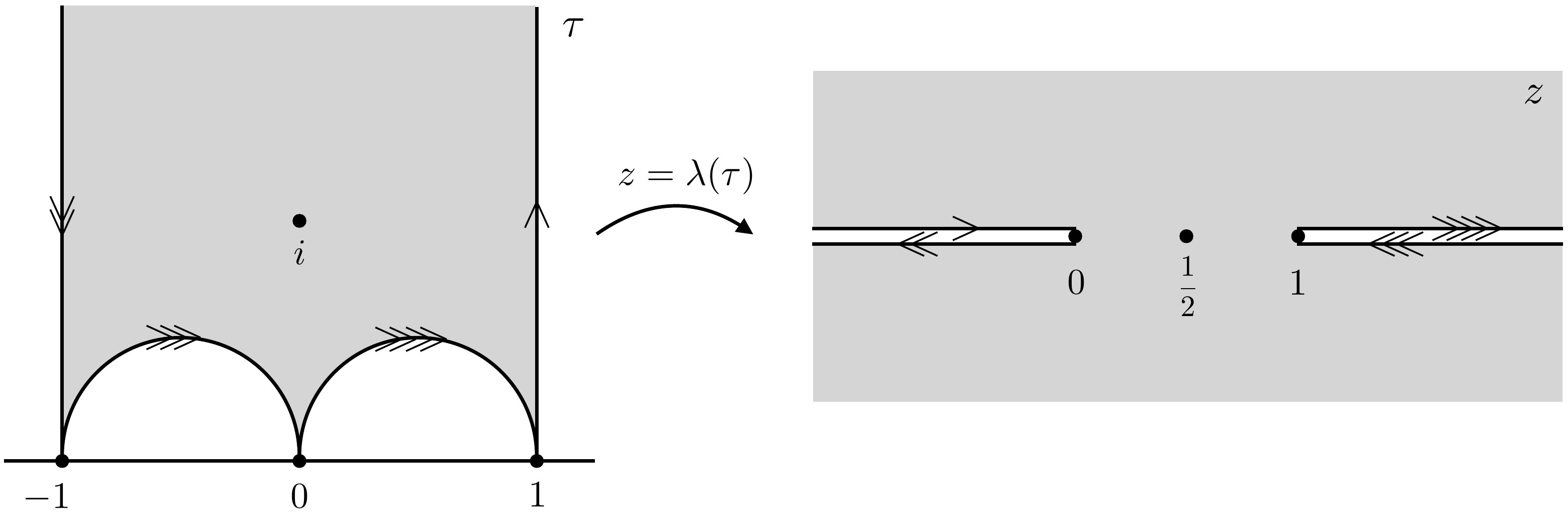}%
\caption{Left: The fundamental domain of $\Gamma(2)$ in the upper half-plane. Right: Its image in the space of the four-point cross-ratio $z$ under the mapping $\tau\mapsto\lambda(\tau)$. The image of the interior of the fundamental domain is the complex plane without the branch cuts $(-\infty,0]$ and $[1,\infty)$. The four boundary segments are mapped as shown by the different arrows.}
\label{fig:domains}%
\end{center}
\end{figure}

Let us describe the mapping $\tau\mapsto z=\lambda(\tau)$ in more detail. Firstly, there is a non-trivial group of conformal automorphisms of $\mathbb{H}_{+}$ which leave $\lambda(\tau)$ invariant. This group consists of all matrices in $SL(2,\mathbb{Z})$ which are congruent to the identity matrix modulo 2. It is denoted  $\Gamma(2)$.
\be
\lambda\left(\mbox{$\frac{a\tau+b}{c\tau+d}$}\right)=\lambda(\tau)\,\wedge\,
\begin{pmatrix}a & b\\c & d\end{pmatrix}\in SL(2,\mathbb{R}) \quad\Leftrightarrow\quad
\begin{pmatrix}a & b\\c & d\end{pmatrix}\in \Gamma(2)\,.
\ee
The fundamental domain for $\Gamma(2)$ can be chosen as the region in $\mathbb{H}_{+}$ bounded by the two lines $\mathrm{Re}(\tau)=1$ and $\mathrm{Re}(\tau)=-1$ and the two semi-circles of radius 1/2 centered at $\tau=1/2$ and $\tau=-1/2$. The interior of this region maps to $\mathcal{R}\equiv \mathbb{C}\backslash(-\infty,0]\cup[1,\infty)$ under $\lambda(\tau)$. The cusps of the fundamental domain get mapped as follows
\ba
&\lambda(i\infty) = 0\\
&\lambda(0)= 1\\
&\lambda(\pm1) =\pm i\infty\,.
\ea
The boundary vertical lines $\mathrm{Re}(\tau)=1$ and $\mathrm{Re}(\tau)=-1$ both map to $z\in(-\infty,0)$, while the boundary semicircles both map to $z\in(1,\infty)$. The fundamental domain as well as the details of the map from $\tau$ to $z$ are illustrated in Figure \ref{fig:domains}.

In other words, the low-temperature limit $\tau\rightarrow i\infty$, $\bar{\tau}\rightarrow-i\infty$ maps to the s-channel OPE limit $z,\bar{z}\rightarrow 0$, while the high-temperature limit $\tau,\bar{\tau}\rightarrow 0$ maps to the t-channel OPE limit $z,\bar{z}\rightarrow 1$. The u-channel OPE limit $z\rightarrow i\infty$, $\bar{z}\rightarrow -i\infty$ corresponds to $\tau,\bar{\tau}\rightarrow 1$.

More generally, $Z(\tau,\bar{\tau})$ is potentially singular for any $\tau,\bar{\tau}\in\mathbb{Q}$. What is the physical interpretation of these singularities in terms of the four-point function $\cG(z,\bar{z})$? If $\tau$ and $\bar{\tau}$ approach the same rational number, this is equivalent to one of the OPE limits above by a $\Gamma(2)$ transformation. What about when $\tau$ and $\bar{\tau}$ approach distinct rational numbers? These can be understood as various interesting limits  of $\cG(z,\bar{z})$ in Lorentzian kinematics. Indeed, if we fix $\bar{\tau}$ and move $\tau$ continuously from the inside to the outside of the fundamental domain of $\Gamma(2)$, $z$ will travel around the branch points at $z=0$ or $z=1$. This corresponds to a situation when one operator in the four-point function crosses the light-cone of another operator. By moving $\tau$ and $\bar{\tau}$ independently on the upper half-plane, we can reach an arbitrary Wightman function of the four twist operators on the Lorentzian cylinder, in any ordering. For example, the Regge limits have the following interpretation on $\mathbb{H}_{+}\times\mathbb{H}_{-}$:
\ba
&\textrm{s-channel Regge:}\quad z,\bar{z}\rightarrow 0\textrm{ after }z\circlearrowleft 1 \quad\Leftrightarrow\quad \tau\rightarrow -1/2,\,\bar{\tau}\rightarrow -i\infty\\
&\textrm{t-channel Regge:}\quad z,\bar{z}\rightarrow 1\textrm{ after }z \circlearrowright 0 \quad\Leftrightarrow\quad \tau\rightarrow 2,\,\bar{\tau}\rightarrow 0\\
&\textrm{u-channel Regge:}\quad z,\bar{z}\rightarrow i\infty \qquad\qquad\quad\Leftrightarrow\quad\tau\rightarrow 1,\,\bar{\tau}\rightarrow -1\\
\ea
In summary, while $\mathcal{G}(z,\bar{z})$ is not a single-valued function of $z,\bar{z}$ unless $\bar{z}=z^*$, by lifting it to $Z(\tau,\bar{\tau})$, we get a single-valued function in $\mathbb{H}_{+}\times\mathbb{H}_{-}$, which is also the full region of analyticity of the correlator.\footnote{The perspective of lifting a general four-point function to a function on $\mathbb{H}_{+}\times\mathbb{H}_{-}$ using \eqref{eq:lambda} was used in \cite{Maldacena:2015iua} to show that correlators in local and unitary 2D CFTs have no `bulk-point' singularities. The argument only works in 2D because the expansion of the correlator in Virasoro conformal blocks converges on the whole upper half-plane, whereas the expansion in global conformal blocks only converges in the fundamental domain of $\Gamma(2)$. It would be interesting to analyze whether four-point functions in general local unitary CFTs are always analytic in $\mathbb{H}_{+}\times\mathbb{H}_{-}$ and whether this perspective can be useful in the conformal bootstrap.}

It is not hard to see that modular invariance of $Z(\tau,\bar{\tau})$ becomes crossing symmetry of $\cG(z,\bar{z})$. Invariance under the $\Gamma(2)$ subgroup of $SL(2,\mathbb{Z})$ is manifest since $\cG(z,\bar{z})$ is single-valued in the Euclidean signature. It remains to understand invariance under the quotient
\be
SL(2,\mathbb{Z})/\Gamma(2) = S_3\,.
\ee
The three transpositions in $S_3$ have representatives $S$, $T$ and $STS$ in $PSL(2,\mathbb{Z})$. The two 3-cycles have representatives $ST$ and $TS$. Under the mapping $\lambda(\tau)$, this $S_3$ simply permutes the punctures at $0,1,\infty$:
\ba
\lambda\circ S(\tau) &= 1-\lambda(\tau)\\
\lambda\circ T(\tau) &= \frac{\lambda(\tau)}{\lambda(\tau)-1}\\
\lambda\circ STS(\tau) &= \frac{1}{\lambda(\tau)}\\
\lambda\circ ST(\tau) &= \frac{1}{1-\lambda(\tau)}\\
\lambda\circ TS(\tau) &= \frac{\lambda(\tau)-1}{\lambda(\tau)}\,.
\label{eq:S3Action}
\ea
In particular, $S$ becomes the usual crossing transformation $z\leftrightarrow1-z$ and $T$ becomes the transformation $z\leftrightarrow z/(z-1)$, which corresponds to switching operators 1 and 2. This is an order-two transformation because $T^2\in\Gamma(2)$.

Recall that we are interested in the spinless modular bootstrap, which amounts to restricting $\tau=-\bar{\tau}\in\mathbb{H}_+$. This maps to the restriction $z=\bar{z}\in\mathcal{R}$ at the level of the four-point function. As above, we write $\mathcal{Z}(\tau) \equiv Z(\tau,-\tau)$ and $\cG(z)\equiv\cG(z,z)$. It follows from \eqref{eq:GFromZ} that
\be
z^{-\frac{c}{4}}\mathcal{G}(z) = \left[2^8z(1-z)\right]^{-\frac{c}{12}}\mathcal{Z}(\tau(z))\,,
\label{eq:GFromZ2}
\ee
where $\tau(z)$ is a left inverse of $\lambda(\tau)$ sending $z\in\mathcal{R}$ to the fundamental domain of $\Gamma(2)$
\be
\tau(z) = i\frac{K(1-z)}{K(z)}\,.
\ee
Here $K(z)$ is the elliptic integral
\be
K(z) \equiv \frac{1}{2}\int\limits_{0}^{1}\frac{dt}{\sqrt{t(1-t)(1-t z)}} = \frac{\pi}{2}\,
{}_2F_1\left(\frac{1}{2},\frac{1}{2};1;z\right)\,.
\ee
The only transformation in the modular group which respects the identification $\tau=-\bar{\tau}$ is $S$. It becomes the crossing symmetry of the four-point function
\be
z^{-\frac{c}{4}}\cG(z) = (1-z)^{-\frac{c}{4}}\cG(1-z)\,.
\label{eq:GfromZ2}
\ee

\subsection{Torus characters and conformal blocks}
The expansion of the torus partition function into characters maps to the OPE of the four-point function of twist operators. However, the torus characters do not simply map to the $sl(2,\mathbb{R})$ conformal blocks appropriate for the 1D bootstrap discussed in Section \ref{sec:ReviewFunctionals}. This is because $\mathrm{Sym}^2(\mathcal{T})$ has a large chiral algebra and the torus characters become the conformal blocks of the full chiral algebra. In general, if $\mathcal{T}$ has a (left-moving) chiral symmetry algebra $ A $, then the left-moving chiral algebra of $\mathrm{Sym}^2(\mathcal{T})$ is $( A \otimes A )/\mathbb{Z}_2$. This means that under the mapping \eqref{eq:lambda} the torus characters appropriate for chiral algebra $ A $ become the sphere conformal blocks for the chiral algebra $( A \otimes A )/\mathbb{Z}_2$ and external twist operators.

It follows that local operators in theory $\mathcal{T}$ which are primary under $ A $ (and its right-moving counterpart) must be in one-to-one correspondence with local operators of $\mathrm{Sym}^2(\mathcal{T})$ which are primary under $( A \otimes A )/\mathbb{Z}_2$ (and its right-moving counterpart) and which appear in the OPE of two twist operators. To see this in a different way, first note that the $\sigma_2\times\sigma_2$ OPE can only contain operators from the untwisted sector. There is a basis for primaries in the untwisted sector consisting of $\mathcal{P}_{ij}=(\cO_i,\cO_j)+(\cO_j,\cO_i)$, where $\cO_i$ span a basis of primaries of $\mathcal{T}$. We claim that $\mathcal{P}_{ij}$ appears in the $\sigma_2\times\sigma_2$ OPE if and only if $i=j$. Indeed, we can compute the three-point function $\langle\sigma_2\sigma_2\mathcal{P}_{ij}\rangle$ by going to the covering space, where it becomes the sphere two-point function $\langle\cO_i \cO_j\rangle$ in theory $\mathcal{T}$. This two-point function is nonzero if and only if $i=j$. The conclusion is that
\be
\sigma_2\times\sigma_2 = \sum\limits_{i} (\cO_i,\cO_i)+\textrm{descendants}\,,
\ee
where the sum runs over primaries of $\mathcal{T}$. Let us recall the expansion of the partition function in the characters of $ A $
\be
\mathcal{Z}(\tau) = \sum\limits_{i}\chi^{ A }_{\Delta_i}(\tau)\,.
\ee
We have explained that this becomes the OPE of $\cG(z)$
\be
\cG(z) = \sum_i G_{2\Delta_i}^{ A }(z)\,,
\ee
where $G_{2\Delta_i}^{ A }(z)$ is the conformal block capturing the contributions of $(\cO_i,\cO_i)$ and all of its descendants under both left- and right-moving $( A \otimes A )/\mathbb{Z}_2$ (recall that we are working on the diagonal locus $z=\bar{z}$). We use $2\Delta$ instead of $\Delta$ as the label of the conformal block since this is the scaling dimension of the primary under the dilatation operator of $\mathrm{Sym}^2(\mathcal{T})$. Using \eqref{eq:GfromZ2}, we get
\be
G^{ A }_{2\Delta}(z) =
\left[\frac{z^2}{2^8(1-z)}\right]^{\frac{c}{12}}\chi^{ A }_{\Delta}(\tau(z))\,.
\label{eq:blockFromChi}
\ee
The global conformal algebra $sl(2,\mathbb{R})$ of the spacelike line to which the twist operators are restricted is a subalgebra of the full $\left[( A \otimes A )/\mathbb{Z}_2\right]_L\times\left[( A \otimes A )/\mathbb{Z}_2\right]_R$. It follows that $G^{ A }_{2\Delta}(z)$ admits an expansion into the $sl(2,\mathbb{R})$ conformal blocks \eqref{eq:sl2Block} considered in the previous section
\be
G^{ A }_{2\Delta}(z) = \sum\limits_{n=0}^{\infty} a_n G_{2\Delta+2n}(z)\,.
\label{eq:gFullExpansion1}
\ee
Only $sl(2,\mathbb{R})$ blocks with dimensions of the form $2\Delta+2m$ appear in the expansion because the contribution of odd-level descandants is fully contained in the individual $sl(2,\mathbb{R})$ blocks. It is instructive to prove this claim using the transformation of $G_{\Delta}(z)$ under $z\mapsto z/(z-1)$, see \eqref{eq:gTr}. First, note that $q(z)=e^{2\pi i \tau(z)}$ admits a Taylor expansion in $z$, starting with $z^2$:
\be
q(z) =\frac{z^2}{256}+\frac{z^3}{256}+\frac{29 z^4}{8192}+\frac{13 z^5}{4096}+\frac{11989 z^6}{4194304} + O(z^7)\,.
\ee
Recall the general form of the torus character for arbitrary $ A $
\be
\chi^{ A }_{\Delta}(\tau) = q^{\Delta-\frac{c}{12}}\left[1+\sum\limits_{j=1}^{\infty}b_j q^j\right]\,.
\ee
It follows from \eqref{eq:blockFromChi} that $G^{ A }_{2\Delta}(z)$ admits an expansion in powers of $z$ of the form $z^{2\Delta+j}$ with $j=0,1,\ldots$. Such series can always be rearranged to a sum over $sl(2,\mathbb{R})$ blocks
\be
G^{ A }_{2\Delta}(z)= \sum\limits_{j=0}^{\infty}c_j G_{2\Delta+j}(z)\,.
\label{eq:gFullExpansion2}
\ee
It remains to be demonstrated that only terms with $j$ even appear in the sum. We will now use the transformation \eqref{eq:gTr} of $G_{2\Delta+j}(z)$:
\be
G_{2\Delta+j}^{\curvearrowleft}\!\left(\zTr\right) = (-1)^j e^{2\pi i \Delta} G_{2\Delta+j}(z)\,.
\ee
At the same time, it is not difficult to show from \eqref{eq:blockFromChi} that $G^{ A }_{2\Delta}(z)$ satisfies
\be
G^{ A \curvearrowleft}_{2\Delta}\!\left(\zTr\right) = e^{2\pi i \Delta} G^{ A }_{2\Delta}(z)\,.
\label{eq:gTr2}
\ee
Indeed, the continuation from $z$ to $\mapsto z/(z-1)$ above $z=0$ maps to the continuation $\tau\mapsto\tau+1$, under which $q^{\Delta}$ picks up a phase $e^{2\pi i \Delta}$. Equation \eqref{eq:gTr2} is only compatible with \eqref{eq:gFullExpansion2} if only terms with even $j$ appear.

Furthermore, the coefficients $a_n$ in \eqref{eq:gFullExpansion1} have to be positive. This is because $G^{ A }_{2\Delta}(z)$ with $z\in(0,1)$ can be interpreted as the norm of a state in radial quantization, and \eqref{eq:gFullExpansion1} expresses this norm as a sum over norms of that state projected to orthogonal subspaces corresponding to irreducible representations of $sl(2,\mathbb{R})$. For concreteness, when $ A $ is respectively $U(1)^c$ and $\mathrm{Vir}_c$, we obtain
\ba
G^{U}_{2\Delta}(z) &= 2^{-8\Delta}\left[G_{2\Delta}(z) + \frac{8 \Delta ^2-6 \Delta + 4\Delta c +c}{64 (4 \Delta +1)}G_{2\Delta+2}(z) + \ldots \right]\\
G^{V}_{2\Delta}(z) &= 2^{-8\Delta}\left[G_{2\Delta}(z) + \frac{16 \Delta ^2-8 \Delta+4\Delta c +c +1}{128 (4 \Delta +1)}G_{2\Delta+2}(z) + \ldots \right]\,.
\label{eq:characterExpansion}
\ea
Recalling $c \geq 1$, $\Delta \geq 0$, the coefficient shown are indeed positive.

In summary, we have explained that the spinless modular bootstrap in the presence of chiral algebra $ A _L\otimes A _R$ takes the form of the four-point function bootstrap with four external operators of dimension $c/8$ restricted to a line, with conformal blocks related to the $sl(2,\mathbb{R})$ blocks by \eqref{eq:gFullExpansion1}. This will allow us to straightforwardly use the analytic extremal functionals reviewed in Section \ref{sec:ReviewFunctionals} for the modular bootstrap problem.

\subsection{Saturation at $c=4$}
Let us consider the modular bootstrap equation in the presence of chiral algebra $ A _L\otimes A _R$. We have explained that it can be cast as a four-point crossing equation
\be
F^{ A }_{\textrm{vac}}(z) + \sum\limits_{i} F^{ A }_{2\Delta_i}(z) = 0\,,
\label{eq:modBootstrapZ}
\ee
where
\ba
F^{ A }_{2\Delta}(z)
&= z^{-\frac{c}{4}}G^{ A }_{2\Delta}(z)-(1-z)^{-\frac{c}{4}}G^{ A }_{2\Delta}(1-z) =\\
&= \left[2^8z(1-z)\right]^{-\frac{c}{12}}\left[\chi^{ A }_{\Delta}(\tau(z))-\chi^{ A }_{\Delta}(-1/\tau(z))\right]\\
F^{ A }_{\textrm{vac}}(z)
&= \left[2^8z(1-z)\right]^{-\frac{c}{12}}\left[\chi^{ A }_{\textrm{vac}}(\tau(z))-\chi^{ A }_{\textrm{vac}}(-1/\tau(z))\right]\,.
\label{eq:FmodZ}
\ea
We will now specialize the discussion to $ A =U(1)^c$ and $ A =\mathrm{Vir}_c$. We would like to learn as much as possible about the optimal upper bounds on the gap $\Delta_{U}(c)$ and $\Delta_{V}(c)$ using our knowledge about the $sl(2,\mathbb{R})$ crossing problem. Firstly, note that a unitary solution of the $U(1)^c$ and $\mathrm{Vir}_c$ modular problem with gap $\Delta$ gives a unitary solution of the $sl(2,\mathbb{R})$ crossing problem with gap $\min(2\Delta,2)$. The gap can not be greater than 2 because the torus vacuum character contains an $sl(2,\mathbb{R})$ block of dimension 2 in its decomposition. Since the optimal upper bound on the gap in the $sl(2,\mathbb{R})$ problem with external dimension $\Df=\frac{c}{8}$ is $2\Df+1=\frac{c}{4}+1$, we find for all $c>1$
\be
\min(\Delta_{U,V}(c),1)\leq\frac{c}{8}+\frac{1}{2}\,.
\ee
The right-hand side is smaller than 1 for $c<4$, so that in that case we must in fact have
\be
\Delta_{U,V}(c) \leq \frac{c}{8}+\frac{1}{2}\quad\textrm{for}\quad c\in(1,4)\,.
\label{eq:boundSmallC}
\ee
Let us now show that for $c=4$, the optimal solutions of the $U(1)^c$, $\mathrm{Vir}_c$ and $sl(2,\mathbb{R})$ bootstrap all coincide, which in particular implies
\be
\Delta_{U}(4) = \Delta_{V}(4) = 1\,.
\ee
$c=4$ maps to crossing with $\Df=\frac{1}{2}$. We already know that in that case the optimal solution of $sl(2,\mathbb{R})$ bootstrap is the fermionic mean field correlator
\be
\mathcal{G}(z) = 1 +\frac{z}{1-z}-z = \frac{1-z+z^2}{1-z}\,,
\ee
which decomposes into $sl(2,\mathbb{R})$ blocks of dimensions $0,2,4,\ldots$. Using \eqref{eq:GFromZ2}, this maps to the partition function
\be
\mathcal{Z}(\tau) = \left[\frac{2^8(1-\lambda(\tau))}{\lambda(\tau)^2}\right]^{\frac{1}{3}}\mathcal{G}(\lambda(\tau)) = \frac{E_4(\tau)}{\eta(\tau)^8}\,,
\label{eq:extremalZ4}
\ee
where $E_4(\tau)$ is an Eisenstein series and $\eta(\tau)$ is the Dedekind eta function, both of which are reviewed in Appendix \ref{s:modforms}. The last equality follows from
\ba
\lambda(\tau) &= \frac{\theta_{2}(\tau)^{4}}{\theta_{3}(\tau)^4}\\
2E_4(\tau) &= \theta_{2}(\tau)^{8}+\theta_{3}(\tau)^{8} +\theta_{4}(\tau)^{8}\\
2^4\eta(\tau)^{12} &= \theta_{2}(\tau)^{4}\theta_{3}(\tau)^{4}\theta_{4}(\tau)^{4}\,.
\ea
The decomposition of $\mathcal{Z}(\tau)$ into $U(1)^4$ or $\mathrm{Vir}_4$ characters contains the vacuum and operators of dimensions $1,2,\ldots$. This is of course consistent with the fact that the decomposition into $sl(2,\mathbb{R})$ conformal blocks contains only dimensions $0,2,4,\ldots$. Furthermore, one can see that the coefficients in the $U(1)^4$ or $\mathrm{Vir}_4$ decompositions are positive. For example, the $U(1)^4$ characters are equal to $\eta(\tau)^{-8}q^{\Delta}$, so that decomposition of $\mathcal{Z}(\tau)$ simply amounts to the power series of $E_4(\tau)$ around $q=0$, which is known to have positive coefficients. Explicitly, we find
\ba
\mathcal{Z}(\tau) &=\chi^{U}_{\textrm{vac}}(\tau) + 240 \chi^{U}_{1}(\tau) + 2160 \chi^{U}_{2}(\tau) + 6720\chi^{U}_{3}(\tau) + \ldots\\
\mathcal{Z}(\tau) &=\chi^{V}_{\textrm{vac}}(\tau) + 248 \chi^{V}_{1}(\tau) + 3626 \chi^{V}_{2}(\tau) + 26258\chi^{V}_{3}(\tau) + \ldots\,.
\ea
One may wonder whether $\mathcal{Z}(\tau)$ arises as a partition function of a full-fledged unitary CFT after specializing to $\tau=-\bar{\tau}$. The answer is yes, the CFT being 8 free fermions with diagonal GSO projection, as pointed out in \cite{Collier:2016cls}. The full partition function reads
\be
Z(\tau,\bar{\tau}) = \frac{\theta_{2}(\tau)^{4}\theta_{2}(-\btau)^{4}+\theta_{3}(\tau)^{4}\theta_{3}(-\btau)^{4}+\theta_{4}(\tau)^{4}\theta_{4}(-\btau)^{4}}{2\eta(\tau)^4\eta(-\bar{\tau})^4}\,.
\ee
From the point of view of the sphere-packing problem, $\mathcal{Z}(\tau)$ corresponds to the $E_8$ lattice packing. Indeed, the theta function of the $E_8$ lattice is precisely the Eisenstein series $E_4(\tau)$
\be
E_4(\tau) = \sum\limits_{x\in\Lambda_8} e^{i\pi\tau x^2}\,,
\ee
where $\Lambda_8$ stands for the $E_8$ lattice.

The nontrivial task is to show that there are no unitary solutions of \eqref{eq:modBootstrapZ} for $c=4$ with gap greater than one. This can be proven using the same extremal functional which also proves extremality of the free fermion for the $sl(2,\mathbb{R})$ problem, i.e. functional $\beta$ reviewed in Section \ref{sec:ReviewFunctionals}. Let us keep $c=8\Df$ general for now and consider the action of $\beta_{\Df}$ on the functions $F_{2\Delta}^{U,V}(z)$ defined in \eqref{eq:FmodZ}. Recall from \eqref{eq:gFullExpansion1} that
\be
F_{2\Delta}^{U,V}(z) = \sum\limits_{n=0}^{\infty} a^{U,V}_n F_{2\Delta+2n}(z)\,,
\label{eq:fVecSeries}
\ee
where $F_{\Delta}(z)$ are the functions entering the $sl(2,\mathbb{R})$ bootstrap and $a^{U,V}_n>0$. Let us also recall the definition of functional $\beta_{\Df}$, i.e. \eqref{eq:funDef} which we repeat here for convenience
\be
\beta_{\frac{c}{8}}[\mathcal{F}] = \int\limits_{\frac{1}{2}}^{1}\!\!dz\,Q^{\beta}_{\frac{c}{8}}(z) \mathcal{F}(z)
+ \frac{1}{2}\!\!\!\int\limits_{\frac{1}{2}}^{\frac{1}{2}+i\infty}\!\!\!dz\,R^{\beta}_{\frac{c}{8}}(z)\mathcal{F}(z)\,,
\ee
where $R^{\beta}_{\frac{c}{8}}(z)=-(1-z)^{\frac{c}{4} - 2} Q^{\beta}_{\frac{c}{8}}\!\left(\zTr\right)$ is given in \eqref{eq:fBeta} for general $c$. When we let $\beta_{\frac{c}{8}}$ act on $F_{2\Delta}^{U,V}(z)$ and use \eqref{eq:fVecSeries}, the sum over $n$ converges uniformly inside both integrals, which means we can swap the sum and action of $\beta_{\frac{c}{8}}$:
\be
\beta_{\frac{c}{8}}[F_{2\Delta}^{U,V}] = \sum\limits_{n=0}^{\infty} a^{U,V}_n \beta_{\frac{c}{8}}[F_{2\Delta+2n}]\,.
\label{eq:betaSeries}
\ee
Now, $\beta_{\frac{c}{8}}[F_{\Delta}(z)]$ is non-negative for $\Delta\geq \frac{c}{4}+1$ with a simple zero at $\Delta =  \frac{c}{4}+1$ and double zeros at $\Delta=\frac{c}{4}+2n+1$ for $n=1,2,\ldots$. It follows that $\beta_{\frac{c}{8}}[F_{2\Delta}^{U,V}]$ is non-negative for $\Delta\geq\frac{c+4}{8}$, has a simple zero at $\Delta=\frac{c+4}{8}$ and double zeros at $\Delta=\frac{c+4}{8}+n$ for $n=1,2,\ldots$. There is also the following more direct way to reach this conclusion. When $\beta_{\frac{c}{8}}$ acts on $F_{2\Delta}^{U,V}$, we can use the same contour deformation as when it acts on $F_{\Delta}$ to show that for $\Delta>\frac{c+4}{8}$
\be
\beta_{\frac{c}{8}}[F_{2\Delta}^{U,V}]
= 2\sin^2\!\left[\pi\left(\Delta-\frac{c+4}{8}\right)\right] \int\limits_{0}^{1}\!\!dz\,Q^{\beta}_{\frac{c}{8}}(z)z^{-\frac{c}{4}}G^{U,V}_{2\Delta}(z)\,.
\label{eq:betaOnFUV}
\ee
Indeed, the contour deformation argument only relies on the transformation property of conformal blocks \eqref{eq:gTr}, which also holds for $G^{U,V}_{2\Delta}(z)$, see \eqref{eq:gTr2}. The integral in the last expression converges for $\Delta>\frac{c+4}{8}$ and both $Q^{\beta}_{\frac{c}{8}}(z)$ and $G^{U,V}_{2\Delta}(z)$ are positive for $z\in(0,1)$, thus manifesting the required positivity of $\beta[F_{2\Delta}^{U,V}]$ and its structure of zeros.

Next, let us show that for $c=4$
\be
\beta_{\frac{1}{2}}[F_{\textrm{vac}}^{U,V}] = 0\,.
\ee
This follows directly from \eqref{eq:betaSeries} and the facts that $F_{\textrm{vac}}^{U,V}$ decomposes into $F_{2n}$ with $n=0,1,\ldots$, which are all annihilated by $\beta_{\Df}$ when $\Df=\frac{1}{2}$. The same conclusion also follows from applying $\beta_{\frac{1}{2}}$ to \eqref{eq:modBootstrapZ} expressing $S$-invariance of the partition function \eqref{eq:extremalZ4} and noting that $\beta_{\frac{1}{2}}$ annihilates all $F_{2n}^{U,V}$ for $n=1,2,\ldots$, as proven in the above.

We conclude that $\beta_{\frac{1}{2}}$ is the extremal functional for gap maximization for both $U(1)^4$ and $\mathrm{Vir}_4$ modular bootstrap. The existence of $\beta_{\frac{1}{2}}$ implies that every unitary solution of these modular bootstrap equations either has a gap above the vacuum smaller than one, or its spectrum consists of the vacuum and a (possibly proper) subset of positive integers. In the latter case, $\eta(\tau)^8\mathcal{Z}(\tau)$ has a Fourier expansion into non-negative integer powers of $q$ and therefore must be a modular form of weight 4. But $E_4(\tau)$ is the unique such modular form up to multiplication, which concludes the proof that $\Delta_U(4) = \Delta_V(4) = 1$.

\subsection{Saturation at $c=12$}
The underlying reason why the optimal solutions of the three bootstrap problems coincide for $c=4$ is that the spectrum of $sl(2,\mathbb{R})$ blocks present in the decomposition of the $U(1)^4$ and $\mathrm{Vir}_4$ vacuum blocks matches the spectrum of $sl(2,\mathbb{R})$ blocks in the free fermion four-point function at $\Df=\frac{1}{2}$. We will now use a small modification of this idea to show that the optimal bounds coincide also for $c=12$, i.e. that
\be
\Delta_U(12)=\Delta_V(12)=2\,.
\ee
$c=12$ maps to $\Df=\frac{3}{2}$, for which the fermionic mean field correlator takes the form
\be
\mathcal{G}(z) = 1 + \left(\mbox{$\frac{z}{1-z}$}\right)^3-z^3\,.
\ee
Again, we can make the four-point function into a partition function using \eqref{eq:GFromZ2}
\be
\mathcal{Z}(\tau) =\frac{2^8(1-\lambda(\tau))}{\lambda(\tau)^2}\mathcal{G}(\lambda(\tau)) = j(\tau)-768\,.
\ee
Here $j(\tau)$ is the modular $j$-function, and one arrives at the last equation by using
\be
j(\tau) = \frac{2^8[\lambda(\tau)^2-\lambda(\tau)+1]^3}{\lambda(\tau)^2[1-\lambda(\tau)]^2}\,.
\ee
Let us recall the Fourier expansion of $j(\tau)$
\be
j(\tau) = \frac{1}{q}+744 + 196884 q + 21493760 q^2+O(q^3)\,.
\ee
When we decompose $\mathcal{Z}(\tau)=j(\tau)-768$ into $U(1)^{12}$ or $\mathrm{Vir}_{12}$ characters, we find a primary of dimension one, whereas we want the gap equal to two. This problem can be easily fixed by adding an appropriate constant to $\mathcal{Z}(\tau)$, which of course does not spoil its $S$-invariance. For $U(1)^{12}$, we get the partition function
\be
\mathcal{Z}_{U}(\tau) = j(\tau) - 720 = \chi^{U}_{\textrm{vac}}(\tau) +196560 \chi^{U}_{2}(\tau) + 16773120 \chi^{U}_{3}(\tau)+\ldots\,.
\label{eq:ZUdecomp}
\ee
For $\mathrm{Vir}_{12}$, we take instead
\be
\mathcal{Z}_{V}(\tau) = j(\tau) - 744 = \chi^{V}_{\textrm{vac}}(\tau) +196882 \chi^{V}_{2}(\tau) + 21099994 \chi^{V}_{3}(\tau)+\ldots\,.
\label{eq:ZVdecomp}
\ee
What is the physical interpretation of these partition functions? $\eta(\tau)^{24}\mathcal{Z}_U(\tau)$ is the theta function of the Leech lattice
\be
\eta(\tau)^{24}\mathcal{Z}_U(\tau) = \sum\limits_{x\in\Lambda_{24}}e^{i\pi\tau x^2}\,.
\ee
This has to be the case since the Leech lattice is the unique even self-dual lattice in $\mathbb{R}^{24}$ with no vector of length $\sqrt{2}$. $\mathcal{Z}_V(\tau)$ is the partition function of a chiral half of the Monster CFT with left- and right-moving central charges $c=24$ and $\bar{c}=0$ \cite{frenkel1984natural}.  These realizations of $\mathcal{Z}_{U,V}(\tau)$ also show that all higher coefficients in their decompositions \eqref{eq:ZUdecomp}, \eqref{eq:ZVdecomp} are non-negative. One may wonder if $\mathcal{Z}_U(\tau)$ can arise as a partition function of a unitary CFT with $c=\bar{c}=12$, specialized to $\tau=-\bar{\tau}$. In fact, this possibility was excluded by the numerical studies of \cite{Collier:2016cls}, which showed that the upper bound on the gap coming from full modular bootstrap at $c=\bar{c}=12$ is strictly less than two.

In order to prove optimality of the above partition functions, we use the exact same logic as we did for $c=4$. We simply apply the extremal functional $\beta_{\frac{3}{2}}$ to the functions $F^{U,V}_{2\Delta}(z)$. It follows from \eqref{eq:betaOnFUV} that $\beta_{\frac{3}{2}}[F^{U,V}_{2\Delta}(z)]$ is non-negative for $\Delta\geq 2$, has a simple zero at $\Delta=2$ and double zeros at $\Delta=3,4,\ldots$. Since we constructed explicit partition functions whose spectrum consists of the vacuum and $\Delta=2,3,\ldots$, it follows from their $S$-invariance that $\beta_{\frac{3}{2}}[F^{U}_{\textrm{vac}}(z)]=\beta_{\frac{3}{2}}[F^{V}_{\textrm{vac}}(z)]=0$. Therefore, $\beta_{\frac{3}{2}}$ is the extremal functional also for the $U(1)^{12}$ and $\mathrm{Vir}_{12}$ problems. While $\beta_{\frac{3}{2}}$ in principle allows for partition functions whose spectrum is a proper subset of the extremal spectrum $\Delta=2,3,\ldots$, it is easy to see they do not exist. Indeed, the difference between such a partition function and $Z_{U,V}(\tau)$ would have to be a modular form for $SL(2,\mathbb{Z})$ of weight zero, with $q$-expansion starting at $q^2$, and so would have to vanish identically.

As a side note, the above results explain why $\beta_{\frac{3}{2}}[F_{\Delta}]$ has a simple zero at $\Delta=2$. This zero may seem accidental since the fermionic mean field correlator does not contain any operator at this dimension. However, the partition functions \eqref{eq:ZUdecomp} and \eqref{eq:ZVdecomp} do contain $G_{2}(z)$ in their $sl(2,\mathbb{R})$ block decomposition and $\beta_{\frac{3}{2}}[F_{2}] = 0$ thus follows from their $S$-invariance, together with $\beta_{\frac{3}{2}}[F_{0}] = 0$ and $\beta_{\frac{3}{2}}[F_{2n}] = 0$ for $n=2,3,\ldots$.

\subsection{Solution of the sphere-packing problem in $\mathbb{R}^{8}$ and $\mathbb{R}^{24}$}\label{ss:mod2sphere}
Let us see compare the above results to the solution of the sphere packing problem in $\mathbb{R}^{8}$ \cite{viazovska8} and $\mathbb{R}^{24}$ \cite{viazovska24}. Recall that the linear programming bounds on sphere packing in $\mathbb{R}^d$ are equivalent to the modular bootstrap bound with $U(1)^{\frac{d}{2}}$ characters. Therefore, the magic functions constructed by Viazovska in $\mathbb{R}^{8}$ and by Cohn et al in $\mathbb{R}^{24}$ should be equal to the action of extremal functionals on the functions $F^{U}_{r^2}(z)$ for $c=4$ and $c=12$, where $r$ is the norm of a vector in $\mathbb{R}^{d}$. Recall that the magic functions take the form of integrals of $e^{i\pi r^2\tau}$ against a judiciously chosen weight functions in the upper half-plane. Since the extremal functionals are written as contour integrals in the $z$ variable, we can prove they are equivalent to the magic functions simply by a change of variables from $z=\lambda(\tau)$ to $\tau$.

Let us start in eight dimensions.\footnote{We use the notation of Section \ref{ssec:ViazovskaProof}, i.e. the Cohn-Elkies function $f(r)$ decomposes as $f(r)=h(r)-g(r)$, where $h(r)$ and $g(r)$ are respectively Fourier-even and Fourier-odd. Viazovska's paper \cite{viazovska8} uses instead functions $a(r)$ and $b(r)$, which are related to our $h(r)$ and $g(r)$ as follows: $h(r) = \frac{i \pi}{8640}a(r)$, $g(r) = \frac{i}{240\pi}b(r)$.} We claim that the Fourier-odd radial function $g(r)$ of Viazovska is a constant multiple of the action of the extremal functional $\beta_{\frac{1}{2}}$ on $F^{U}_{r^2}(z)$
\be
g(r) = \frac{16}{15} \beta_{\frac{1}{2}}[F^{U}_{r^2}]\,.
\label{eq:bFromBeta4}
\ee
This claim can be most easily checked using the form of $g(r)$ which manifests the $\sin^2(\pi r^2/2)$ prefactor, i.e. equation \eqref{vansatz}
\be
g(r) = i \sin^2(\pi r^2/2) \int\limits_{0}^{i\infty} G(\tau)e^{i\pi r^2\tau} d\tau\,,
\label{eq:functionalTauPlane}
\ee
where
\be
G(\tau) = - \frac{32 \theta_4^4(5\theta_3^8-5 \theta_3^4 \theta_4^4 + 2 \theta_4^8)}{15\pi \theta_3^8 \theta_2^8}\,.
\ee
This should be compared with \eqref{eq:betaOnFUV}, which gives
\ba
\beta_{\frac{1}{2}}[F^U_{r^2}] &= 2\sin^2(\pi r^2/2) \int\limits_{0}^{1}Q^{\beta}_{\frac{1}{2}}(z)[2^8z(1-z)]^{-\frac{1}{3}}\chi^{U}_{r^2/2}(\tau(z)) dz=\\
&= 2\sin^2(\pi r^2/2) \int\limits_{0}^{1}Q^{\beta}_{\frac{1}{2}}(z)[2^8z(1-z)]^{-\frac{1}{3}}\eta(\tau(z))^{-8}e^{i\pi r^2\tau(z)}dz
\label{eq:functionalZPlane}
\ea
with $Q^{\beta}_{\frac{1}{2}}(z)$ given in the first line of \eqref{eq:gBetaHalfs}
\be
Q^{\beta}_{\frac{1}{2}}(z) = \frac{(1-z) \left(2 z^2+z+2\right)}{\pi ^2 z^2}\,.
\ee
In order to show the equivalence of \eqref{eq:functionalTauPlane} and \eqref{eq:functionalZPlane}, it remains to change coordinates in the latter from $z=\lambda(\tau)$ to $\tau$. The integration contour $z\in(0,1)$ maps to $\tau\in i\mathbb{R}_{>0}$ as needed. The measure transforms as follows
\be
dz = i\pi\frac{\theta_{2}(\tau)^4\theta_{4}(\tau)^4}{\theta_{3}(\tau)^4} d\tau\,.
\ee
Therefore, the claim \eqref{eq:bFromBeta4} is equivalent to
\be
G(\tau) =
-\frac{32\pi\theta_{2}(\tau)^4\theta_{4}(\tau)^4}{15\theta_{3}(\tau)^4}[2^8\lambda(\tau)(1-\lambda(\tau))]^{-\frac{1}{3}}\eta(\tau)^{-8}Q^{\beta}_{\frac{1}{2}}(\lambda(\tau))\,,
\ee
which is a true identity.

\begin{figure}[ht]%
\begin{center}
\includegraphics[width=\textwidth]{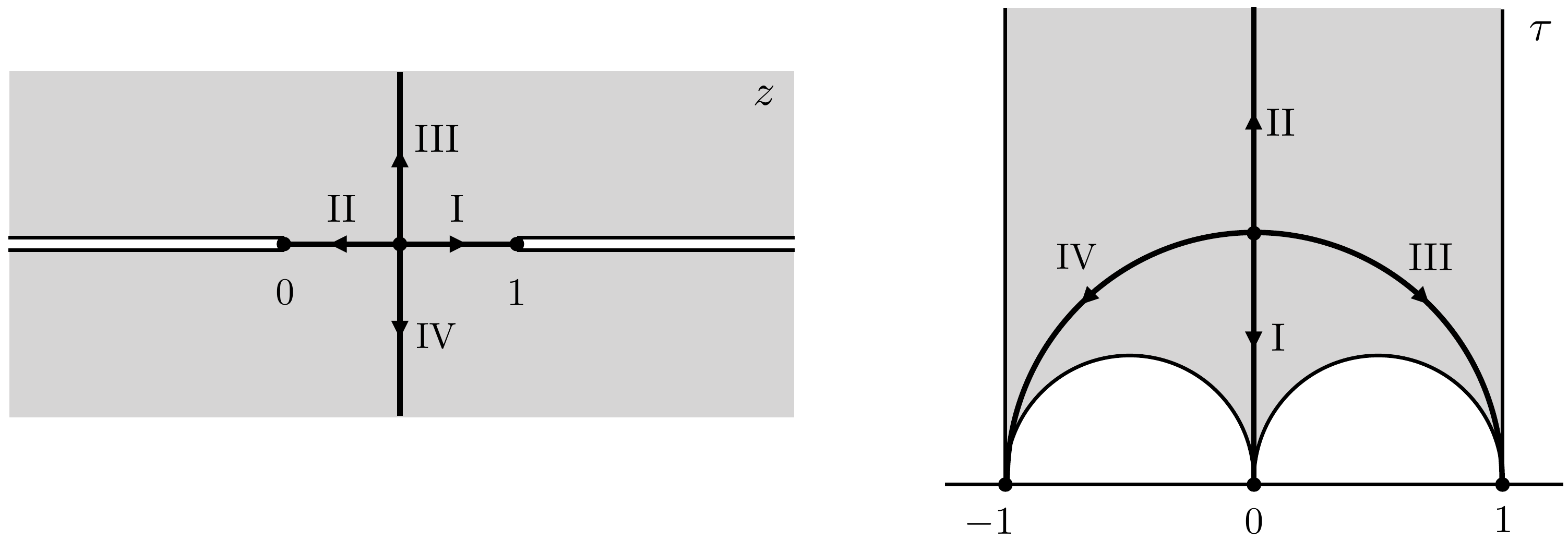}%
\caption{Left: Contour integral definition of the analytic extremal functional in the $z$-plane, see \eqref{eq:funDef}, \eqref{eq:ourContours}. Right: Viazovska's contour integral definition of the magic functions, see \eqref{eq:bViazovska}. The two definitions are related by the transformation $z=\lambda(\tau)$. The contours labelled by the same Roman numerals get mapped to each other.}
\label{fig:contours}%
\end{center}
\end{figure}

What is the counterpart of our formula \eqref{eq:funDef}, which manifests that $\beta_{\frac{1}{2}}$ can be written as a linear functional? This is equation (52) in \cite{viazovska8}, which manifests the fact that $g(r)$ is odd under the Fourier transform:
\ba
g(r) = &-\frac{i}{2}\int\limits_{i}^{0} G(\tau)e^{i\pi r^2\tau} d\tau+\frac{i}{2}\int\limits_{i}^{i\infty} G_{S}(\tau)e^{i\pi r^2\tau} d\tau+\\
&+\frac{i}{4}\int\limits_{i}^{1} G_{T}(\tau)e^{i\pi r^2\tau} d\tau +\frac{i}{4}\int\limits_{i}^{-1} G_{T}(\tau)e^{i\pi r^2\tau} d\tau\,,
\label{eq:bViazovska}
\ea
where
\ba
G_{S}(\tau) &\equiv \tau^{2}G(-1/\tau)\\
G_{T}(\tau) &\equiv G(\tau+1)\,.
\ea
Indeed, we can start from our equation \eqref{eq:funDef} with $\mathcal{F}(z) = F^{U}_{r^2}(z)$, and write $F^{U}_{r^2}(z)$ as the difference of the character and its $S$-transformation to arrive at
\ba
\beta_{\frac{1}{2}}[F^{U}_{r^2}] =
 &\left[
 \int\limits_{\frac{1}{2}}^{1}\!\!dz\,Q_{\frac{1}{2}}^{\beta}(z)+
 \int\limits_{\frac{1}{2}}^{0}\!\!dz\,Q_{\frac{1}{2}}^{\beta}(1-z)+
\frac{1}{2}\!\!\!\int\limits_{\frac{1}{2}}^{\frac{1}{2}+i\infty}\!\!\!dz R_{\frac{1}{2}}^{\beta}(z)+
\frac{1}{2}\!\!\!\int\limits_{\frac{1}{2}}^{\frac{1}{2}-i\infty}\!\!\!dz R_{\frac{1}{2}}^{\beta}(z)\right]\times\\
 &\times[2^8z(1-z)]^{-\frac{1}{3}}\eta(\tau(z))^{-8}e^{i\pi r^2\tau(z)}\,,
 \label{eq:ourContours}
\ea
where we performed the change of variables $z\mapsto1-z$ in the terms involving the $S$-transformed character. The square bracket is meant to be distributed over the integrand on the second line. If we  now change the integration variable from $z$ to $\tau$ in \eqref{eq:ourContours}, we arrive precisely at Viazovska's formula \eqref{eq:bViazovska}. Indeed, the four straight contours $z\in(1/2,1)$, $z\in(1/2,0)$, $z\in(1/2,1/2+i\infty)$ and $z\in(1/2,1/2-i\infty)$ map to $\tau\in(i,0)$, $\tau\in(i,i\infty)$, $\tau\in(i,1)$ and $\tau\in(i,-1)$, as shown in Figure \ref{fig:contours}. The integrands can be checked to match as well.

Finally, it can be seen that the functional constraints which $G(\tau)$ needs to satisfy for Viazovska's contruction to work are precisely equivalent to the constraints that can be used to fix $Q_{\frac{1}{2}}^{\beta}(z)$. Indeed, the constraint \eqref{eq:fSymmetry} is equivalent to \eqref{gcond2}
\be
G_{T}(\tau) = -\tau^2G_{T}(-1/\tau)\,,
\ee
while the constraint \eqref{eq:dDiscZero} is equivalent to \eqref{gcond1}
\be
G(\tau) - G_S(\tau) - G_T(\tau) = 0\,.
\ee

This completes the proof of equivalence of the Fourier-odd magic function $g(r)$ in $\mathbb{R}^{8}$ with the extremal functional $\beta_{\frac{1}{2}}$. The Fourier-even magic function $h(r)$ arises in exactly the same manner from the action of the extremal functional $\beta_{\frac{1}{2}}^{+}$, discussed in Section \ref{ssec:omegaPlus}, on the symmetric combination
\ba
F^{U+}_{r^2}(z) &=
z^{-\frac{c}{4}}G^{U}_{r^2}(z)+(1-z)^{-\frac{c}{4}}G^{U}_{r^2}(1-z)=\\
&= \left[2^8z(1-z)\right]^{-\frac{c}{12}}\left[\chi^{U}_{r^2/2}(\tau(z))+\chi^{U}_{r^2/2}(-1/\tau(z))\right]\,,
\ea
where we should take $c=4$. More precisely, we find
\be
h(r) = -\frac{16}{15}\beta_{\frac{1}{2}}^{+}[F^{U+}_{r^2}]\,.
\label{eq:aFromOmega4}
\ee
This follows from the identity
\be
H(\tau) = -\frac{32\pi\theta_{2}(\tau)^4\theta_{4}(\tau)^4}{15\theta_{3}(\tau)^4}[2^8\lambda(\tau)(1-\lambda(\tau))]^{-\frac{1}{3}}\eta(\tau)^{-8}Q^{\beta+}_{\frac{1}{2}}(\lambda(\tau))\,,
\ee
where $H(\tau)$ is described in Section \ref{sssec:magicFunction8} and $Q^{\beta+}_{\frac{1}{2}}(z)$ is given in equation \eqref{eq:gOmegaPlusHalf}. Incidentally, this identity gives a simple proof of the fact
\be
H(\tau) < 0\quad\textrm{for}\quad \tau\in i\mathbb{R}_{>0}\,,
\ee
which is useful in proving positivity properties of the magic functions. Finally, we can combine $h(r)$ and $g(r)$ into the magic function $f(r)$
\be
f(r) = h(r) - g(r)\,.
\ee
Using \eqref{eq:bFromBeta4} and \eqref{eq:aFromOmega4}, we can express $f(r)$ and its Fourier transform $\widehat{f}(r)$ in terms of actions of the extremal functionals
\ba
f(r) &= -\frac{16}{15}\left\{\beta_{\frac{1}{2}}^{+}[F^{U+}_{r^2}]+\beta_{\frac{1}{2}}[F^{U}_{r^2}]\right\}\\
\widehat{f}(r) &= -\frac{16}{15}\left\{\beta_{\frac{1}{2}}^{+}[F^{U+}_{r^2}]-\beta_{\frac{1}{2}}[F^{U}_{r^2}]\right\}\,.
\ea
These are the right linear combinations of $\beta_{\frac{1}{2}}$ and $\beta_{\frac{1}{2}}^{+}$ with the required structure of zeros and positivity properties.\footnote{The prefactor $\frac{16}{15}$ sets the normalization to $f(0) = \widehat{f}(0)=1$, while the functionals are normalized by $\partial_{\Delta}\beta_{\frac{1}{2}}[F_{\Delta}]=\partial_{\Delta}\beta_{\frac{1}{2}}^{+}[F_{\Delta}]=1$ at $\Delta=2$.}

The magic functions for the sphere packing problem $\mathbb{R}^{24}$ can be recovered from the extremal functionals for $\Df=\frac{3}{2}$ in exactly the same way as we did for $\mathbb{R}^{8}$ and $\Df=\frac{1}{2}$. This time, we find the following proportionality constants\footnote{The functions $a(r)$ and $b(r)$ of reference \cite{viazovska24} are related to $h(r)$ and $g(r)$ as follows: $h(r) = -\frac{\pi i}{113218560}a(r)$, $g(r) = \frac{i}{262080\pi}b(r)$.}
\ba
h(r) &=-\frac{2048}{4095}\,\beta_{\frac{3}{2}}^{+}[F^{U+}_{r^2}]\\
g(r) &= \frac{2048}{4095} \,\beta_{\frac{3}{2}}[F^{U}_{r^2}]\,.
\label{eq:bFromBeta12}
\ea

% !TEX root = ../ModularBootstrapV3.tex

\section{Bounds at large central charge}

\subsection{Upper bounds for $c\neq4,12$}
Having understood the exact bounds $\Delta_U(c)$ and $\Delta_V(c)$ for $c=4,12$, we will now analyze them away from these special points -- first for general $c$, and then focussing on the regime of large $c$.

It is natural to apply the extremal functional $\beta_{\frac{c}{8}}$ to the modular bootstrap equations for general $c$. We have seen from equation \eqref{eq:betaOnFUV} that, for any $c>1$, $\beta_{\frac{c}{8}}[F^{ A }_{2\Delta}]$ is non-negative for $\Delta\geq\frac{c+4}{8}$ and vanishes at\footnote{The symbol $ A $ is a placeholder for $U$ or $V$, but many results in this section hold also for general chiral algebra $ A $.}
\be
\Delta_n = \frac{c+4}{8}+n\quad\textrm{for}\quad n=0,1,\ldots\,.
\ee
Therefore, $\frac{c+4}{8}$ is a valid upper bound on the gap whenever $\beta_{\frac{c}{8}}[F^{ A }_{\textrm{vac}}]>0$. In order to study the sign of $\beta_{\frac{c}{8}}[F^{ A }_{\textrm{vac}}]$, we can first expand it in terms of action of $\beta_{\frac{c}{8}}$ on the $sl(2,\mathbb{R})$ blocks of dimensions $2,4,6,\ldots$
\be
\beta_{\frac{c}{8}}[F^{ A }_{\textrm{vac}}] = \sum\limits_{n=1}^{\infty}b^{ A }_n\beta_{\frac{c}{8}}[F_{2n}]\,,
\ee
where $b^{ A }_n > 0$ and we used $\beta_{\frac{c}{8}}[F_0] = 0$. We can immediately conclude $\beta_{\frac{c}{8}}[F^{ A }_{\textrm{vac}}]>0$ for $c\in(1,4)$ since then all terms in the sum over $n$ are positive. In this way we recover the bound \eqref{eq:boundSmallC}, derived here in a different way.

\begin{figure}%
\begin{center}
\includegraphics[width=0.65\textwidth]{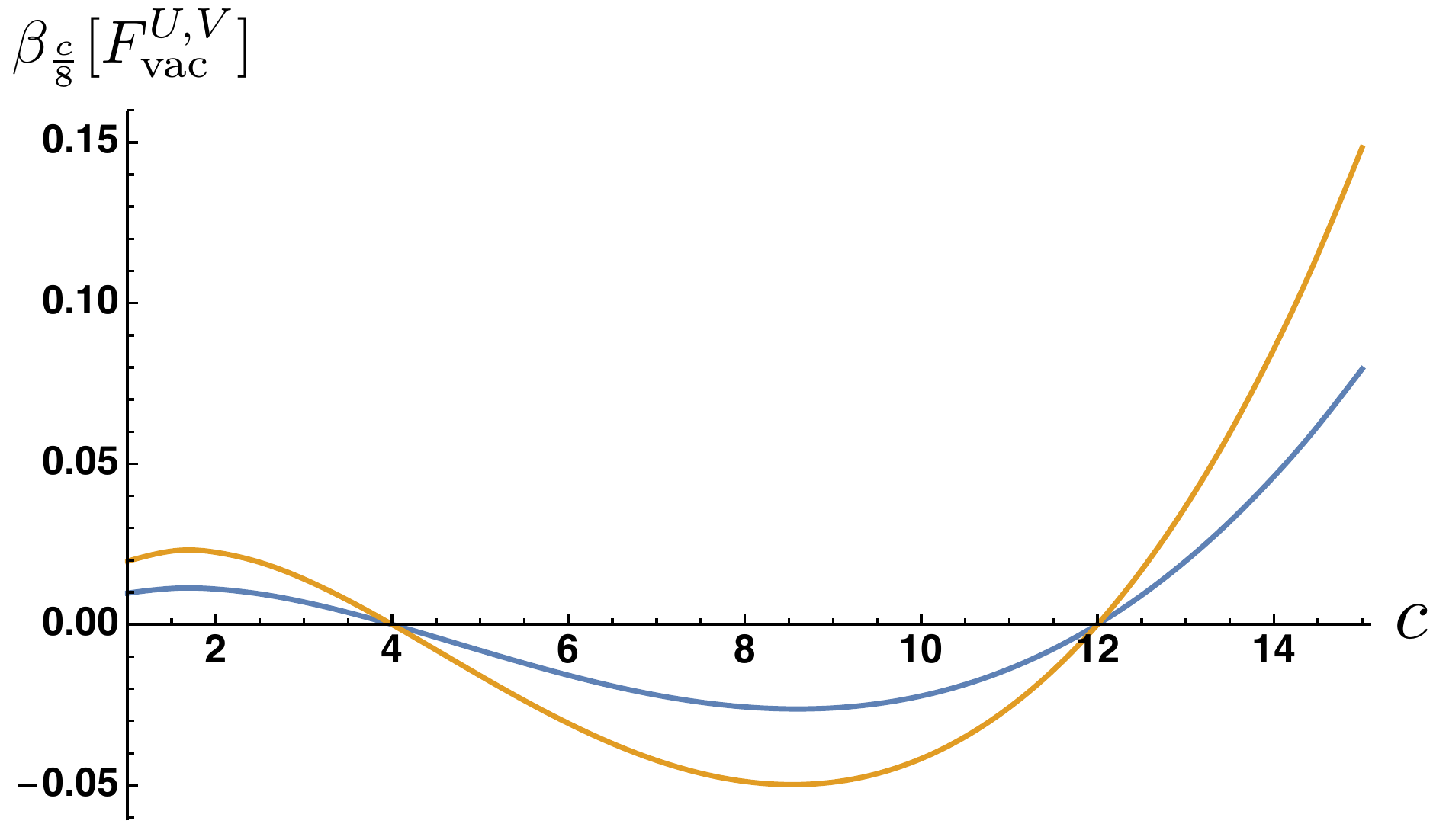}%
\caption{Action of the functional $\beta_{\frac{c}{8}}$ on the vacuum modular bootstrap vectors $F_{\textrm{vac}}^{U,V}$. The orange curve corresponds to the $U(1)^c$ vacuum and the blue one to the $\mathrm{Vir}_c$ vacuum. The action on both kinds of vacuum characters vanishes at $c=4$ and $c=12$, where $\beta_{\frac{c}{8}}$ becomes the optimal functional. The plot also illustrates that the vacuum action is positive for $c\in(1,4)\cup(12,\infty)$. For $c>12$, the vacuum action increases exponentially with $c$.}
\label{fig:vacuumAction}%
\end{center}
\end{figure}

What is the sign of $\beta_{\frac{c}{8}}[F^{ A }_{\textrm{vac}}]$ for $c\geq 4$? We already know that $\beta_{\frac{c}{8}}[F^{ A }_{\textrm{vac}}]$ vanishes at $c=4,12$ and it is a reasonable guess that these are the only zeros of $\beta_{\frac{c}{8}}[F^{ A }_{\textrm{vac}}]$ and thus
\ba
\beta_{\frac{c}{8}}[F^{ A }_{\textrm{vac}}]&<0\quad\textrm{for}\quad c\in(4,12)\\
\beta_{\frac{c}{8}}[F^{ A }_{\textrm{vac}}]&>0\quad\textrm{for}\quad c\in(12,\infty)\,.
\ea
This guess turns out to be correct, which can be established beyond reasonable doubt for example by numerically evaluating the contour integrals defining $\beta_{\frac{c}{8}}$. Figure \ref{fig:vacuumAction} illustrates this fact in the range $c\in(1,15)$. We will also soon give a rigorous argument for the second inequality at large $c$. It follows that the existence of $\beta_{\frac{c}{8}}$ implies the rigorous inequality
\be
\Delta_{U,V}(c) < \frac{c+4}{8}\quad\textrm{for}\quad c\in(12,\infty)\,.
\ee
That is, every unitary 2D CFT of central charge $c\geq12$ contains a Virasoro primary other than identity of scaling dimension at most $c/8+1/2$.

\subsection{Saddle-point evaluation of the functional actions}
Let us consider the regime of large central charge. The key insight is that the action of functionals $\beta_{\frac{c}{8}}$ and $\widetilde{\alpha}_{\frac{c}{8}}$ can be evaluated using the saddle-point approximation. The discussion of this subsection is analogous to the computation of the action of $\beta_{\Df}$ on $sl(2,\mathbb{R})$ conformal blocks in the limit of a large external dimension, explained in Appendix B of \cite{Mazac:2018mdx}. We will first derive the saddle-point approximation and use it to show that $\beta[F^{U,V}_{\textrm{vac}}]>0$ for large enough $c$. This will establish $\Delta_{U,V}(c)>c/8+1/2$ for large enough $c$. In the next subsection, we will improve the asymptotic bound by considering a suitable linear combination of functionals $\beta_{\frac{c}{8}}$ and $\widetilde{\alpha}_{\frac{c}{8}}$.

Let us analyze the functional action on operators whose dimension is proportional to $c$. We set $\Delta = \mu c$, keep $\mu$ fixed and look at
\be
\beta_{\frac{c}{8}}[F^{ A }_{2 \mu c}]
\ee
for $c\rightarrow \infty$. We will focus on the regime $\mu\in[0,1/8)$. In this range, the contour deformation leading to \eqref{eq:betaOnFUV} is invalid and we need to resort to the contour prescription \eqref{eq:funDef}
\be
\beta_{\frac{c}{8}}[F^{ A }_{2 \mu c}] = \int\limits_{\frac{1}{2}}^{1}\!\!dz\,Q_{\frac{c}{8}}^{\beta}(z) F^{ A }_{2 \mu c}(z)
+ \frac{1}{2}\!\!\!\int\limits_{\frac{1}{2}}^{\frac{1}{2}+i\infty}\!\!\!dz\,R_{\frac{c}{8}}^{\beta}(z)F^{ A }_{2 \mu c}(z)\,.
\ee
As we will now explain, the contour integrals localize to a saddle point when $c\rightarrow\infty$. We will need the large-$c$ behaviour of the functional kernels. It turns out to be very simple:
\ba
R_{\frac{c}{8}}^{\beta}(z) &\stackrel{c\rightarrow\infty}{\sim} -\frac{\sqrt{c}}{2 \pi ^{3/2} } \frac{2 z-1}{[z(z-1)]^{3/2}}\\
R_{\frac{c}{8}}^{\widetilde{\alpha}}(z) &\stackrel{c\rightarrow\infty}{\sim}
\frac{2}{\pi ^{3/2} \sqrt{c} }\frac{(z-2) (z+1) (2 z-1)}{[z(z-1)]^{5/2}}\,.
\label{eq:fKerAs}
\ea
To proceed, let us first use the definition of $F^{ A }_{2 \mu c}(z)$ in terms of modular characters:
\ba
\beta_{\frac{c}{8}}[F^{ A }_{2 \mu c}] &=
 \int\limits_{\frac{1}{2}}^{1}\!\!dz\,Q_{\frac{c}{8}}^{\beta}(z) [2^8z(1-z)]^{-\frac{c}{12}}\left[\chi^{ A }_{\mu c}(\tau(z))-\chi^{ A }_{\mu c}(\tau(1-z))\right]+\\
&+ \frac{1}{2}\!\!\!\int\limits_{\frac{1}{2}}^{\frac{1}{2}+i\infty}\!\!\!dz R_{\frac{c}{8}}^{\beta}(z)[2^8z(1-z)]^{-\frac{c}{12}}\left[\chi^{ A }_{\mu c}(\tau(z))-\chi^{ A }_{\mu c}(\tau(1-z))\right]\,.
\ea
Let us change coordinates from $z$ to $1-z$ in the terms involving the crossed-channel character to arrive at
\ba
\beta_{\frac{c}{8}}[F^{ A }_{2 \mu c}] =
 &\left[
 \int\limits_{\frac{1}{2}}^{1}\!\!dz\,Q_{\frac{c}{8}}^{\beta}(z)+
 \int\limits_{\frac{1}{2}}^{0}\!\!dz\,Q_{\frac{c}{8}}^{\beta}(1-z)+
\frac{1}{2}\!\!\!\int\limits_{\frac{1}{2}}^{\frac{1}{2}+i\infty}\!\!\!dz R_{\frac{c}{8}}^{\beta}(z)+
\frac{1}{2}\!\!\!\int\limits_{\frac{1}{2}}^{\frac{1}{2}-i\infty}\!\!\!dz R_{\frac{c}{8}}^{\beta}(z)\right]\times\\
 &\times[2^8z(1-z)]^{-\frac{c}{12}}\chi^{ A }_{\mu c}(\tau(z))\,,
\ea
where the square bracket is meant to be distributed on the term on the second line and we used $R_{\frac{c}{8}}^{\beta}(z) = R_{\frac{c}{8}}^{\beta}(1-z)$. Now, we use the following trick. Since the double discontinuity of $Q_{\frac{c}{8}}^{\beta}(z)$ vanishes (see \eqref{eq:dDiscZero}), we can simultaneously shift the starting point of all four integrals in the square bracket from $z=1/2$ to an arbitrary $z_0\in(0,1)$:
\ba
\beta_{\frac{c}{8}}[F^{ A }_{2 \mu c}] =
 &\left[
 \int\limits_{z_0}^{1}\!\!dz\,Q_{\frac{c}{8}}^{\beta}(z)+
 \int\limits_{z_0}^{0}\!\!dz\,Q_{\frac{c}{8}}^{\beta}(1-z)+
\frac{1}{2}\!\!\!\int\limits_{z_0}^{\frac{1}{2}+i\infty}\!\!\!dz R_{\frac{c}{8}}^{\beta}(z)+
\frac{1}{2}\!\!\!\int\limits_{z_0}^{\frac{1}{2}-i\infty}\!\!\!dz R_{\frac{c}{8}}^{\beta}(z)\right]\times\\
 &\times[2^8z(1-z)]^{-\frac{c}{12}}\chi^{ A }_{\mu c}(\tau(z))\,.
 \label{eq:LargeCShifted}
\ea
The idea is to choose $z_0$ to be the $c\rightarrow\infty$ saddle-point of the integrals involving $R_{\frac{c}{8}}^{\beta}(z)$. Let us isolate the exponential dependence of the integrand at large $c$ by writing
\be
[2^8z(1-z)]^{-\frac{c}{12}}\chi^{ A }_{\mu c}(\tau(z)) = 
h_0(z)e^{c\,h_1(z)}\,,
\ee
where $h_{0,1}(z)$ are independent of $c$. The saddle point of the integrals involving $R_{\frac{c}{8}}^{\beta}(z)$ is located at the stationary point of $h_1(z)$. For $ A =U(1)^c$, one finds
\ba
h_0(z) &= 1\\
\quad h_1(z) &= 
-\frac{1}{12}\log[2^8z(1-z)]
-2\log[\eta(\tau(z))]
+2\pi i \tau(z) \mu
\label{eq:h01U}
\ea
and for $ A =\mathrm{Vir}_{c}$, we have
\ba
h_0(z) &=\frac{e^{i \pi \tau(z)/6}}{\eta(\tau(z))^2}\\
h_1(z) &= 
-\frac{1}{12}\log[2^8z(1-z)]
+2\pi i\tau(z)\left(\mu-\frac{1}{12}\right)\,.
\label{eq:h01V}
\ea
Let $z_0(\mu)$ be the minimum of $h_1(z)$ in the interval $z\in(0,1)$. $z_0(\mu)$ depends on both $\mu$ and algebra $ A $. One can check that $z_0(1/8)=0$ for both algebras of interest. Furthermore, $z_0(\mu)$ is monotonic decreasing. As we decrease $\mu$ from 1/8 to 0, $z_0(\mu)$ increases to
\ba
z_0(0) &= 0.826115\ldots \quad\textrm{for}\quad A =U(1)^c\\
z_0(0) &= 0.887578\ldots \quad\textrm{for}\quad A =\textrm{Vir}_c\,.
\label{eq:z00}
\ea
Going back to the expression for the functional action \eqref{eq:LargeCShifted}, we see that the path of the steepest descent of the integrals including $R_{\frac{c}{8}}^{\beta}(z)$ is parallel to the imaginary axis at $z=z_0(\mu)$. At the same time, the integrals involving $Q_{\frac{c}{8}}^{\beta}(z)$ are exponentially subleading in the $c\rightarrow\infty$ limit because $Q_{\frac{c}{8}}^{\beta}(z)$ includes the extra (exponentially small) factor $(1-z)^{\frac{c}{4}}$ with respect to $R_{\frac{c}{8}}^{\beta}(z)$. In order to evaluate the integrals involving $R_{\frac{c}{8}}^{\beta}(z)$, let us expand $h_1(z)$ around $z=z_0(\mu)$
\be
h_1(z) = h_1(z_0(\mu)) + \frac{1}{2}h^{''}_1(z_0(\mu))(z-z_0(\mu))^2 + \ldots\,.
\ee
The saddle-point evaluation then leads to
\be
\beta_{\frac{c}{8}}[F^{ A }_{2 \mu c}] \stackrel{c\rightarrow\infty}{\sim}
-\mathrm{Im}\!\!\left[\frac{R_{\frac{c}{8}}^{\beta}(z_0(\mu)+i0^+)}{\sqrt{c}}\right]
\sqrt{\frac{\pi}{2 h^{''}_1(z_0(\mu))}}h_0(z_0(\mu))e^{c\,h_1(z_0(\mu))}\,.
\label{eq:betaAs}
\ee
If we insert the explicit form of the large-$c$ limit of $R_{\frac{c}{8}}^{\beta}(z)$ from \eqref{eq:fKerAs}, we arrive at
\be
\beta_{\frac{c}{8}}[F^{ A }_{2 \mu c}] \stackrel{c\rightarrow\infty}{\sim}
 \frac{2 z_0(\mu)-1}{[z_0(\mu)(1-z_0(\mu))]^{3/2}}
\frac{h_0(z_0(\mu))}{\sqrt{8\pi^2 h^{''}_1(z_0(\mu))}}e^{c\,h_1(z_0(\mu))}\,.
\ee
We are interested in the sign of the above expression. All terms except for $2z_0(\mu)-1$ in the numerator are manifestly positive, and thus the sign of $\beta_{\frac{c}{8}}[F^{ A }_{2 \mu c}]$ at large $c$ agrees with the sign of $2z_0(\mu)-1$. We can see from \eqref{eq:z00} that
\ba
2z_0(0)-1 &> 0\\
2z_0(1/8)-1 &< 0\,.
\ea
In particular, the action of $\beta_{\frac{c}{8}}$ on $F^{ A }_{0}(z)$ is positive at large $c$. Thus we immediately conclude
\be
\beta_{\frac{c}{8}}[F^{U}_{\textrm{vac}}]>0
\ee
for sufficiently large $c$. For $ A =\mathrm{Vir}_c$, the vacuum block differs from the $\mu=0$ block by factor $(1-e^{2\pi i \tau(z)})^2$, which is positive for $z=z_0(0)$, so the same conclusion follows in this case too:
\be
\beta_{\frac{c}{8}}[F^{V}_{\textrm{vac}}]>0
\ee
for sufficiently large $c$. As $\mu$ rises from $0$ to $1/8$, the expression $2z_0(\mu)-1$ monotonically decreases, showing that at large $c$, $\beta_{\frac{c}{8}}[F^{ A }_{2 \mu c}]$ has a single zero as a function of $\mu$ in the region $\mu\in(0,1/8)$, located at the solution of equation $z_0(\mu_0)=1/2$. In fact, the value of $\mu_0$ can be found analytically:
\ba
\mu_0 &= \frac{1}{4\pi} \quad\textrm{for}\quad A =U(1)^c\\
\mu_0 &= \frac{1}{12} \quad\textrm{for}\quad A =\textrm{Vir}_c\,.
\ea
Note that when we study the action $\widetilde{\alpha}_{\frac{c}{8}}[F^{ A }_{2 \mu c}]$ at large $c$, equation \eqref{eq:betaAs} still applies, with the appropriate replacement $\beta\rightarrow\widetilde{\alpha}$, i.e. the only difference is in the asymptotic form of $R(z)$, given in \eqref{eq:fKerAs}.

\subsection{Improved bound at large $c$}\label{ssec:ImprovedLargeC}
We have seen that $\beta$ implies upper bounds with asymptotic form $\Delta_{U,V}(c)<c/8+O(1)$ as $c\rightarrow\infty$. Now we will obtain a better bound from an appropriate linear combination of $\beta$ and $\widetilde{\alpha}$. The underlying idea is that $\beta$ is not the optimal functional at large $c$ because its action on vacuum is positive. On the other hand, the action of $\widetilde{\alpha}$ on the vacuum is negative. By taking a linear combination of $\beta$ and $\widetilde{\alpha}$ which annihilates the vacuum at large $c$, we get a functional which is non-negative from $\Delta_{*}(c)<\frac{c+4}{8}$, thus obtaining a stronger bound than when using only $\beta$.

In order to get an interesting answer, the linear combination of $\beta$ and $\widetilde{\alpha}$ should be such that as $c\rightarrow\infty$, $\beta_{\frac{c}{8}}[F^{ A }_{2 \mu c}]$ and $\widetilde{\alpha}_{\frac{c}{8}}[F^{ A }_{2 \mu c}]$ contribute at the same order in the $c^{-1}$ expansion. Looking at \eqref{eq:fKerAs} and \eqref{eq:betaAs}, it follows we want
\be
\widetilde{\beta}_{\frac{c}{8}} = \beta_{\frac{c}{8}} + a\,c\,\widetilde{\alpha}_{\frac{c}{8}}\,,
\ee
where $c$ is the central charge and $a$ is a constant which we want to fix. We can see from \eqref{eq:betaAs} that if $\mu\in(0,1/8)$, then the sign of $\widetilde{\beta}_{\frac{c}{8}}[F^{ A }_{2 \mu c}]$ at large $c$ agrees with the sign of
\be
-\mathrm{Im}\!\!\left[\frac{R^{\beta}_{\frac{c}{8}}(z_0(\mu)+i0^+)}{\sqrt{c}}+a \sqrt{c} R_{\frac{c}{8}}^{\widetilde{\alpha}}(z_0(\mu)+i0^+)\right]
\ee
so let us define for $z\in(0,1)$
\ba
K(z) &= -2 \pi ^{3/2} [z(1-z)]^{5/2}\lim_{c\rightarrow \infty}\mathrm{Im}\!\!\left[\frac{R^{\beta}_{\frac{c}{8}}(z+i0^+)}{\sqrt{c}}+a \sqrt{c} R^{\widetilde{\alpha}}_{\frac{c}{8}}(z+i0^+)\right]=\\
&=
(2 z-1)\left[-z(z-1)+4a(z+1)(z-2)\right]\,,
\ea
so that the sign of $\widetilde{\beta}_{\frac{c}{8}}[F^{ A }_{2 \mu c}]$ at large $c$ agrees with the sign of $K(z_0(\mu))$. Now, functional $\widetilde{\beta}$ will imply an asymptotic upper bound of the form $\Delta_{*}(c)< \mu_* c+O(1)$ for some $\mu_*<1/8$ only if
\ba
K(z_0(0)) &> 0\\
K(z_0(\mu)) &> 0\quad\textrm{for all}\quad \mu\in(\mu_*,1/8)\,.
\label{eq:kConditions}
\ea
It will be more convenient to parametrize the undetermined constant $a$ using $x$ as follows
\be
a\equiv \frac{(1-x) x}{4 (2-x) (x+1)}
\ee
so that
\be
K(z) = \frac{2 (2 z-1) (x-z) (x+z-1)}{(2-x) (x+1)}\,.
\ee
For $x=0$, we recover the original extremal functional $\beta$. Now, let us deform $\beta$ by increasing $x$ to some value $x\in(0,1/2)$. We find the following signs for $K(z)$ in the unit interval:
\ba
K(z)&>0\quad\textrm{for}\quad0<z<x\\
K(z)&<0\quad\textrm{for}\quad x<z<1/2\\
K(z)&>0\quad\textrm{for}\quad1/2<z<1-x\\
K(z)&<0\quad\textrm{for}\quad1-x<z<1\,.
\label{eq:kSigns}
\ea
Recall from equation \eqref{eq:z00} that $1/2<z_0(0)<1$. It then follows from the third line of \eqref{eq:kSigns} that the first line of \eqref{eq:kConditions} will be satisfied provided $x<1-z_0(0)$. Since $z_0(\mu)$ is monotonic decreasing, it follows from the first line of \eqref{eq:kSigns} that the second line of \eqref{eq:kConditions} will be satisfied provided $z_0(\mu_*)<x$. Since we want to minimize $\mu_*$, we must maximize $x$ subject to $x<1-z_0(0)$. In the optimal case, we get $x=1-z_0(0)$ and $z_0(\mu_*)=x$. In other words, the slope $\mu_*$ of the best upper bound that can be derived using $\widetilde{\beta}$ is the solution of
\be
z_0(\mu_*) = 1-z_0(0)\,.
\ee
There exists precisely one solution since $z_0(0)>1/2$, $z_0(1/8)=0$ and $z_0(\mu)$ is decreasing. The solution can be found to an arbitrary precision starting from the expressions for $h_1(z)$ in \eqref{eq:h01U} and \eqref{eq:h01V}. We find
\ba
\mu_* &= 0.1129140\ldots = \frac{1}{8.856295\ldots}\quad\textrm{for}\quad  A =U(1)^c\\
\mu_* &= 0.1176008\ldots = \frac{1}{8.503345\ldots}\quad\textrm{for}\quad  A =\mathrm{Vir}_c\,.
\ea
Finally, we should check that $\widetilde{\beta}_{\frac{c}{8}}[F^{ A }_{2 \mu c}]\geq 0$ for $\mu>1/8$. This follows from the inequality $Q^{\beta}_{\frac{c}{8}}(z) + a\,c\,Q^{\widetilde{\alpha}}_{\frac{c}{8}}(z)\geq 0$ for $0<z<1$, which indeed holds for the optimal choice of $a$.

\subsection{Comments on the optimal solution for general $c$}
It is natural to ask what are the optimal partition functions $\mathcal{Z}^{U}_c(\tau)$, $\mathcal{Z}^{V}_c(\tau)$ saturating the upper bounds on the gap for general $c$. One can find an approximation to the optimal solution using numerical bootstrap. In the limit of a large number of derivatives, this approximate spectrum approaches the true optimal spectrum. The following observations will be based on the results of the numerical bootstrap analysis of \cite{Afkhami-Jeddi:2019zci}, which were obtained for the Virasoro case. We expect similar comments apply also in the $U(1)^c$ case.

$\mathcal{Z}^{V}_c(\tau)$ can be expanded in the appropriate characters as follows
\ba
\mathcal{Z}^{V}_c(\tau) = \chi^{V}_{\textrm{vac}}(\tau) + \sum\limits_{n=0}^{\infty}\rho_n(c)\chi^{V}_{\Delta_n(c)}(\tau)\,,
\ea
where $\Delta_n(c)$ for $n=0,1,\ldots$ is the optimal spectrum and $\rho_n(c)$ are the corresponding degeneracies. If we were expanding instead in $sl(2,\mathbb{R})$ blocks and asking for the maximal gap above identity, the spectrum of the optimal solution would be that coming from the free fermion OPE
\be
\Delta^{\textrm{free}}_{n}(c) = \frac{c+4}{8}+n\,.
\label{eq:spectrumFree}
\ee
This is not be the optimal spectrum for the $U(1)^c$ and $\mathrm{Vir}_c$ problems unless $c=4,12$, which follows immediately from applying $\beta_{\frac{c}{8}}$ to the optimal solution and noting that the action of $\beta_{\frac{c}{8}}$ on the vacuum is non-vanishing.

Nevertheless, there is a sense in which the optimal solutions are close to the free fermion solution for any $c>12$. Indeed, we find that for any fixed $c$ the optimal spectrum asymptotes to the free spectrum \eqref{eq:spectrumFree} as $n\rightarrow\infty$. Define the anomalous dimensions by
\be
\gamma_n(c) = \Delta_n(c) - \Delta^{\textrm{free}}_n(c)\,.
\ee
Based on the numerics, we conjecture that for fixed $c>12$, $\gamma_n(c)$ is negative for all $n$, and approaches zero \emph{exponentially} fast as $n\rightarrow\infty$. More precisely, we find $\gamma_n(c)\sim - \exp\{-t \sqrt{n}[1+o(1)]\}$ as $n\rightarrow\infty$ with $c$ fixed, where $t$ is a $c$-dependent coefficient.

Since $-\gamma_n(c)$ is monotonic decreasing, the gaps between consecutive primaries $\Delta_{n+1}(c)-\Delta_{n}(c)$ are bigger than 1 and approach 1 as $n\rightarrow\infty$. It was recently proven in \cite{Mukhametzhanov:2019pzy} that the gap between consecutive Virasoro primaries is asymptotically bounded by $2\sqrt{\frac{3}{\pi^2}}\approx 1.1$, which is consistent with this behavior of $\mathcal{Z}^{V}_c(\tau)$. Since the optimal spectrum is asymptotically free at high energy, one can try to bootstrap it in perturbation theory at large $n$. There may also exist a perturbative expansion at large $c$ with $n/c$ fixed, which would help us derive the true asymptotic slope of the optimal bounds. We leave this interesting direction for future work.

Finally, note that the optimal partition functions $\mathcal{Z}^{V,U}_c(\tau)$ presumably do not correspond to physical CFTs for $c>4$ since the bootstrap bound on the gap coming from modular invariance of the full $\tau,\bar{\tau}$-dependent partition function is expected to be strictly stronger than the spinless bound discussed in the present paper (for $c>4$).

% !TEX root = ../ModularBootstrapV3.tex

\section{Complete Sets of Functionals and Fourier Interpolation in any Dimension}
\subsection{A complete set for the $sl(2,\mathbb{R})$ problem}
It is natural to ask whether the functional $\beta$, which played a central role in this work, is a part of a more general framework. An affirmative answer was provided in \cite{Mazac:2018ycv}, where it was explained that there is a distinguished basis for the space of functionals, and $\beta$ is simply one element of this basis. Remarkably, the analogous basis was recently constructed independently in the literature related to the sphere-packing problem \cite{InterpolationMath}, where it was shown to lead, among other impressive results, to Fourier interpolation formulas in $d=8,24$. The purpose of this section is to explain the connection between these sets of results, and sketch how to generalize them to arbitrary $d$.

We start by recalling some of the results of \cite{Mazac:2018ycv}. There, a basis of functionals acting on crossing-antisymmetric functions was constructed. We will find it useful to generalize the discussion to include functionals acting on crossing-symmetric functions too. Thus, define
\ba
F^{+}_{\Delta}(z) &= z^{-2\Df} G_{\Delta}(z) + (1-z)^{-2\Df} G_{\Delta}(1-z)\,,\\
F^{-}_{\Delta}(z) &= z^{-2\Df} G_{\Delta}(z) - (1-z)^{-2\Df} G_{\Delta}(1-z)\,,
\ea
where $G_{\Delta}(z)$ is the $sl(2,\mathbb{R})$ block \eqref{eq:sl2Block}. Set $\Delta_n=2\Df+2n+1$. The claim is that there exist linear functionals $\alpha^{+}_n$, $\beta^{+}_n$ for $n\in\mathbb{N}$ acting on functions satisfying $\mathcal{F}(z) = \mathcal{F}(1-z)$ and functionals $\alpha^{-}_n$, $\beta^{-}_n$ acting on functions satisfying $\mathcal{F}(z) =- \mathcal{F}(1-z)$, such that\footnote{$\mathbb{N}$ stands for the set of \emph{non-negative} integers. To avoid cluttered notation, we drop the $\Df$ label from all functionals.}
\ba
\alpha^{+}_n[F^+_{\Delta_m}] &= \delta_{m n}\qquad \alpha^{+}_n[\partial_{\Delta}F^+_{\Delta_m}] = 0\\
\beta^{+}_n[F^+_{\Delta_m}] &= 0\qquad\quad\, \beta^{+}_n[\partial_{\Delta}F^+_{\Delta_m}] = \delta_{m n}
\ea
and
\ba
\alpha^{-}_n[F^-_{\Delta_m}] &= \delta_{m n}\qquad \alpha^{-}_n[\partial_{\Delta}F^-_{\Delta_m}] = 0\\
\beta^{-}_n[F^-_{\Delta_m}] &= 0\qquad\quad\, \beta^{-}_n[\partial_{\Delta}F^-_{\Delta_m}] = \delta_{m n}
\ea
for all $m,n\in\mathbb{N}$. The functionals used earlier in this paper are specific elements of the basis. In particular: $\beta=\beta^{-}_0$, $\alpha=\alpha^{-}_0$, $\beta^+=\beta^{+}_0$. The functionals for higher values of $n$ are constructed in a similar way as the $n=0$ functionals. They always take the form \eqref{eq:funDef}, \eqref{eq:omegaPlus}, and the kernels satisfy essentially the same constraints discussed in Sections \ref{ssec:funConstruction}, \ref{ssec:alpha} and \ref{ssec:omegaPlus} for all $n\in\mathbb{N}$. Indeed, the functionals $\alpha^{\pm}_{n}$, $\beta^{\pm}_{n}$ can be thought of as a complete set of solutions of these constraints.

This basis of functionals can be used for example to prove interesting theorems about the distribution of primary operators in unitary solutions to crossing and for perturbative calculations around mean field theory. The existence of the basis is closely connected to a Lorentzian inversion formula for $sl(2,\mathbb{R})$ \cite{Mazac:2018qmi}.

It follows from the completeness of the functional basis that the functions $F^{\pm}_{\Delta}(z)$ with general $\Delta$ can be expanded in the basis consisting of functions $F^{\pm}_{\Delta_n}(z)$ and $\partial_{\Delta}F^{\pm}_{\Delta_n}(z)$
\ba
F^{+}_{\Delta}(z) &= \sum\limits_{n=0}^{\infty}
\left[\alpha^{+}_n[F^{+}_{\Delta}] F^{+}_{\Delta_n}(z) + \beta^{+}_n[F^{+}_{\Delta}] \partial_{\Delta}F^{+}_{\Delta_n}(z)\right]\\
F^{-}_{\Delta}(z) &= \sum\limits_{n=0}^{\infty}
\left[\alpha^{-}_n[F^{-}_{\Delta}] F^{-}_{\Delta_n}(z) + \beta^{-}_n[F^{-}_{\Delta}] \partial_{\Delta}F^{-}_{\Delta_n}(z)\right]\,.
\label{eq:completeness}
\ea

For most practical purposes, we need some control over the functional actions $\alpha^{\pm}_n[F^{\pm}_{\Delta}]$, $\beta^{\pm}_n[F^{\pm}_{\Delta}]$. For that and other reasons, it is useful to package these functions of $\Delta$ into generating functions, dubbed the Polyakov blocks
\ba
P^{+}_{\Delta}(z) &= G_{\Delta}(z) - \sum\limits_{n=0}^{\infty}
\left[\alpha^{+}_n[F^{+}_{\Delta}] G_{\Delta_n}(z) + \beta^{+}_n[F^{+}_{\Delta}] \partial_{\Delta}G_{\Delta_n}(z)\right]\\
P^{-}_{\Delta}(z) &= G_{\Delta}(z) - \sum\limits_{n=0}^{\infty}
\left[\alpha^{-}_n[F^{-}_{\Delta}] G_{\Delta_n}(z) + \beta^{-}_n[F^{-}_{\Delta}] \partial_{\Delta}G_{\Delta_n}(z)\right]\,.
\label{eq:polyakovP}
\ea
The Polyakov blocks contain the single-trace contribution $G_{\Delta}(z)$ as well as an infinite tower of double-trace contributions, whose coefficients are computed precisely by the basis functionals. The advantage of this presentation is that there is a constructive way to compute $P^{\pm}_{\Delta}(z)$ without knowing the integral kernels defining the individual functionals. Indeed, $P^{\pm}_{\Delta}(z)$ are computed by sums of appropriate exchange Witten diagrams in $AdS_2$ in the s-, t- and u-channel with fermionic external legs and exchange dimension $\Delta$. The functional actions can then be read off by expanding $P^{\pm}_{\Delta}(z)$ in conformal blocks.

Finally, note that equations \eqref{eq:completeness} are equivalent to saying that $P_{\Delta}^{+}(z)$ and $P_{\Delta}^{-}(z)$ are respectively antisymmetric and symmetric under crossing
\ba
&z^{-2\Df}P_{\Delta}^{+}(z) + (1-z)^{-2\Df}P_{\Delta}^{+}(1-z) = 0\\
&z^{-2\Df}P_{\Delta}^{-}(z) - (1-z)^{-2\Df}P_{\Delta}^{-}(1-z) = 0\,.
\ea

\subsection{A complete set for the $U(1)^c$ problem}
It is relatively straightforward to adjust the above discussion to the context of the modular bootstrap. We will focus on the modular bootstrap with $U(1)^c$ characters. In order to emphasize the connection to the sphere-packing literature, we will set $c=d/2$ and $\Delta=r^2/2$, where $d$ is the dimension of space and $r$ is the distance in $\mathbb{R}^{d}$. We start by defining the analogue of the functions\footnote{We change notation slightly with respect to \eqref{defF} and label the $\Phi(\tau)$ functions by $r$ rather than by $\Delta=r^2/2$.} $F^{\pm}_{\Delta}(z)$
\ba
\Phi^{+}_{r}(\tau) &= \frac{e^{i \pi r^2 \tau}}{\eta(\tau)^d} + \frac{e^{-i\pi\,r^2/\tau}}{\eta(-1/\tau)^d}\,,\\
\Phi^{-}_{r}(\tau) &= \frac{e^{i\pi r^2 \tau}}{\eta(\tau)^d} - \frac{e^{-i\pi\,r^2/\tau}}{\eta(-1/\tau)^d}\,.
\ea
Note that $\Phi^{+}_{|x|}(\tau)$ and $\Phi^{-}_{|x|}(\tau)$ are respectively even and odd under the $d$-dimensional Fourier transform in $x$. Recall from \eqref{eq:gFullExpansion1} that the characters can be expanded in the $sl(2,\mathbb{R})$ blocks
\be
\nu_{\frac{d}{2}}(\tau)\frac{e^{i \pi r^2 \tau}}{\eta(\tau)^d} = \sum\limits_{j=0}^{\infty}s_j(d,r)\lambda(\tau)^{-\frac{d}{8}} G_{r^2+2j}(\lambda(\tau))\,,
\label{eq:decompositionChar}
\ee
where $\nu_{c}(\tau)\equiv\left[2^8\lambda(\tau)(1-\lambda(\tau))\right]^{-\frac{c}{12}}$ and $s_j(d,r)$ are computable rational functions. It follows that $\Phi^{\pm}_{r}(\tau)$ can be expanded in $F^{\pm}_{\Delta}(\lambda(\tau))$
\ba
\nu_{\frac{d}{2}}(\tau)\Phi^{+}_{r}(\tau) &=
\sum\limits_{j=0}^{\infty}s_j(d,r) F^{+}_{r^2+2j}(\lambda(\tau))\\
\nu_{\frac{d}{2}}(\tau)\Phi^{-}_{r}(\tau) &=
\sum\limits_{j=0}^{\infty}s_j(d,r) F^{-}_{r^2+2j}(\lambda(\tau))\,.
\label{eq:changeOfBasisF}
\ea
Recall that $F^{\pm}_{\Delta_n}(z)$ and $\partial_{\Delta}F^{\pm}_{\Delta_n}(z)$ form a basis for the space of functions of $z$. Equations \eqref{eq:changeOfBasisF} then give a change of basis to $\Phi^{\pm}_{r_n}(\tau)$ and $\partial_{r}\Phi^{\pm}_{r_n}(\tau)$, where $r_n$ is the set of radii given by
\be
r^2_n = \Delta_n = \frac{d}{8}+2n+1\quad\textrm{for}\quad n\in\mathbb{N}\,,
\label{eq:radii}
\ee
where we used $\Df=d/16$. Explicitly, we find
\ba
\Phi^{\pm}_{r_n}(\tau) &=
\nu_{\frac{d}{2}}(\tau)^{-1}\sum\limits_{j=0}^{\infty}s_j(d,r_n) F^{\pm}_{\Delta_{n+j}}(\lambda(\tau))\\
\partial_r\Phi^{\pm}_{r_n}(\tau) &=
\nu_{\frac{d}{2}}(\tau)^{-1}\sum\limits_{j=0}^{\infty}\left[\partial_r s_j(d,r_n) F^{\pm}_{\Delta_{n+j}}(\lambda(\tau))+
2r_n s_j(d,r_n) \partial_{\Delta}F^{\pm}_{\Delta_{n+j}}(\lambda(\tau))
\right]\,.
\label{eq:PhiFromF}
\ea
Analogously to the previous section, we can now define a set of linear functionals $A^{\pm}_n$ and $B^{\pm}_n$ as the dual basis of $\Phi^{\pm}_{r_n}(\tau)$ and $\partial_{r}\Phi^{\pm}_{r_n}(\tau)$. Thus $A^{+}_n$ and $B^{+}_n$ act on functions $\tau$ satisfying $\mathcal{F}(\tau) = \mathcal{F}(-1/\tau)$ and $A^{-}_n$ and $B^{-}_n$ act on functions satisfying $\mathcal{F}(\tau) = -\mathcal{F}(-1/\tau)$, and we have
\ba
A^{+}_n[\Phi^{+}_{r_m}] &= \delta_{m n}\qquad A^{+}_n[\partial_{r}\Phi^{+}_{r_m}] = 0\\
B^{+}_n[\Phi^{+}_{r_m}] &= 0\qquad\quad\, B^{+}_n[\partial_{r}\Phi^{+}_{r_m}] = \delta_{m n}
\label{eq:dualityPlus}
\ea
and
\ba
A^{-}_n[\Phi^{}_{r_m}] &= \delta_{m n}\qquad A^{-}_n[\partial_{r}\Phi^{-}_{r_m}] = 0\\
B^{-}_n[\Phi^{-}_{r_m}] &= 0\qquad\quad\, B^{-}_n[\partial_{r}\Phi^{-}_{r_m}] = \delta_{m n}\,.
\label{eq:dualityMinus}
\ea
It follows from \eqref{eq:PhiFromF} that $A^{\pm}_n$ and $B^{\pm}_n$ are \emph{finite} linear combinations of the $sl(2,\mathbb{R})$ functionals $\alpha^{\pm}_n$ and $\beta^{\pm}_n$.\footnote{Here and in the following, we will use the tacit convention that when $\alpha^{\pm}_n$ and $\beta^{\pm}_n$ act on a function of $\tau$, we first need to multiply the function by the conversion factor $\nu_{\frac{d}{2}}(\tau)$ and then act with the functional.} More precisely, $B_n^{+}$ is a computable linear combination of $\beta_m^{+}$ with $0\leq m\leq n$ and $A_n^{+}$ is a linear combination of $\alpha_m^{+}$ and $\beta_m^{+}$ with $0\leq m\leq n$ (and similarly for $A_n^{-}$, $B_n^{-}$).

Analogously to \eqref{eq:completeness}, we can expand $\Phi^{\pm}_{r}(\tau)$ for general $r$ in the basis as follows
\ba
\Phi^{+}_{r}(\tau) &= \sum\limits_{n=0}^{\infty}
\left[A^{+}_n[\Phi^{+}_{r}] \Phi^{+}_{r_n}(\tau) + B^{+}_n[\Phi^{+}_{r}] \partial_{r}\Phi^{+}_{r_n}(\tau)\right]\\
\Phi^{-}_{r}(\tau) &= \sum\limits_{n=0}^{\infty}
\left[A^{-}_n[\Phi^{-}_{r}] \Phi^{-}_{r_n}(\tau) + B^{-}_n[\Phi^{-}_{r}] \partial_{r}\Phi^{-}_{r_n}(\tau)\right]\,.
\label{eq:completeness2}
\ea
Again by analogy with \eqref{eq:polyakovP}, we can define the Polyakov blocks for the $U(1)^{\frac{d}{2}}$ problem as the generating functions for the functional actions
\ba
\Pi^{+}_{r}(\tau) &= e^{i\pi r^2\tau} - \sum\limits_{n=0}^{\infty}
\left[A^{+}_n[\Phi^{+}_{r}] e^{i\pi r_n^2\tau} +2\pi i r_n\tau B^{+}_n[\Phi^{+}_{r}] e^{i\pi r_n^2\tau}\right]\\
\Pi^{-}_{r}(\tau) &= e^{i\pi r^2\tau} - \sum\limits_{n=0}^{\infty}
\left[A^{-}_n[\Phi^{-}_{r}] e^{i\pi r_n^2\tau} +2\pi i r_n\tau B^{-}_n[\Phi^{-}_{r}] e^{i\pi r_n^2\tau}\right]\,.
\label{eq:polyakovPI}
\ea
It follows from the completeness of the two functional bases that the Polyakov blocks for $U(1)^\frac{d}{2}$ can be decomposed into the Polyakov blocks for $sl(2,\mathbb{R})$ with exactly the same terms and coefficients that appear in the decomposition of $U(1)^{\frac{d}{2}}$ characters into conformal blocks, i.e. \eqref{eq:decompositionChar}
\be
\Pi^{\pm}_{r}(\tau) = \eta(\tau)^{d}\lambda(\tau)^{-\frac{d}{8}}\nu_{\frac{d}{2}}(\tau)^{-1}\sum\limits_{j=0}^{\infty}s_j(d,r) P^{\pm}_{r^2+2j}(\lambda(\tau))\,.
\ee
In particular, $\Pi^{\pm}_{r}(\tau)$ is still computed by exchange Witten diagrams in $AdS_2$, where the usual bulk to bulk propagator is replaced with an exchange of infinitely many particles of dimensions $r^2+2j$. Note that equations \eqref{eq:completeness2} are equivalent to
\ba
&\Pi^{+}_{r}(\tau) + (-i\tau)^{-\frac{d}{2}}\Pi^{+}_{r}(-1/\tau) = 0\\
&\Pi^{-}_{r}(\tau) - (-i\tau)^{-\frac{d}{2}}\Pi^{-}_{r}(-1/\tau) = 0\,.
\ea

\subsection{Fourier interpolation in any $d$}
The functionals $A^{\pm}_n$ and $B^{\pm}_n$ can be used to generalize the Fourier interpolation formula of \cite{InterpolationMath} to any $d$. It will be convenient to introduce shorthand notation for the actions of the basis functionals on the functions $\Phi^{\pm}_{r}(\tau)$ for general $r$
\ba
a^{+}_n(r) &= A^{+}_n[\Phi^{+}_{r}]\qquad b^{+}_n(r) = B^{+}_n[\Phi^{+}_{r}]\\
a^{-}_n(r) &= A^{-}_n[\Phi^{-}_{r}]\qquad b^{-}_n(r) = B^{-}_n[\Phi^{-}_{r}]\,.
\ea
$a^{\pm}_n(x)$ and $b^{\pm}_n(x)$ are radial Schwartz functions on $\mathbb{R}^d$. Here and in the following $a^{\pm}_n(x)$, $b^{\pm}_n(x)$ really means $a^{\pm}_n(|x|)$ and $b^{\pm}_n(|x|)$. $a^{+}_n(x)$ and $b^{+}_n(x)$ are even under the Fourier transform in $\mathbb{R}^d$, while $a^{-}_n(x)$ and $b^{-}_n(x)$ are Fourier-odd. Equations \eqref{eq:dualityPlus} and \eqref{eq:dualityMinus} translate into a specific structure of double zeros of the functions $a^{\pm}_n(r)$, $b^{\pm}_n(r)$ on the radii $r_m=\frac{d}{8}+2m+1$. Let us also introduce the functions
\ba
a_n(r) &= \frac{a_n^+(r)+ a_n^-(r)}{2}\qquad b_n(r) = \frac{b_n^+(r)+ b_n^-(r)}{2}\\
\widetilde{a}_n(r) &= \frac{a_n^+(r)- a_n^-(r)}{2}\qquad \widetilde{b}_n(r) = \frac{b_n^+(r)- b_n^-(r)}{2}\,.
\ea
These functions have the following structure of double zeros
\ba
a_n(r_m) &= \delta_{m n} \qquad a'_n(r_m) = 0\\
b_n(r_m) &= 0 \qquad\quad\; b'_n(r_m) = \delta_{m n}\\
\widetilde{a}_n(r_m) &= 0 \qquad\quad\;  \widetilde{a}'_n(r_m) = 0\\
\widetilde{b}_n(r_m) &= 0 \qquad\quad\;  \widetilde{b}'_n(r_m) = 0\,
\ea
where the prime denotes the radial derivative. Furthermore, $\widetilde{a}_n(x)$ and $\widetilde{b}_n(x)$ are the Fourier transforms of $a_n(x)$ and $b_n(x)$ respectively.

The key observation is that by taking the average of the two lines in \eqref{eq:completeness2}, we can write $e^{i\pi r^2\tau}$ for an arbitrary $r$ and $\tau$ as the linear combination
\ba
e^{i\pi r^2\tau} &=\sum\limits_{n=0}^{\infty}a_n(r)e^{i\pi r_n^2\tau} +
2\pi i \tau \sum\limits_{n=0}^{\infty}r_nb_n(r)e^{i\pi r_n^2\tau}+\\
&+\sum\limits_{n=0}^{\infty}\widetilde{a}_n(r)\frac{e^{-i\pi r_n^2/\tau}}{(-i\tau)^{\frac{d}{2}}}-
\frac{2\pi i}{\tau} \sum\limits_{n=0}^{\infty}r_n\widetilde{b}_n(r)\frac{e^{-i\pi r_n^2/\tau}}{(-i\tau)^{\frac{d}{2}}}\,.
\label{eq:EXPansion}
\ea
Consider now a linear functional $\omega$ acting on holomorphic functions of $\tau$ in the upper half-plane. Set
\be
f(x) = \omega[e^{i\pi x^2\tau}]\,.
\ee
$f(x)$ is a radial function on $\mathbb{R}^{d}$ and (under a suitable restriction on $\omega$) is in the Schwartz space.\footnote{For example, if $\omega$ is a finite linear combination of derivatives at $\tau=i$, then $f(x)$ is a Schwartz function.} Furthermore, every radial Schwartz function on $\mathbb{R}^d$ arises in this way from some $\omega$. Let us apply $\omega$ to \eqref{eq:EXPansion}. If $f(x)$ is Schwartz then $\omega$ can be chosen such that it commutes with the infinite sums on the RHS. We find
\ba
f(x) &=\sum\limits_{n=0}^{\infty}f(r_n)a_n(x) +
\sum\limits_{n=0}^{\infty}f'(r_n)b_n(x)+\\
&+\sum\limits_{n=0}^{\infty}\widehat{f}(r_n)\widetilde{a}_n(x)+
\sum\limits_{n=0}^{\infty}\widehat{f}'(r_n)\widetilde{b}_n(x)\,,
\ea
where $\widehat{f}(x)$ is the Fourier transform of $f(x)$. This is the Fourier interpolation theorem, proven rigorously in \cite{InterpolationMath} for $d=8$ and $d=24$. The theorem reconstructs an arbitrary radial Schwartz function $f(x)$ from the values $f(r_n)$, $f'(r_n)$, $\widehat{f}(r_n)$ and $\widehat{f}'(r_n)$ using the interpolating functions $a_n(x)$, $b_n(x)$, $\widetilde{a}_n(x)$ and $\widetilde{b}_n(x)$. Furthermore, the existence of the individual interpolating functions shows that there is no universal linear relation among $f(r_n)$, $f'(r_n)$, $\widehat{f}(r_n)$ and $\widehat{f}'(r_n)$ valid for all radial Schwartz functions.

We sketched how to generalize the theorem to arbitrary $d$. One of the main remaining technical challenges is a more explicit construction of the interpolating functions. As mentioned after equation \eqref{eq:polyakovPI}, the interpolating functions in any $d$ are in principle computed by infinite sums of exchange Witten diagrams in $AdS_2$. It will be important to study and simplify this prescription further, and fill in the remaining details of our sketch.

% !TEX root = ../ModularBootstrapV3.tex

\section{Discussion}\label{sec:Discussion}

We have described a connection between sphere packing in $\mathbb{R}^{d}$ and  modular bootstrap. The bootstrap problem with chiral algebra $U(1)^c$ maps to the Cohn-Elkies linear programming method for sphere packing in $d=2c$ dimensions. The bootstrap upper bound on the first primary operator, $\DeltaU(c)$, is related to the linear programming bound on the sphere packing density by equation \eqref{generalbound}.

More accurately, the usual modular bootstrap problem maps to the Fourier-odd part of the sphere packing method. According to a conjecture of Cohn and Elkies, supported by numerics and for which we have supplied new analytical examples, the bounds on the Fourier-odd part are identical to the full bounds.

It is worth emphasizing that our result is not exactly a map between CFTs and sphere packings, or even between the modular bootstrap and sphere packings. Rather it is a map between the modular bootstrap and the linear programming bounds on sphere packing. There are known to be additional constraints on sphere packings, and away from $d=1,2,8,24$, the linear programming bounds are not believed to be saturated or even close to saturated by actual packings (see for example \cite{viazovska2018sharp}). The situation with CFTs is similar; the spinless modular bootstrap is just one of an infinite family of consistency conditions, including also the spinning modular bootstrap and the crossing equations for correlation functions. We do not know whether the additional constraints on CFTs, beyond the spinless modular bootstrap, have any relationship to the additional constraints on sphere packings that go beyond the Cohn-Elkies method. It would clearly be of great interest, for example, to find a version of the modular bootstrap that maps to sphere packing in $d=3$, {\it i.e.} the Kepler problem.

The magic functions of sphere packing are related to the extremal functionals of the bootstrap. Acting with the extremal functional on the $U(1)^c$ characters for $c=4$ and $c=12$ produces the magic functions for sphere packing in 8 and 24 dimensions, which were found numerically by Cohn and Elkies \cite{cohn1} and analytically by Viazovska \cite{viazovska8} and Cohn et~al. \cite{viazovska24}. The optimality of the $E_8$ lattice for sphere packing in 8 dimensions maps to the statement that a CFT consisting of 8 free fermions with a diagonal GSO projection saturates the modular bootstrap bound, $\Delta_U(4) = 1$. In 24 dimensions, the optimality of sphere packing on the Leech lattice maps to the statement that the modular $j$-function, up to a constant, saturates the modular bootstrap bound $\Delta_U(12) = 2$. In this case, although this function satisfies all the requirements of the spinless modular bootstrap, it does not correspond to a full-fledged $c=12$ CFT. For both $c=4$ and $c=12$, the bootstrap bounds for the $U(1)^c$ algebra are identical to the bounds obtained using just the Virasoro algebra, appropriate to general 2D CFTs.

The extremal functionals for the spinless modular bootstrap at central charge $c=4,12$ are essentially the same as those for the four-point function bootstrap on a line, with external scaling dimension $\Df=\frac{1}{2},\frac{3}{2}$. Remarkably, an analytic construction of these functionals appeared in the conformal bootstrap literature independently in the same year when Viazovska constructed the magic functions \cite{Mazac:2016qev}. We have explained the equivalence of the two approaches in the main text.

The generalization of the four-point bootstrap functionals to arbitrary $\Df>0$ found in \cite{Mazac:2016qev,Mazac:2018mdx} can be applied to the modular bootstrap equation for any $c \geq 1$. This gives bounds on sphere packing in higher dimensions, generalizing Viazovska's construction to all $d\geq 2$. In bootstrap language, this proves the upper bounds $\Delta_{U,V}(c) < \frac{c+4}{8}$ for $ c\in(1,4)\cup(12,\infty)$. However, for these values of $c$ the functional is positive rather than zero on the vacuum term, so this bound is suboptimal. We constructed an improved functional that leads to a better bound at large $c$, but it is still suboptimal.

In the modular bootstrap, the limit of large central charge $c \to \infty $ is the regime relevant to quantum gravity in macroscopic $AdS_3$ space. The asymptotics of the sphere packing problem as $d \to \infty$ are also of interest to mathematicians. This limit may even have practical applications: Dense sphere packings in high dimensions correspond to highly effective classical error-correcting codes, with a large number of codewords \cite{thompson1983error}. To make the connection, we can view each codeword in a message as a vector in $\mathbb{R}^d$. If we send this message through a noisy channel, the receiver has the best chance of decoding it if the allowed codewords are spaced sufficiently far apart in $\mathbb{R}^d$. Thus the problem of constructing effective codes maps to the problem of packing non-overlapping spheres. 

\begin{figure}
\begin{center}
\includegraphics[scale=1.0]{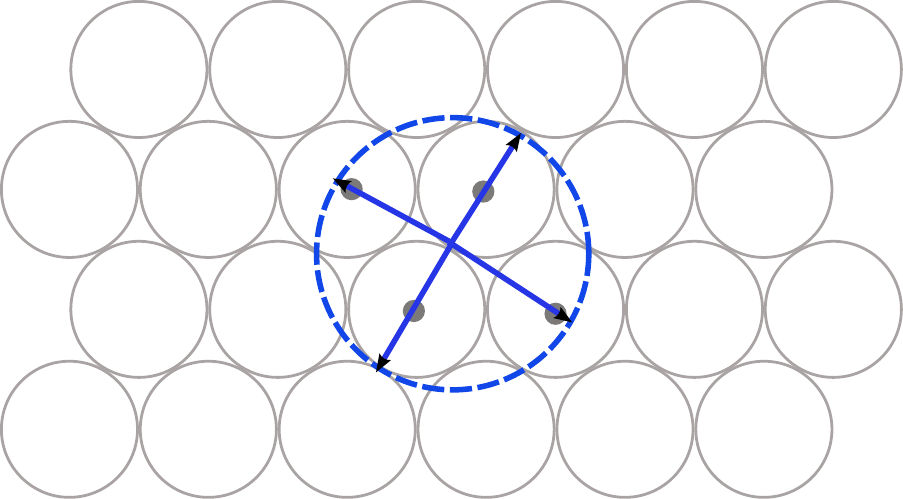}
\end{center}
\caption{
A dense sphere packing (solid gray) can be used to construct a large spherical code (dashed blue) by surrounding some of the centers with an auxiliary sphere, then projecting the enclosed centers onto that sphere.
\label{fig:sphericalcodes}
}
\end{figure}

Codes are, in fact, the origin of the best known upper bounds on the sphere packing density as $d \to \infty$. The starting point is a linear programming bound on spherical codes. A spherical code of minimal angle $\theta$ is a set of points on the unit sphere in $\mathbb{R}^d$, with no two points closer than $\theta$. An upper bound on the size of a spherical code also places an upper bound on the density of a sphere packing, by a geometric argument illustrated in Figure \ref{fig:sphericalcodes} \cite{kabatiansky1978bounds,ConwaySloane,zong2008sphere,cohn2014sphere}. The result, which can also be obtained directly from the Cohn-Elkies linear program \cite{cohn2014sphere}, is the Kabatiansky-Levenshtein bound reviewed in Section \ref{sec:Relation}. It implies the analytic bootstrap bound $\DeltaU \lesssim c/9.796$ as $c\to\infty$, significantly better than our bound from the four-point extremal functionals, $\DeltaU \lesssim c/8.856$.

This leaves open two very interesting questions. The first is whether the Kabatiansky-Levenshtein bound is the best asymptotic bound that can be obtained from linear programming. The second is whether something similar can be achieved for the Virasoro modular bootstrap, where the numerics suggest it should be possible to improve the bound to $\DeltaVir \lesssim c/9.08$ \cite{Afkhami-Jeddi:2019zci}. For the Virasoro bootstrap, we do not know of any analogue of spherical codes or the geometric argument above. 

Another approach that leads to strong but suboptimal bounds on sphere packing, or $U(1)^c$ bootstrap, uses functionals with compact support in Fourier space \cite{levkissing,cohn1}. This could be another route to improving the Virasoro bound at large $c$. There are some intriguing similarities between this approach and the very recent modular bootstrap results of \cite{Mukhametzhanov:2019pzy}.

Parisi and Zamponi have also described a connection between large-$d$ sphere packing and the physics of glassy systems \cite{Parisi2006,RevModPhys.82.789}.

It would also be interesting to explore whether our results can be extended to the full, spinning modular bootstrap. It is not clear whether this generalization has an analogue in sphere packing. Little is known about analytic extremal functionals for problems with spin. One rather trivial exception is the scalar gap maximization in the spinning modular bootstrap at $c=4$. In this case, the optimal bound and optimal theory are the same as for the spinlees modular bootstrap, {\it i.e.} eight free fermions, with scalar gap $\Delta^{s=0}_0 = 1$. The optimal functional for the spinning problem is identical to the one for the spinless problem {\it i.e.} $\beta_{\frac{1}{2}}$ acting on modular characters specialized to $\btau=-\tau$, so its action is spin-independent. In particular, the simple zero of $\beta_{\frac{1}{2}}$ at $\Delta=1$ sits at the scalar gap in the scalar sector, at the unitarity bound in the spin-one sector, and below the unitarity bound for all higher spins, so this is indeed the correct extremal functional.

It would be interesting to translate the recent advances in understanding of CFTs in Lorentzian signature to the context of the torus partition function and the modular bootstrap \cite{Hartman:2015lfa,Caron-Huot:2017vep,Kravchuk:2018htv}. Our work suggests a clear candidate for the analogue of the double discontinuity of the torus partition function, namely the double commutator of twist operators in the $\mathbb{Z}_2$ symmetric product orbifold.

Finally, returning to the question of whether sphere packings are related to interacting CFTs beyond the linear programming constraints discussed here, let us point out one more similarity. Free bosons compactified on a lattice have a perfectly regular spectrum, determined by the distances between pairs of lattice points. Interacting CFTs, on the other hand, have a quasi-regular structure of spins and scaling dimensions. The low-lying dimensions are irregular, depending strongly on the interactions, but regularity begins to appear at high spin \cite{Fitzpatrick:2012yx,Komargodski:2012ek} and scaling dimension \cite{Mukhametzhanov:2018zja}. This is due to the pattern of composite operators, enforced by locality and the bootstrap constraints. These features have a qualitative parallel in sphere packing. Lattice packings are perfectly regular, while general packings can be well approximated by a periodic packing, in which the low-lying spectrum is arbitrary --- defining the unit cell --- while eventually, the spectrum has regularities set by the underlying lattice. Perhaps this suggests a more complete, quantitative mapping between sphere packing and CFT. 

\bigskip

\acknowledgments

We thank Nima Afkhami-Jeddi, Ryan Bilotta, Miguel Paulos, Amir Tajdini, and Xinan Zhou for useful discussions. This work is supported in part by the Simons Foundation, as part of the Simons Collaboration on the Non-perturbative Bootstrap.  The work of LR is supported in part by NSF grant \# PHY1620628.

\bigskip

\appendix
% !TEX root = ../ModularBootstrapV3.tex

\section{Modular forms}\label{s:modforms}
For an introduction to modular forms, see \cite{zagier2008elliptic}.
$\mathbb{H}_+$ denotes the upper half complex plane. The modular group $\Gamma(1) \equiv PSL(2,\mathbb{Z})$ acts on $\tau \in \mathbb{H}_+$ as 
\be
\gamma \tau = \frac{a \tau + b}{c \tau + d} , \qquad
\gamma = 
\begin{pmatrix}
a & b\\
c& d
\end{pmatrix}  \in \Gamma(1) \ .
\ee
It is generated by
\be
T: \tau \to \tau + 1 , \quad S : \tau \to -1/\tau \ .
\ee
 The  level $N$ principal congruence subgroup is defined
\be
\Gamma(N) = \{ \gamma \in \Gamma(1)\ \  | \ \  \gamma = \mathbf{1}  \ \ \mbox{mod}\  N \} \ .
\ee
For example, $\Gamma(2)$ is generated by $T^2$ and $ST^2S$.

A \textit{modular form} of weight $k$  and congruence subgroup $\Gamma$ is a holomorphic function $f(\tau)$ that for $\gamma \in \Gamma$ transforms as
\be
f\left(a \tau + b \over c \tau + d\right) = (c\tau + d)^k f(\tau) , 
\ee
and for any $\gamma \in \Gamma(1)$ has a Fourier expansion
\be
f\left(a \tau + b \over c \tau + d\right)  = (c\tau + d)^k \sum_{n=0}^\infty c_f(\gamma, \frac{n}{n_\gamma} ) e^{2\pi i n \tau/ n_\gamma}
\ee
with  $n_\gamma \in \mathbb{N}$. There are no negative Fourier modes, so $f$ is holomorphic at $\tau = i\infty$. If this condition is relaxed so negative modes $n<0$ are allowed, $f$ is called \textit{weakly holomorphic}.

A \textit{quasimodular form} of weight $k$ and depth $s$ transforms as
\be
f\left(a \tau + b \over c \tau + d\right)  = (c\tau + d)^k \sum_{j=0}^s f_j(\tau) \left( c\over c\tau + d\right)^j \ .
\ee
These arise naturally by taking derivatives of modular forms.

The \textit{Eisenstein series}, for even $k \geq 2$, is 
\be
E_k(\tau) = \frac{1}{2\zeta(k)} \sum_{(c,d) \in \mathbb{Z}^2\backslash (0,0) }(c\tau + d)^{-k} \ .
\ee
For $k \geq 4$, this is a modular form of weight $k$ for $\Gamma(1)$. These are useful for building a basis for the finite-dimensional space of weight-$p$ modular forms for $\Gamma(1)$.

The case $k=2$ is special, because the sum does not converge absolutely. This leads to the anomalous transformation law
\be
E_2(-1/\tau) = \tau^2 E_2(\tau) - \frac{6 i \tau}{\pi} \ ,
\ee
so it is a weight-2, depth-1 quasimodular form. 

Any modular form for $\Gamma(1)$ can be written as a polynomial in $E_4$ and $E_6$. Similarly, any quasimodular form for $\Gamma(1)$ is a polynomial in $E_2,E_4,E_6$, with degree in $E_2$ equal to its depth.

An important example of a weakly holomorphic modular form for $\Gamma(1)$ is the modular $j$-function,
\be
j = \frac{1728E_4^3}{E_4^3 - E_6^2} = q^{-1} + 744 + 196884 q + \cdots
\ee
with $q = e^{2\pi i \tau}$. Both numerator and denominator have weight 12, so $j$ has weight zero. The denominator is proportional to the modular discriminant,
\be
\Delta = \frac{64 \pi^{12}}{27}(E_4^3-E_6^2) = (2\pi)^{12} \eta(\tau)^{24} \ ,
\ee
with $\eta = q^{1/24}\prod_{m=1}^\infty (1-q^m)$ the Dedekind $\eta$, which satisfies $\eta(\tau+1) = e^{i \pi /12}\eta(\tau)$ and $\eta(-1/\tau)=(-i\tau)^{1/2}\eta(\tau)$. $\Delta$ is the unique weight-12 form which vanishes at $q=0$.

We will also make use of the theta functions
\begin{align}
\theta_2 &= \theta_{10} = \sum_{n \in \mathbb{Z}}e^{i \pi(n + \half)^2\tau } \\
\theta_3 &=\theta_{00} =  \sum_{n \in \mathbb{Z}}e^{i \pi n^2 \tau} \\
\theta_4 &= \theta_{01} = \sum_{n \in \mathbb{Z}}(-1)^n e^{i \pi n^2 \tau}\ .
\end{align}
These satisfy the Jacobi identity
\be
\theta_3^4 = \theta_4^4 + \theta_2^4 
\ee
and transform as
\be
\theta_3^4(\frac{-1}{\tau}) = -\tau^2\theta_3^4 \ ,  \quad
\theta_2^4(\frac{-1}{\tau}) = -\tau^2\theta_4^4  \ ,  \quad
\theta_4^4(\frac{-1}{\tau}) = -\tau^2\theta_2^4  \ .
\ee
\be
\theta_2^4(\tau+1) = - \theta_2^4 \ , \quad
\theta_3^4(\tau+1) = \theta_4^4(\tau) , \quad
\theta_4^4(\tau+1) = \theta_3^4 \ .
\ee
$\theta_{i=1,2,3}^4$ are modular forms of weight 2 for $\Gamma(2)$. In particular, they are periodic under  $\tau \to \tau + 2$.

\section{Details of Analytic Extremal Functionals}\label{sec:kernelEven}
Let us explain in more detail how we used the constraints 1--4 after equation \eqref{eq:omegaPlus} to arrive at the solution for the weight-function $R_{\Df}^{\beta+}(z)$, shown in equation \eqref{eq:fEven}. Our discussion parallels Section 4.2 and Appendix A.2 of reference \cite{Mazac:2018mdx}, which can be consulted for more details. The first step is to note that when $\Df\in\mathbb{N}$, the constraints admit a solution of the form
\be
R_{\Df}^{\beta+}(z)=\frac{1}{w^{2}}\left[\left(\sum_{k=1}^{\Df+1} a_k w^k\right)\log(\mbox{$\frac{z-1}z$})+(2z-1)\sum_{k=0}^{\Df} b_k w^k\right],
\ee
where $w=z(z-1)$. In fact, the constraints uniquely fix the coefficients $a_k$ and $b_k$. For example for $\Df=1$, we find
\be
R_{1}^{\beta+}(z) = \frac{4 \log \left(\frac{z-1}{z}\right)}{\pi ^2}+\frac{2 (2 z-1) \left(z^2-z+1\right)}{\pi ^2 (z-1)^2 z^2}\,.
\ee
A general formula for $R_{\Df}^{\beta+}(z)$ and all $\Df$ can be found by first performing the following Mellin transform
\be
M_{\Df}(s) = -\frac{1}{2\cos(\pi s)}\int\limits_{0}^1\!\!dz \,[z(1-z)]^s(2z-1)\mathrm{Re}[R_{\Df}^{\beta+}(z)]\,.
\ee
By computing $M(s)$ for many low-lying values of $\Df\in\mathbb{N}$, we find experimentally that it always takes the form
\be
M_{\Df}(s) =-\frac{2^{-2 (\Df+s)} \Gamma (\Df+3) \Gamma \left(\frac{1}{2}-s\right) \Gamma (s-1) \Gamma (s+1) \Gamma (2 \Df+s+2)}{\pi ^2 \Gamma \left(\Df+\frac{3}{2}\right) \Gamma (\Df+s+1) \Gamma (\Df+s+2)}\,.
\label{eq:mellinGeneral}
\ee
It is natural to conjecture that this is the correct Mellin transform for general $\Df>0$. The transform can be inverted using the formula
\be
R_{\Df}^{\beta+}(z) = \int\limits_{\Gamma}\!\!\frac{ds}{2\pi i}\,[z(z-1)]^{1-s}M_{\Df}(s)\,,
\label{eq:mellinInversion}
\ee
where the contour $\Gamma$ goes from $s=-i\infty$ to $s=i\infty$ passing to the left of the poles of $M_{\Df}(s)$ at $s=1/2+n$ with $n=0,1,\ldots$ and to the right of all other poles, including $s=1$. We arrive at the general formula \eqref{eq:fEven} by inserting \eqref{eq:mellinGeneral} into \eqref{eq:mellinInversion}. The integral becomes a sum over residues at positive half-integer $s$. We checked that the resulting $R_{\Df}^{\beta+}(z)$ satisfies the constraints 1--4 for general $\Df$, and not just $\Df\in\mathbb{N}$.

\bibliography{sphere} 
\bibliographystyle{utphys}

\end{document}